\documentclass{article}
\usepackage{arxiv_style,times}
\AtBeginDocument{\renewcommand{\headrulewidth}{0pt}}
\usepackage[colorlinks=true,linkcolor=blue!60!black,citecolor=blue!60!black,urlcolor=blue!60!black]{hyperref}
\usepackage{url}
\usepackage[utf8]{inputenc}
\usepackage[T1]{fontenc}
\usepackage{amsmath,amssymb,amsfonts,amsthm}
\usepackage{microtype}
\usepackage[table]{xcolor}
\usepackage{booktabs}
\usepackage{multirow}
\usepackage{array}
\usepackage{graphicx}
\usepackage{caption}
\usepackage{subcaption}
\usepackage{adjustbox}
\usepackage[ruled,vlined,linesnumbered]{algorithm2e}
\usepackage{wrapfig}
\usepackage{enumitem}

\theoremstyle{plain}
\newtheorem{proposition}{Proposition}
\newtheorem{theorem}{Theorem}

\theoremstyle{definition}

\newtheorem{remark}{Remark}

\usepackage{xspace}
\newcommand{\sanai}{\textbf{\texttt{SANAI}}\xspace}

\definecolor{hdrblue}{RGB}{222,235,247}
\definecolor{rowgrey}{RGB}{244,244,244}
\definecolor{hlgreen}{RGB}{228,241,228}
\newcommand{\blk}{\rowcolor{hdrblue}}
\definecolor{oursgrey}{RGB}{235,235,235}
\newcommand{\hl}{\rowcolor{oursgrey}}

\title{\textbf{SANAI}: Secure ECC Distributed Keyed Parity Embedded in Weights for Protecting Neural Networks against Bit-Flip Attacks}

\author{\textbf{Habibur Rahaman, Daniel Lobo, Md Rezoan Ferdous, Swarup Bhunia} \\
Department of Electrical and Computer Engineering (ECE) \\
University of Florida \\
Gainesville, FL 32611, USA \\
\texttt{\{rahaman.habibur,daniel.lobo,mdrezoan.ferdous\}@ufl.edu} \\
\texttt{swarup@ece.ufl.edu}}

\iclrfinalcopy

\begin{document}
\maketitle

\begin{abstract}

Quantized neural networks store weights in dense memory, where bit flips caused by voltage scaling, aging, radiation, or deliberate attacks such as RowHammer can severely degrade model behavior. A single corrupted high-order bit can lead to erroneous classifications or nonsensical language-model outputs. Embedded error-correcting codes (ECC) avoid dedicated parity storage by hiding parity within model weights. However, existing schemes use largely uniform embedding across layers and bit planes, leave parity unsecured, and tolerate only low bit-error rates (BERs), typically up to 6\%. Moreover, an adversary aware of sensitive weight locations and the ECC scheme can bypass protection by deliberately exceeding the correction capability of targeted codewords. We propose \sanai, a sensitivity-guided, bit-plane-selective scheme that robustly protects critical model weights against bit flips with zero additional parity-storage overhead. \sanai secures ECC parity through keyed embedding across a two-dimensional weight space. Spread-transform dither modulation binds parity to a secret key, while keyed interleaving disperses concentrated attacks across codewords. We derive a survival law showing that robustness is governed by the correctable fraction $t/n$, rather than correction strength $t$ alone, and establish a closed-form bound on the protectable fraction of a model. Implemented as a streaming checker on an $8\times8$ systolic array, \sanai verifies weights on the load path with only 0.41\% area overhead and no additional memory. Across ten vision backbones, three quantization schemes, and ten language models ranging from 82M to 13B parameters, \sanai maintains accuracy within approximately one percentage point of the clean baseline at BERs up to 12\%, twice the maximum BER supported by prior embedded-ECC approaches. It also survives fine-tuning bit-exactly, neutralizes progressive and targeted bit-flip attacks, and exactly corrects AttentionBreaker and SilentStriker attacks.

\end{abstract}

\section{Introduction}
\label{sec:intro}

Deep networks increasingly run on energy-constrained edge accelerators \citep{jouppi2017datacenter,rahaman2025secure}, where weights quantized to eight bits or fewer sit in dense memory. Reduced-voltage operation buys efficiency and pays in bit errors \citep{chandramoorthy2019resilient,koppula2019eden,zhang2018thundervolt,stutz2021bit}; aging and radiation add more \citep{reagen2018ares,baumann2005radiation}, memory errors are ordinary in deployed fleets \citep{schroeder2011dram}, and RowHammer lets an adversary produce them on purpose \citep{kim2014flipping,yao2020deephammer}. Because a quantized weight packs its value into few bits, one high-order flip in a sensitive layer can collapse the whole model \citep{hong2019terminal,rakin2019bit}, and such fault-injection attacks are now a central threat to deployed AI systems \citep{rahaman2024secure,rahaman2026fortifying}, alongside tampering inside the ML infrastructure that loads and serves the model \citep{rahaman2026ispy}.

Uniform ECC guards every stored bit at a standing storage and energy price, although most bits barely move the loss. Embedded schemes such as E3-CODE \citep{kamineni2026e3} fold check bits into the weights themselves, but apply one policy to every layer and bit plane, leave the parity readable by anyone who holds the weights, and stop at a six percent BER. We instead concentrate the budget on the few fragile layers and damaging bits, hide the parity at zero added memory in bits the robust layers do not need, bind it to a key, and carry resilience well past six percent.

\begin{figure}[!t]
\centering
\includegraphics[width=0.85\linewidth]{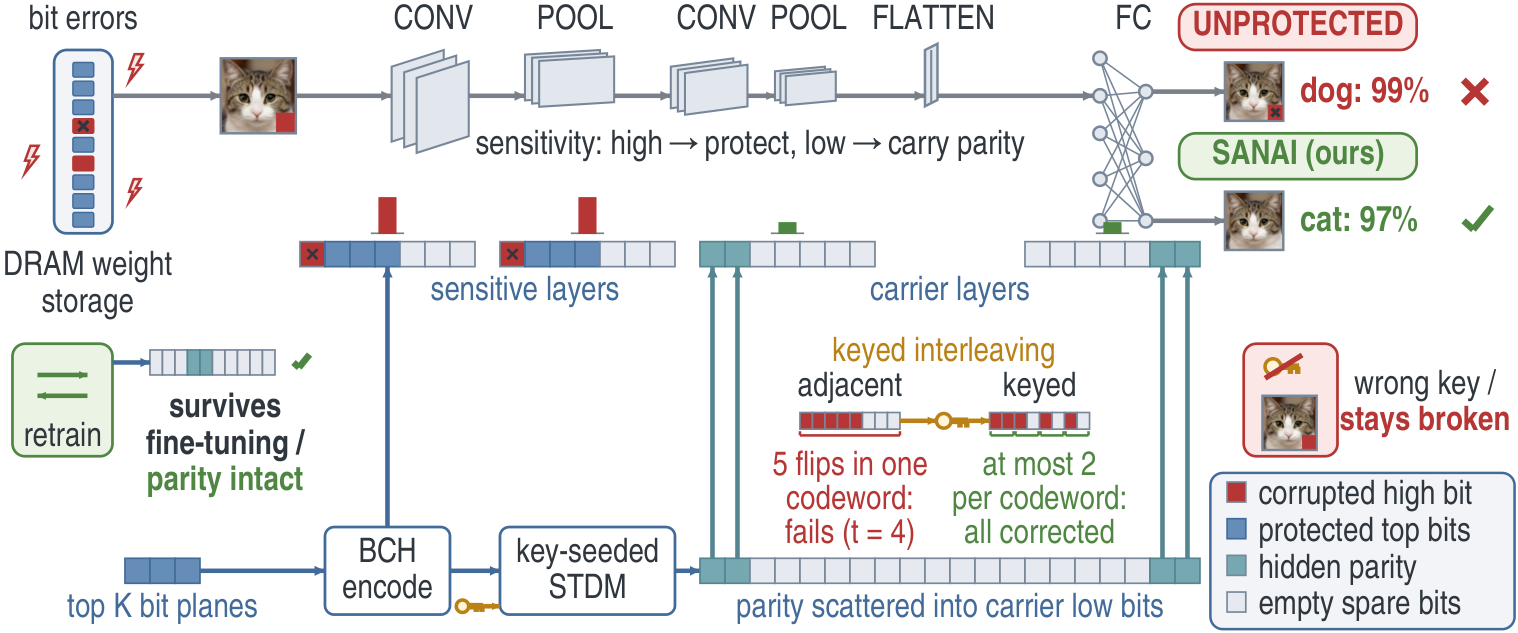}
\caption{\sanai overview. Bit errors in weight memory flip high-order bits and the unprotected model misclassifies (top). \sanai probes layer sensitivity (bars), BCH-encodes the top $K$ bit planes of the sensitive few, assigns weights to codewords by a keyed permutation, and hides the parity with key-seeded STDM in the low-order bits of robust carrier layers at zero added memory (bottom). Inset: the same attack flips overload one codeword under adjacent grouping but stay within $t$ under keyed interleaving. The parity survives fine-tuning and is unreadable without the key.}
\label{fig:pipeline}
\end{figure}

\sanai (Sensitivity-Aware Neural-network protection using Adaptive cross-layer STDM embeddIng, Figure~\ref{fig:pipeline}) identifies sensitive layers by direct fault injection, protects only their most significant bit planes with a BCH code, and embeds the parity into low-order bits of robust carrier layers using spread-transform dither modulation (STDM) seeded by a secret key. At read time the key regenerates the projection and BCH decoding repairs the high-order bits; a keyed permutation decides which weights share a codeword, so concentrated flips cannot overload any codeword. Because recovery is a pure read-path operation, the construction admits a streaming hardware checker on the weight bus of a systolic accelerator.

In summary, the paper makes the following key contributions.
\begin{itemize}[leftmargin=1.2em,itemsep=3pt,topsep=3pt]
\item \textbf{Sensitivity-guided ECC at zero added memory.} Flip damage in a quantized network is doubly concentrated: across layers, the top 1\% of weights by attack damage sit in layers holding 24.8\% of ResNet-18, and protecting only those layers keeps 95.0\% accuracy at 6\% BER; within a weight, damage decays geometrically with bit position. \sanai measures layer sensitivity by direct fault injection, replaces the integer program of prior embedded ECC with one $O(L\log L)$ sort, BCH-encodes only the top $K$ bit planes of the fragile layers, and moves the parity cross-layer into the low-order bits of robust carrier layers. It needs no retraining, no training data, and no extra storage.
\item \textbf{Secure ECC: parity bound to a key and codewords hidden by a key.} Key-seeded spread-transform dither modulation makes the hidden parity both the error-correcting carrier and an access control (Proposition~\ref{prop:access}): with a wrong key, ResNet-18 on CIFAR-10 falls to 44.6\% at 3\% BER and to chance by 8\%, and an ImageNet model is unrecoverable from the first error level. Keyed codeword interleaving disperses concentrated attacks with a balls-into-bins guarantee (Theorem~\ref{thm:interleave}): under the same 500 progressive-bit-search flips, uncorrectable codewords drop from 9 to 49 under adjacent grouping to zero on five of six settings, and the largest fault count per codeword drops to 2 or 3, inside $t=4$.
\item \textbf{A survival law and a protection ceiling.} Robustness is governed by the correctable fraction $t/n$ rather than by $t$ (Proposition~\ref{prop:survival}), which turns code selection into a design rule (Remark~\ref{rem:t}), and a closed-form supply and demand bound caps the protectable fraction of a model at $\rho^{\star}\approx48\%$ at zero added memory (Eq.~\ref{eq:ceiling}).
\item \textbf{Broad evaluation under random and adversarial faults.} Across ten vision backbones, three quantizers, and ten language models from 82M to 13B parameters, \sanai stays within about one point of clean accuracy up to 12\% BER, twice the range of prior embedded ECC. Under the burst protocol of prior work it retains 0.97 of its clean accuracy at 6\% BER against 0.94 for the strongest baseline, and its critical BER on ResNet-50/ImageNet is $4\times$ that baseline and $300{,}000\times$ the unprotected model. A 500-flip progressive bit search that leaves ResNet-18 on CIFAR-10 at 10.1\% is corrected to 94.9\%; targeted T-BFA success drops from 93 to 98\% to 0 to 8\%; AttentionBreaker, including an adaptive variant flipping $1.17\times10^{6}$ bits on an 8B model, is corrected to the exact clean perplexity and MMLU; a code-aware attacker needs $23\times$ more errors to match the undefended damage; and SilentStriker on Llama-3-8B-Instruct (MMLU 70.0 to 37.0\%) is detected and restored to 70.0\%. The parity survives fine-tuning bit-exactly at realistic learning rates.
\item \textbf{A load-path hardware checker.} The STDM embedding reduces exactly to a keystream-whitened least-significant-bit write (Proposition~\ref{prop:hwreduce}). Synthesized on an $8\times8$ systolic array in 45\,nm, the checker detects every single-bit flip including the sign bit at 0.41\% area, 0.63\% dynamic power, zero stall cycles, and zero added memory, and closes at 1.27\,ns. Against a measured conventional checker it needs no parity array (1.22\,MB for ResNet-18) and is $2.6\times$ faster than the correcting variant.
\end{itemize}

Section~\ref{sec:related} reviews fault tolerance, embedded ECC, bit-flip attacks, and data hiding. Section~\ref{sec:method} presents \sanai: the fault model, sensitivity-guided selection, bit-plane BCH encoding and the survival law, key-seeded STDM parity hiding, keyed interleaving, and the zero-memory ceiling. Section~\ref{sec:setup} gives the models, quantizers, fault models, and operating points. Section~\ref{sec:results} reports random bit errors on vision models (Section~\ref{sec:vision}), the comparison with prior embedded ECC and the key-gated access control (Section~\ref{sec:sota}), progressive and targeted attacks (Section~\ref{sec:attacks}), language models under random errors, AttentionBreaker, and SilentStriker (Section~\ref{sec:llm}), and fine-tuning and ablations (Section~\ref{sec:ablation}). Section~\ref{sec:hardware} presents the hardware checker, Section~\ref{sec:discussion} discusses scope and limitations, and Section~\ref{sec:conclusion} concludes. The appendices give the master tables behind every reported number (Appendix~\ref{app:master}), the algorithm and embedding details (Appendix~\ref{app:algo}), proofs (Appendix~\ref{app:proofs}), the qualitative comparison (Appendix~\ref{app:related}), experimental details (Appendix~\ref{app:setup}), ONNX portability (Appendix~\ref{app:onnx}), storage accounting (Appendix~\ref{app:storage}), extended vision, attack, and language-model results (Appendices~\ref{app:vision} to \ref{app:llm}), fine-tuning (Appendix~\ref{app:finetune}), ablations (Appendix~\ref{app:ablation}), and hardware details (Appendix~\ref{app:hardware}), each referenced from the main text where it is summarized.

\section{Background and Related Work}
\label{sec:related}

\paragraph{Fault tolerance for neural networks.}
Conventional ECC keeps extra check bits and corrects reliably at a large memory and energy cost \citep{reagen2018ares}. Activation clipping and error-aware retraining \citep{hong2019terminal,he2020defending} harden the model rather than the stored weights, need training data, and cannot recover the original parameters; selective protection \citep{koppula2019eden} leans on retraining. \sanai protects stored weights directly with no added memory and no retraining.

\paragraph{Embedded weight ECC.}
Weight nulling \citep{qin2017robustness}, opportunistic parity \citep{burel2021zero}, value-aware parity \citep{lee2022value}, and E3-CODE \citep{kamineni2026e3} conceal parity inside the weight bits, but apply one policy to every layer and bit plane, keep parity next to the data it protects, and stop at 6\% BER. \sanai differs on all three counts: placement follows a measured sensitivity ranking, protection is confined to the high-order planes of the fragile few, and the parity moves cross-layer into tolerant carriers (Table~\ref{tab:qual}, Appendix~\ref{app:related}).

\paragraph{Bit-flip attacks and detection.}
Progressive bit search (PBS) \citep{rakin2019bit}, targeted attacks \citep{rakin2020tbt,rakin2021tbfa}, and, on LLMs, AttentionBreaker \citep{das2024attentionbreaker} and SilentStriker \citep{xu2025silentstriker} show that a handful of chosen flips suffices. Detection defenses, including signature checks and runtime monitoring through built-in AI performance counters \citep{li2021radar,javaheripi2021hashtag,liu2023neuropots,rahaman2026bare}, raise an alarm rather than repair.

\paragraph{Data hiding.}
Quantization index modulation \citep{chen2001quantization} and its spread-transform variant \citep{malvar2003improved} embed information into a host by quantization, and weight watermarking \citep{uchida2017embedding,adi2018turning,darvish2019deepsigns} embeds ownership signatures; we repurpose STDM as a key-seeded carrier for functional parity.

Table~\ref{tab:qual} (Appendix~\ref{app:related}) sets these schemes side by side on the properties that decide deployability on an edge accelerator; \sanai is the only one that keeps the parity in-weight, moves it cross-layer, and binds it to a key.

\section{Method}
\label{sec:method}

\subsection{Problem Formulation and Fault Model}
\label{sec:problem}

A trained network has quantized weight tensors $\theta=\{\theta_\ell\}_{\ell=1}^{L}$ with $W=\sum_\ell|\theta_\ell|$ weights at $b$ bits each, so the stored bitstring $s(\theta)$ has $Wb$ bits. Weights are quantized per output channel, $q=\mathrm{clip}(\mathrm{round}(w/s),-2^{b-1},2^{b-1}-1)$ with $s=\max|w|/(2^{b-1}-1)$, in two's complement, so a flip in bit plane $j$ perturbs the integer by $2^{j}$ and a sign flip moves the weight by half its range. Memory is a cause-agnostic binary symmetric channel $\mathcal{F}_p$ flipping each stored bit independently with probability $p$, which is also the substrate for the burst protocol of prior work and the adversarial attackers of Sections~\ref{sec:attacks} and \ref{sec:llm}. We adopt a localized fault model in which errors land in the protected layers' memory regions while carrier regions stay clean (revisited in Section~\ref{sec:discussion}).

The defender applies an offline encoder $E:\theta\mapsto\theta'$ and a keyed read-time decoder $D_\kappa$. Writing $A(\cdot)$ for task accuracy, the design problem is
\begin{equation}
\max_{E,\,D_\kappa}\;\mathbb{E}_{\mathcal{F}_p}\big[A\big(D_\kappa(\mathcal{F}_p(E(\theta)))\big)\big]
\label{eq:objective}
\end{equation}
subject to (i) zero added memory, $|s(\theta')|=|s(\theta)|$; (ii) no retraining, so $E$ may not move a weight beyond its quantization step; and (iii) key-gated recovery, so for any wrong key the decoder output is no better than the unprotected corrupted model. \sanai exploits a measured double asymmetry. Across layers, damage is concentrated: scoring every weight of ResNet-18/CIFAR-10 by its attack-damage proxy puts the damage in the top 1\% of weights, and protecting the layers that hold them (24.8\% of the model) keeps 95.0\% at 6\% BER, while extending protection to the next 8\% of weights doubles coverage and parity for no gain (Master Table~\ref{tab:master_v1b}, rows V1.34 and V1.35); the remaining 91\% is the carrier pool. Within a weight, the damage of a flip decays geometrically with bit position.

\subsection{Sensitivity-Guided Layer Selection}
\label{sec:select}

For each layer $\ell$ we inject flips at a small probe rate $p_0$ into that layer alone and record the accuracy drop
\begin{equation}
S_\ell \;=\; A_{\mathrm{clean}}-\mathbb{E}_{\epsilon\sim\mathrm{Bern}(p_0)}\big[A(\theta_\ell\oplus\epsilon)\big].
\label{eq:sens}
\end{equation}
Layers are sorted by $S_\ell$ and the protected set $\mathcal{P}$ is grown greedily, most fragile first, to a coverage target $\rho$:
\begin{equation}
\mathcal{P}=\{\ell_{(1)},\dots,\ell_{(j)}\},\qquad j=\max\Big\{j:\textstyle\sum_{i\le j}|\theta_{\ell_{(i)}}|\le\rho W\Big\}.
\label{eq:select}
\end{equation}
The remaining layers form the carrier set $\mathcal{C}$, least sensitive first. In every network we studied a few early or narrow layers are fragile but hold few weights, while the large late layers are robust and hold most parameters, so the fragile layers are cheap to protect and the robust ones supply carrier bits. For LLM attacks we score layers with a gradient-informed probe $\sum|g\odot w|$, since magnitude is blind to the small attention projections attacks target (Section~\ref{sec:llm-attacks}); random-BER runs use magnitude. Selection is one sort, $O(L\log L)$ (Appendix~\ref{app:algo}).

\subsection{Bit-Plane BCH Encoding and a Survival Law}
\label{sec:bch}

Within each protected layer we encode only the top $K$ bit planes: groups of $G$ weights form $GK$-bit messages for a binary BCH code \citep{bose1960class} over $\mathrm{GF}(2^7)$ ($x^7+x+1$, parent length $n=127$), shortened to the message. The strong code is shortened from the parent BCH$(127,50,t{=}12)$ and stores 48 message bits plus 88 parity bits per codeword, so $K=4$ gives $G=12$; the light code is shortened from BCH$(127,99,t{=}4)$ with 96 message and 32 parity bits, $G=24$ at $K=4$. The $G\le t$ property that makes any sign-bit attack correctable (Section~\ref{sec:interleave}) holds for the $K=4$, $G=12$, $t=12$ configuration used in all LLM experiments, not for $K=3$ ($G=16$) or the light code ($G=24$). The governing quantity for survival is not $t$ but the correctable fraction $f=t/n$ of the parent.

\begin{proposition}[Codeword survival]
\label{prop:survival}
If the channel flips each of the $n$ codeword bits independently with probability $p$, the decoder fails exactly when $X\sim\mathrm{Bin}(n,p)$ exceeds $t$, so $P_{\mathrm{fail}}=\sum_{i=t+1}^{n}\binom{n}{i}p^{i}(1-p)^{n-i}$, and for any $p<f$, $P_{\mathrm{fail}}\le\exp\!\big(-2n(f-p)^2\big)$.
\end{proposition}

Survival is exponentially reliable whenever $p<f$ (Remark~\ref{rem:t} turns this into a design rule). Protecting $K>1$ planes widens the flat region further, since residual errors below the corrected planes perturb a weight by at most $2^{\,b-K-1}$ steps. Under an adversary the relevant quantity is the per-codeword count $f_c$: a codeword survives iff $f_c\le t$, random faults keep $\mathbb{E}[f_c]$ small, and a targeted attack maximizes $\max_c f_c$ (Appendix~\ref{app:proofs}).

\subsection{Key-Seeded STDM Parity Hiding}
\label{sec:stdm}

The parity is embedded into the low-order bits of carrier weights by STDM seeded by a secret key $\kappa$. To embed one bit $\beta$ into a group of $L_s$ carrier weights $w$, the key generates a direction $v\in\{-1,+1\}^{L_s}$ and a dither $d$; the projection $p=w\cdot v+d$ is snapped to one of two interleaved lattices of step $\Delta$ selected by $\beta$ and written back with the minimum-norm update:
\begin{equation}
p^{\star}=\Delta\Big\lceil\tfrac{p-d-\beta\Delta/2}{\Delta}\Big\rceil+d+\tfrac{\beta\Delta}{2},\qquad
w'=w+\tfrac{p^{\star}-p}{L_s}\,v,\qquad
\hat\beta=\Big\lfloor\tfrac{w'\cdot v+d}{\Delta/2}\Big\rfloor\bmod 2 .
\label{eq:stdm}
\end{equation}
Each carrier weight moves by at most $\Delta/2$, one low-bit step, so the embedding is invisible in accuracy; it generalizes to $B$ bits per symbol (Appendix~\ref{app:algo}).

\begin{proposition}[Recovery and access control]
\label{prop:access}
(a) With the correct key, $v$ and $d$ cancel exactly, so $\hat\beta=\beta$ whenever the read-time perturbation of the projection stays below $\Delta/2$. (b) With a wrong key $\kappa'\neq\kappa$ the regenerated direction satisfies $\mathbb{E}[v\cdot v']=0$, the projection is uncorrelated with $\beta$, each parity bit is recovered with probability $\tfrac12$, and all $P_{\mathrm{bits}}$ parity bits are recovered with probability $2^{-P_{\mathrm{bits}}}$, negligible at the operating points of this paper ($P_{\mathrm{bits}}>10^{6}$).
\end{proposition}

Property (a) is the recovery guarantee and (b) the access-control guarantee: without the key the parity is noise. At read time the key regenerates $(v,d)$, the parity is de-projected, and the BCH decoder corrects the top $K$ planes (Algorithm~\ref{alg:main}, Appendix~\ref{app:algo}).

\begin{figure}[!t]
\centering
\includegraphics[width=0.72\textwidth]{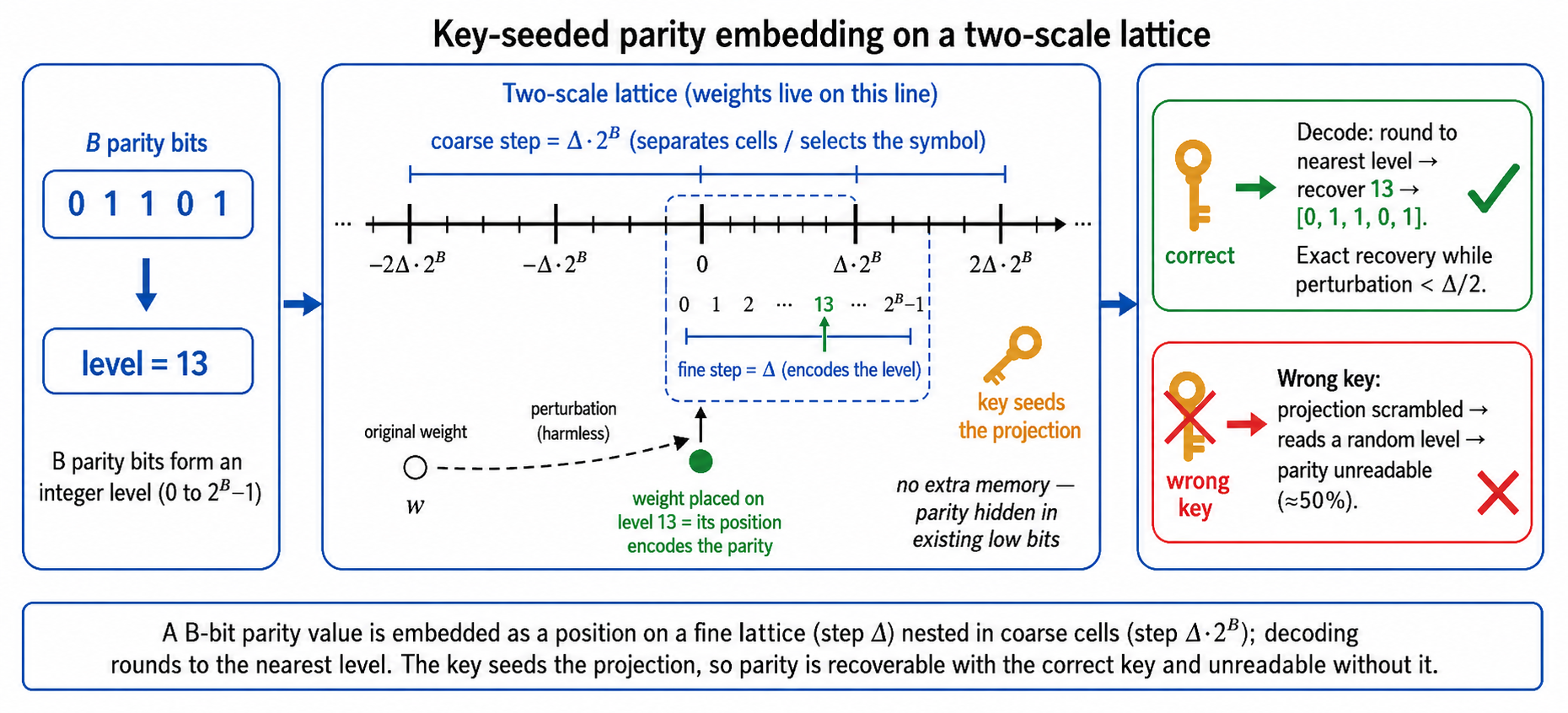}
\caption{Key-seeded parity embedding on a two-scale lattice. Recovery is exact while the perturbation stays below $\Delta/2$; without the key the parity reads at chance.}
\label{fig:embedding}
\end{figure}

Figure~\ref{fig:embedding} shows the lattice geometry for the multi-bit case.

\subsection{Keyed Codeword Interleaving}
\label{sec:interleave}

In the default layout a codeword covers $G$ adjacent weights, the weak point against a targeted attacker: PBS concentrates flips on sensitive weights that cluster in a few early layers and drive $f_c$ far above $t$. We break adjacency with a keyed permutation $\pi$ of each layer's protected indices, assigning weight $i$ to codeword
\begin{equation}
\mathrm{cw}(i)=\lfloor\pi(i)/G\rfloor,\qquad \mathrm{pos}(i)=\pi(i)\bmod G .
\label{eq:interleave}
\end{equation}
Adjacent weights land in unrelated codewords, encoder and decoder share the mapping through $\kappa$, and no parity or memory is added.

\begin{figure}[!t]
\centering
\includegraphics[width=0.8\textwidth]{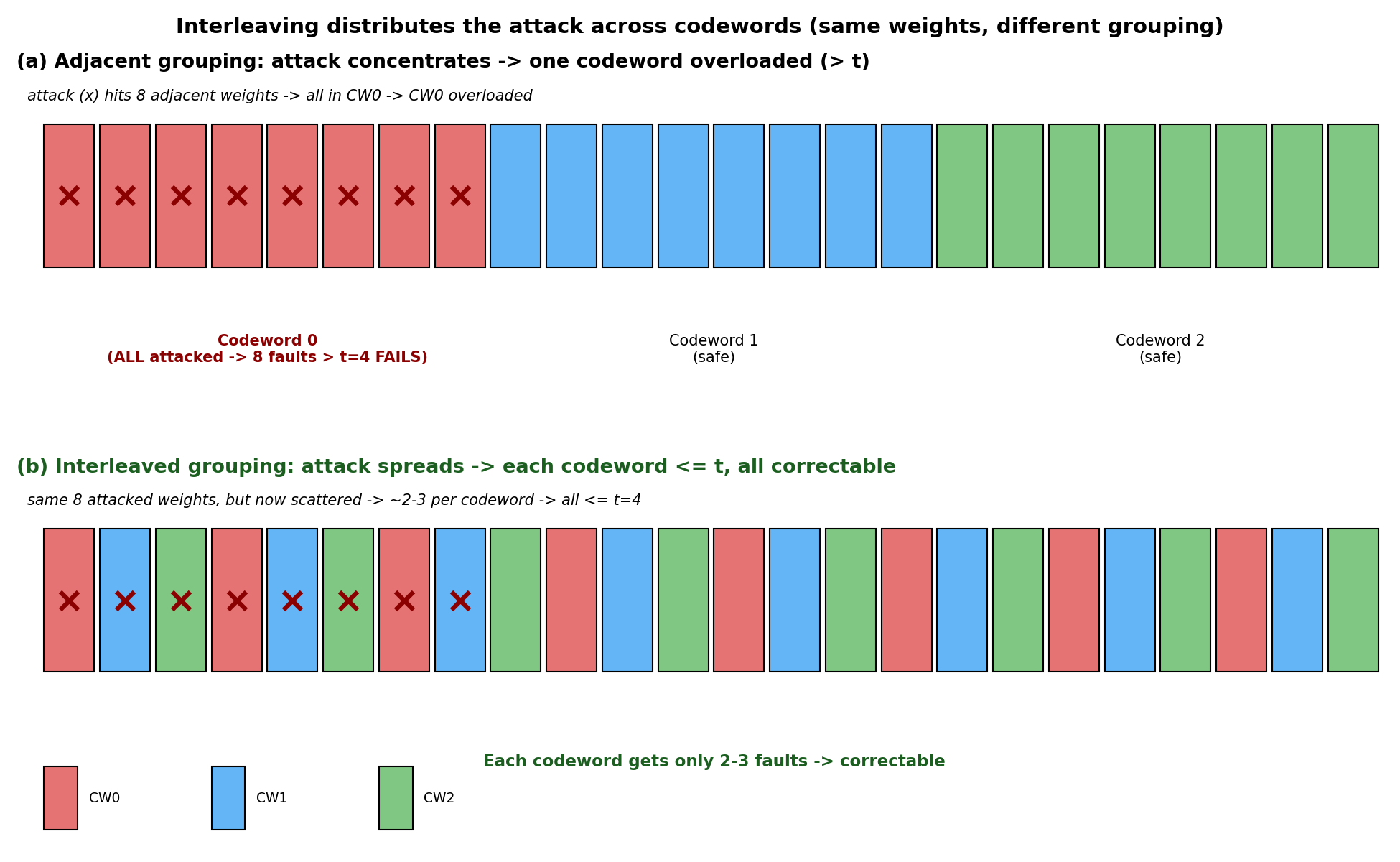}
\caption{Interleaving mechanism. Top, adjacent grouping: the attacked weights fall into one codeword and overload it ($f_c>t$). Bottom, keyed interleaving: the same weights are scattered by the keyed permutation so that each codeword holds at most a few faults, all within $t$. The measured fault maps for a real 500-flip attack are in Figures~\ref{fig:faultmap} and \ref{fig:bitbox}.}
\label{fig:interleave}
\end{figure}

Figure~\ref{fig:interleave} contrasts the two layouts on the same attacked weights.

\begin{theorem}[Interleaving bound]
\label{thm:interleave}
Let $\pi$ be a uniformly random keyed permutation and let the attacker choose $F$ flips without knowledge of $\pi$. Over $M$ codewords, $\mathbb{E}[\max_c f_c]=O\big(F/M+\log M/\log\log M\big)$, which is $O(1)$ for $F\ll M\log M$, so with high probability $\max_c f_c\le t$.
\end{theorem}

A second guarantee needs no key: a sign-bit attack flips one plane per weight and each weight belongs to one codeword, so it injects at most $G$ errors per codeword, and $G\le t$ makes every sign-bit attack of any size exactly correctable on protected layers (Section~\ref{sec:llm-attacks}).

\subsection{Zero-Memory Feasibility and the Protection Ceiling}
\label{sec:capacity}

The code is systematic, so only the parity, $p$ bits per $k_s$-bit message ($88/48$ strong, $32/96$ light), must be evicted to carriers. With $W_p=\rho W$ protected weights, replication $R$ ($R=1$ by default), $W_c=(1-\rho)W$ carriers with $h$ spare host bits each, and STDM at $B\le h$ bits per group of $L_s$ carriers, demand and supply are
\begin{equation}
D=W_pKR\,\tfrac{p}{k_s},\quad C_{\mathrm{STDM}}=\tfrac{W_cB}{L_s}\le W_ch,\quad
D\le C_{\mathrm{STDM}}\iff\rho\le\rho^{\star}=\tfrac{Bk_s}{Bk_s+L_sKRp}.
\label{eq:ceiling}
\end{equation}
The parity cost is $r=\rho Kp/k_s$ bits per weight ($1.10$ predicted versus 1.05 to 1.10 measured, Table~\ref{tab:storage}, Appendix~\ref{app:storage}), with zero added memory regardless of $r$, and $\rho^{\star}\approx0.48$: at most 48\% of a model can be protected before the parity no longer fits. Above $\rho^{\star}$ the encoder places the remaining parity in the low-order bits of the protected layers themselves; those bits sit inside the fault region but belong to the codeword and are corrected like any other, so coverage can exceed $\rho^{\star}$ (52.8\% in Section~\ref{sec:ablation}) until every layer's host bits are exhausted (INT4, Appendix~\ref{app:storage}).

\begin{remark}[Choice of $t$]
\label{rem:t}
Proposition~\ref{prop:survival} and Eq.~\ref{eq:ceiling} pull in opposite directions: a larger $t$ widens the safe region $p<t/n$, but each additional correctable error costs roughly $m$ more parity bits, which lowers $\rho^{\star}$. The right $t$ is the largest one whose parity still fits at the coverage the fragile layers need. At $m=7$, $t=12$ gives $f=9.4\%$ with 88 parity bits and $\rho^{\star}\approx48\%$, above the 20 to 40\% coverage used here, whereas $t=18$ would raise $f$ to $14.2\%$ at a parity cost that pushes $\rho^{\star}$ below that range; enlarging the field never helps, since at $m=8$ even $t=20$ gives a smaller $f$ ($7.8\%$) at higher cost.
\end{remark}

\section{Experimental Setup}
\label{sec:setup}

\paragraph{Vision.}
CIFAR-10, CIFAR-100 \citep{krizhevsky2009cifar}, and ImageNet-1K \citep{deng2009imagenet}. Ten backbones from 3.5M to 85.8M parameters: ResNet-18/50 and VGG-16 \citep{he2016deep,simonyan2014very}, MobileNet-V2 and EfficientNet-B0 \citep{sandler2018mobilenetv2,tan2019efficientnet}, Wide ResNet-50-2 \citep{zagoruyko2016wide}, RegNet-Y-16GF \citep{radosavovic2020designing}, and Swin-T, MaxViT-T, and ViT-B/16 \citep{liu2021swin,tu2022maxvit,dosovitskiy2020image} (sizes in Table~\ref{tab:arch}, Appendix~\ref{app:setup}). We use post-training quantization (PTQ) with per-output-channel scales and two quantization-aware methods, Jacob \citep{jacob2018quantization} and LSQ \citep{esser2019learned}, at INT8 and INT6 precision.

\paragraph{Language.}
Ten decoder-only transformers from 82M to 13B parameters: DistilGPT-2 and GPT-2 \citep{sanh2019distilbert,radford2019gpt2}, Qwen2-0.5B/1.5B, Qwen2.5-7B \citep{yang2024qwen2,yang2024qwen25}, Llama-3.2-1B/3B, Llama-3-8B-Instruct, Llama-3.1-8B \citep{dubey2024llama3}, and Llama-2-13B \citep{touvron2023llama2}, at signed INT8 (per-tensor below 7B, per-channel above); embedding and output-head matrices are excluded from injection, protection, and carriers. We report perplexity on WikiText-2/103 \citep{merity2017wikitext} and 5-shot accuracy on MMLU \citep{hendrycks2021mmlu}, ARC \citep{clark2018arc}, HellaSwag \citep{zellers2019hellaswag}, and TruthfulQA \citep{lin2022truthfulqa} (2,000 questions, three seeds).

\paragraph{Fault models and operating points.}
\emph{Random BER} flips bits uniformly at a target rate; \emph{burst} injection follows the E3-CODE protocol; \emph{PBS} \citep{rakin2019bit} is gradient-guided and untargeted (500 flips); \emph{T-BFA} \citep{rakin2021tbfa} forces a source class to a target class, measured by attack success rate (ASR); AttentionBreaker and SilentStriker are re-implemented on \sanai's own INT8 tensors. The mechanism is instantiated at three operating points (Table~\ref{tab:ops}, Appendix~\ref{app:setup}): the strong $t=12$ code at $K=3$ ($K=4$ for depthwise networks and ImageNet backbones, $K=5$ for ViT-B/16), 20\% coverage (40\% on ImageNet) for random BER; the light $t=4$ code at $K=4$ ($G=24$), 10\% coverage, keyed interleaving for concentrated attacks; and the strong code at $K=4$ ($G=12\le t$) for LLMs (gradient ranking and keyed interleaving for attacks; magnitude ranking, adjacent grouping under random BER). STDM uses $B=5$, $L_s=1$, $R=1$, and $p_0=2\%$ unless noted.

Experiments run on four NVIDIA RTX 6000 Ada GPUs; protecting the largest vision model takes a few minutes, dominated by the sensitivity probes. The hardware checker is verified with Synopsys VCS and synthesized with Synopsys Design Compiler on the gscl45nm library.

\section{Results}
\label{sec:results}

\subsection{Random Bit Errors Across Architectures and Datasets}
\label{sec:vision}

\begin{figure}[!t]
\centering
\begin{minipage}[t]{0.32\linewidth}\centering
\includegraphics[width=\linewidth]{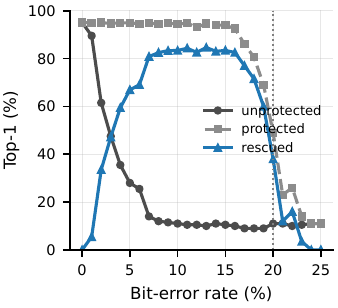}
\end{minipage}\hfill
\begin{minipage}[t]{0.32\linewidth}\centering
\includegraphics[width=\linewidth]{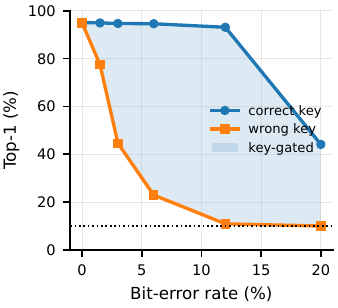}
\end{minipage}\hfill
\begin{minipage}[t]{0.32\linewidth}\centering
\includegraphics[width=\linewidth]{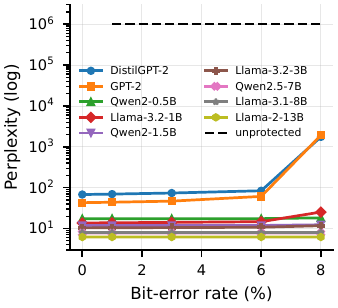}
\end{minipage}
\caption{Left: unprotected, protected, and rescued accuracy over a 0 to 25\% BER sweep (ResNet-18, CIFAR-10); the rescue plateaus and falls only near the code limit at 20\%. Middle: recovery with the correct versus a wrong key; the shaded gap is recoverable only with the key. Right: protected LLM perplexity versus BER on WikiText-2 (log scale); the unprotected models sit at the $10^{6}$ ceiling.}
\label{fig:crossover}
\end{figure}

Table~\ref{tab:main} (block 1) reports protected top-1 accuracy; the embedding itself is invisible, since at 0\% BER every protected CIFAR checkpoint matches the unprotected INT8 model to two decimals (Master Table~\ref{tab:master_v1}). On CIFAR-10 the three classic architectures stay within about one point of clean through a twelve percent BER; on CIFAR-100 the residual and plain models hold within one to two points and EfficientNet-B0 through ten percent. Jacob QAT matches PTQ; LSQ's learned scales are more fragile, so a matched budget collapses at 12\% (27.4\%) while $2.5\times$ the coverage restores 85.8\% (rows V1.6, V1.7). Unprotected ResNet-18 is at random by an eight percent BER while the protected model holds near 95\% until the code limit near twenty percent (Figure~\ref{fig:crossover}, left). Proposition~\ref{prop:survival} predicts this: with parity in clean carriers a codeword exposes 48 bits, so the failure probability is $1.3\times10^{-4}$ at 3\%, $3.4\times10^{-6}$ at 6\%, and 0.15 at 20\%, where the rescue curve falls.

\begin{table}[!t]
\centering
\caption{Vision results at INT8 and zero added memory. Block 1: protected top-1 (\%) under random BER, strong code ($K=3$; $K=4$ for EfficientNet-B0 and ImageNet backbones other than ResNet-18; $K=5$ for ViT-B/16). Block 2: burst protocol of \citet{kamineni2026e3}, prior embedded ECC as reported on their reference models versus \sanai on ours, so retention normalized to each method's own clean accuracy is the comparable metric. Block 3: adversarial bit flips, light code, $K=4$, 10\% coverage (T-BFA rows: strong code, high-bin selection). Every row traces to Master Tables~\ref{tab:master_v1}, \ref{tab:master_v2}, and \ref{tab:master_a1}.}
\label{tab:main}\label{tab:baseline}\label{tab:attacks}
\footnotesize
\setlength{\tabcolsep}{4pt}
\renewcommand{\arraystretch}{1.1}
\begin{adjustbox}{max width=\textwidth}
\begin{tabular}{l l c *{6}{>{\centering\arraybackslash}p{0.1\textwidth}}}
\toprule
\addlinespace[2.5pt]
\blk \multicolumn{9}{l}{\textbf{Block 1. Random BER: protected top-1 (\%)}} \\[2pt]
\textbf{Model} & \textbf{Data} & \textbf{Cov.} & \textbf{1\%} & \textbf{3\%} & \textbf{6\%} & \textbf{8\%} & \textbf{10\%} & \textbf{12\%} \\
\midrule
\addlinespace[2pt]
ResNet-18 & CIFAR-10  & 19.3\% & 94.95 & 94.66 & 94.54 & 94.46 & 94.36 & 93.06 \\
ResNet-50 & CIFAR-10  & 19.2\% & 94.87 & 94.79 & 94.51 & 94.54 & 94.10 & 93.70 \\
VGG-16    & CIFAR-10  & 20.0\% & 90.12 & 89.98 & 89.87 & 89.72 & 89.76 & 89.19 \\
ResNet-18 & CIFAR-100 & 14.0\% & 77.30 & 77.10 & 76.82 & 76.69 & 76.11 & 75.86 \\
ResNet-50 & CIFAR-100 & 36.1\% & 75.64 & 75.36 & 75.32 & 75.41 & 75.36 & 74.71 \\
EffNet-B0 & CIFAR-100 & 30.0\% & 63.29 & 62.49 & 61.96 & 61.46 & 59.97 & 53.22 \\
ResNet-18 & ImageNet  & 40.4\% & 76.05 & 76.14 & 75.95 & 75.70 & 75.78 & 75.70 \\
ResNet-50 & ImageNet  & 40.9\% & 82.72 & 82.52 & 82.41 & 82.06 & 81.31 & 80.88 \\
MaxViT-T  & ImageNet  & 40.9\% & 88.14 & 88.06 & 88.05 & 87.97 & 87.69 & 87.55 \\
Swin-T    & ImageNet  & 40\%   & 86.50 & 86.28 & 85.42 & 85.34 & 84.92 & 83.44 \\
Wide RN-50-2 & ImageNet & 40\% & 83.91 & 83.97 & 84.06 & 83.91 & 83.58 & 83.58 \\
RegNet-Y-16GF & ImageNet & 40\% & 85.66 & 85.50 & 85.05 & 84.64 & 84.62 & 84.48 \\
ViT-B/16 & ImageNet & 43.5\% & 85.50 & 85.42 & 85.28 & 84.44 & 83.25 & 77.89 \\
\addlinespace[2pt]
\midrule
\addlinespace[3.5pt]
\blk \multicolumn{9}{l}{\textbf{Block 2. Burst errors versus prior embedded ECC: top-1 (\%) at 1.5 / 3 / 6\% BER}} \\[2pt]
\textbf{Method} & \multicolumn{2}{c}{\textbf{ResNet-18 C10}} & \multicolumn{2}{c}{\textbf{EffNet-B0 C10}} & \multicolumn{2}{c}{\textbf{ResNet-50 IN}} & \multicolumn{2}{c}{\textbf{MobileNet-V2 IN}} \\
\midrule
\addlinespace[2pt]
Weight nulling & \multicolumn{2}{c}{82.0 / 19.0 / 0.05} & \multicolumn{2}{c}{82.0 / 2.9 / 0.0} & \multicolumn{2}{c}{75.9 / 18.8 / 0.4} & \multicolumn{2}{c}{69.9 / 5.5 / 0.0} \\
Opportunistic parity & \multicolumn{2}{c}{81.7 / 81.4 / 10.9} & \multicolumn{2}{c}{80.3 / 64.8 / 9.8} & \multicolumn{2}{c}{75.8 / 73.8 / 0.0} & \multicolumn{2}{c}{65.8 / 58.1 / 0.1} \\
Value-aware parity & \multicolumn{2}{c}{82.0 / 81.2 / 77.5} & \multicolumn{2}{c}{82.0 / 77.9 / 22.6} & \multicolumn{2}{c}{75.9 / 75.1 / 65.1} & \multicolumn{2}{c}{69.9 / 66.0 / 20.9} \\
E3-CODE & \multicolumn{2}{c}{81.8 / 81.8 / 78.8} & \multicolumn{2}{c}{82.3 / 78.6 / 55.6} & \multicolumn{2}{c}{75.9 / 75.8 / 70.0} & \multicolumn{2}{c}{69.9 / 66.7 / 36.9} \\
E3-CODE-H & \multicolumn{2}{c}{82.0 / 81.9 / 81.3} & \multicolumn{2}{c}{82.7 / 82.3 / 80.1} & \multicolumn{2}{c}{76.1 / 75.9 / 74.7} & \multicolumn{2}{c}{71.2 / 70.5 / 65.5} \\
\hl \textbf{\sanai (ours)} & \multicolumn{2}{c}{94.9 / 94.7 / \textbf{94.5}} & \multicolumn{2}{c}{83.8 / 83.0 / \textbf{81.4}} & \multicolumn{2}{c}{82.2 / 81.6 / \textbf{80.5}} & \multicolumn{2}{c}{78.4 / 78.1 / \textbf{77.8}} \\
\addlinespace[2pt]
\multicolumn{9}{l}{Retention at 6\% BER, normalized to each method's own baseline over eight settings: 0.01, 0.06, 0.58, 0.72, 0.94, \textbf{0.97}.} \\
\addlinespace[2pt]
\midrule
\addlinespace[3.5pt]
\blk \multicolumn{9}{l}{\textbf{Block 3a. Progressive bit search, 500 flips, keyed interleaving, light code: top-1 (\%)}} \\[2pt]
\textbf{Model} & \textbf{Data} & \textbf{Cov.} & \multicolumn{3}{c}{\textbf{Unprotected}} & \multicolumn{3}{c}{\textbf{\sanai}} \\
\midrule
\addlinespace[2pt]
ResNet-18 / ResNet-50 & CIFAR-10 & 10\% & \multicolumn{3}{c}{10.1 / 35.0} & \multicolumn{3}{c}{\textbf{94.9} / \textbf{94.8}} \\
ResNet-18 / ResNet-50 & CIFAR-100 & 10\% & \multicolumn{3}{c}{1.2 / 0.9} & \multicolumn{3}{c}{\textbf{75.5} / \textbf{75.8}} \\
\addlinespace[2pt]
\midrule
\addlinespace[3.5pt]
\blk \multicolumn{9}{l}{\textbf{Block 3b. Same flipped bits, adjacent / keyed grouping, ResNet-18 / CIFAR-10}} \\[2pt]
\textbf{Flips} & & & \multicolumn{2}{c}{\textbf{cw $>t$}} & \multicolumn{2}{c}{\textbf{max/cw}} & \multicolumn{2}{c}{\textbf{Top-1 (\%)}} \\
\midrule
\addlinespace[2pt]
100 & & & \multicolumn{2}{c}{3 / 0} & \multicolumn{2}{c}{12 / 2} & \multicolumn{2}{c}{66.0 / \textbf{95.1}} \\
500 & & & \multicolumn{2}{c}{22 / 0} & \multicolumn{2}{c}{14 / 3} & \multicolumn{2}{c}{10.1 / \textbf{94.9}} \\
\addlinespace[2pt]
\midrule
\addlinespace[3.5pt]
\blk \multicolumn{9}{l}{\textbf{Block 3c. Targeted attack (T-BFA): attack success rate (\%)}} \\[2pt]
\textbf{Model} & \multicolumn{2}{l}{\textbf{Data, pair}} & \multicolumn{3}{c}{\textbf{Unprotected}} & \multicolumn{3}{c}{\textbf{\sanai}} \\
\midrule
\addlinespace[2pt]
ResNet-18 & \multicolumn{2}{l}{CIFAR-10, $3\!\to\!5$} & \multicolumn{3}{c}{95.7} & \multicolumn{3}{c}{\textbf{4.9}} \\
ResNet-18 & \multicolumn{2}{l}{CIFAR-100, $90\!\to\!70$ / $90\!\to\!80$} & \multicolumn{3}{c}{93.0 / 94.0} & \multicolumn{3}{c}{\textbf{0.0} / \textbf{0.0}} \\
ResNet-50 & \multicolumn{2}{l}{ImageNet, 132-pair mean} & \multicolumn{3}{c}{$\sim$98} & \multicolumn{3}{c}{$\sim$8} \\
\addlinespace[3pt]
\bottomrule
\end{tabular}
\end{adjustbox}
\end{table}

The same pipeline applies unchanged to ImageNet with $K=4$ at a 40\% budget (Table~\ref{tab:main}, block 1): ResNet-50, MaxViT-T, RegNet-Y-16GF, Wide ResNet-50-2, and Swin-T hold a normalized accuracy of 0.96 or above at twelve percent, and the plain ViT-B/16 needs a five-plane budget and holds 0.97 through ten percent and 0.91 at twelve.

\subsection{Comparison with State-of-the-Art Embedded ECC}
\label{sec:sota}

\begin{figure}[!t]
\centering
\begin{minipage}[t]{0.32\textwidth}\centering
\includegraphics[width=\linewidth]{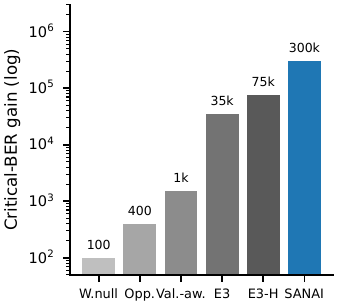}
\end{minipage}\hfill
\begin{minipage}[t]{0.32\textwidth}\centering
\includegraphics[width=\linewidth]{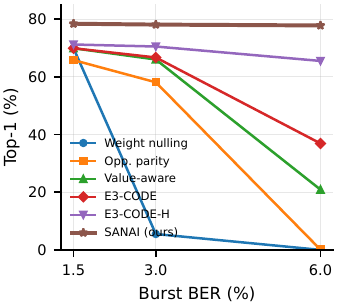}
\end{minipage}\hfill
\begin{minipage}[t]{0.32\textwidth}\centering
\includegraphics[width=\linewidth]{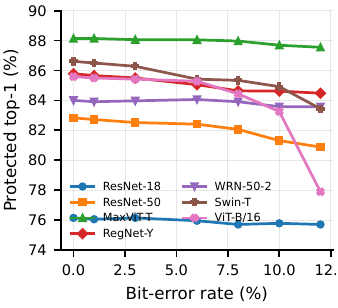}
\end{minipage}
\caption{Left: critical-BER improvement ratio over the unprotected model, ResNet-50 on ImageNet (log scale); baseline values as reported by \citet{kamineni2026e3}. Middle: MobileNet-V2 on ImageNet under the burst protocol against prior embedded ECC; the strongest prior method bends downward while ours stays flat. Right: modern architectures over a 0 to 10\% sweep (ImageNet, INT8); ViT-B/16 omitted for scale.}
\label{fig:critical}
\end{figure}

Table~\ref{tab:baseline} (block 2; full table in Table~\ref{tab:baseline_full}, Appendix~\ref{app:vision}) compares \sanai with five prior embedded ECC schemes under their burst-error protocol: weight nulling and opportunistic parity collapse by six percent (normalized retention 0.01 and 0.06), value-aware parity and E3-CODE retain 0.58 and 0.72, and E3-CODE-H, the strongest, 0.94. Because the reported baselines use lower-accuracy reference models (82.0\% clean for ResNet-18 on CIFAR-10 versus our 95\%), the comparable metric is normalized retention: \sanai reaches 0.99 at 3\% and 0.97 at 6\%, above E3-CODE-H, so the advantage is retention, not a higher start. With $K=4$, MobileNet-V2 and EfficientNet-B0 reach 0.95 and 0.97 at six percent. On the critical-BER metric, the highest rate at which accuracy stays near clean, \sanai reaches $6\times10^{-2}$ on ResNet-50/ImageNet, $300{,}000\times$ the unprotected $2\times10^{-7}$ and $4\times$ beyond E3-CODE-H at $1.5\times10^{-2}$ (Figure~\ref{fig:critical}, left). On MobileNet-V2/ImageNet the strongest prior method bends downward with the burst rate while \sanai stays flat (middle), and the modern ImageNet backbones hold their clean accuracy over the full 0 to 10\% sweep (right).

\paragraph{Access control: parity is unreadable without the key.}
\label{sec:security}
We protect a model once and attempt recovery with the correct host key and a wrong key across a 0 to 20\% sweep (rows V2.14 and V2.15 of Master Table~\ref{tab:master_v2}; Figure~\ref{fig:crossover}, middle). With the correct key ResNet-18 on CIFAR-10 stays above 94\% to ten percent; with the wrong key the de-projected parity is noise and accuracy collapses to 44.6\% by three percent and to chance by eight. On ImageNet the wrong key yields essentially zero from the first error level, so a copied model cannot be restored under faults and the protection doubles as a lightweight anti-piracy control.

\subsection{Progressive and Targeted Bit-Flip Attacks on Vision Models}
\label{sec:attacks}

PBS concentrates on the most sensitive weights, which forces it into the protected set: 76 to 100\% of its flips land on \sanai-protected weights (Table~\ref{tab:attacks}, block 3a). Under the adjacent layout the same 500 flips overload the code: 9 to 49 codewords exceed $t=4$ across six ResNet settings, peaking at 10 to 15 faults in one, and accuracy collapses to 0 to 35\%; more coverage does not help, so the problem is grouping, not budget. The compact networks are harder: at $t=4$ MobileNet-V2 reaches only 17\%, and even the strong code leaves them at 64 and 10\% on CIFAR-10 (Master Table~\ref{tab:master_a1}).

\begin{figure}[!t]
\centering
\includegraphics[width=0.98\textwidth]{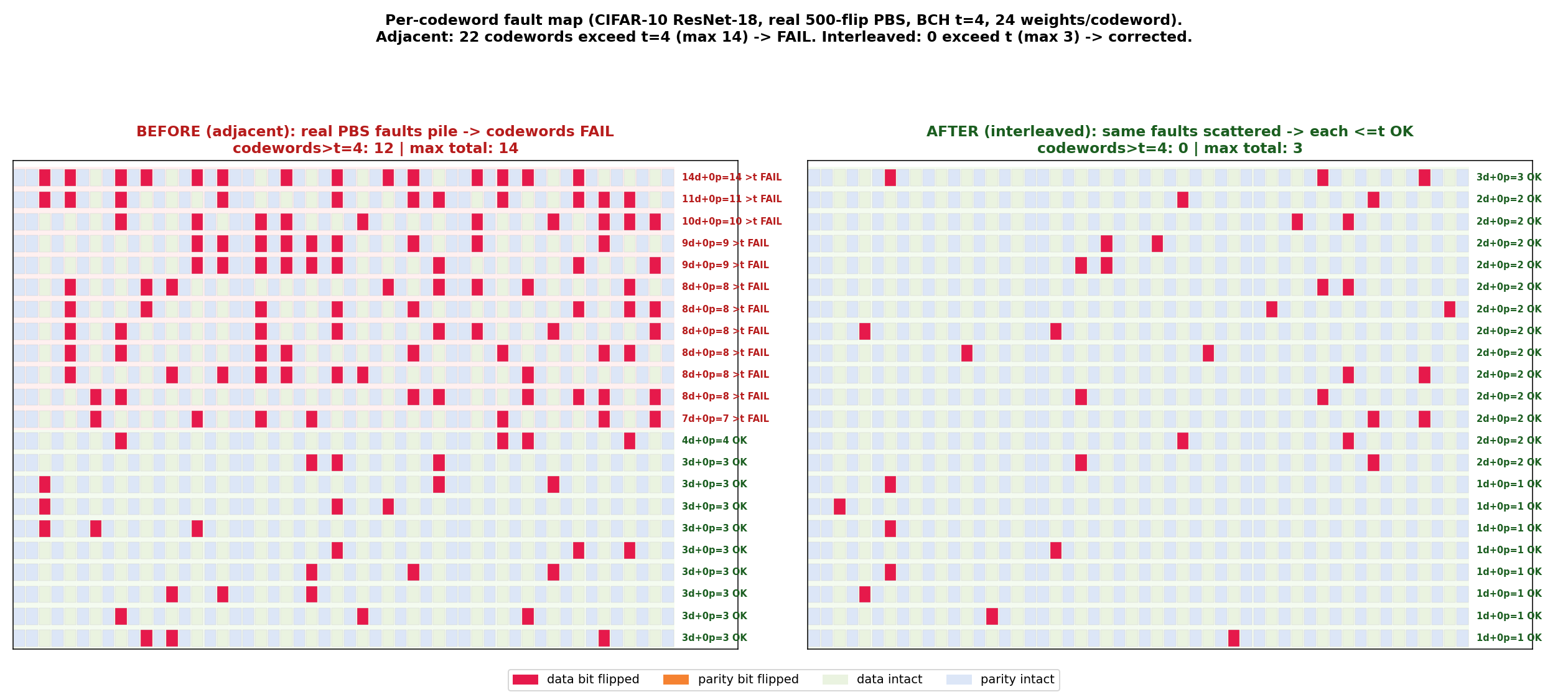}
\caption{Per-codeword fault map under a real 500-flip PBS attack (ResNet-18, CIFAR-10, light code $t=4$, 24 weights per codeword). Each row is one codeword and each red cell a flipped bit; the panels show the most heavily hit codewords. Left, adjacent grouping: the attack piles faults into a few codewords, 22 of which exceed $t=4$ (up to 14 faults) and fail. Right, keyed interleaving: the identical faults are scattered by the keyed permutation so that no codeword exceeds 3 faults and every codeword is corrected. The attacker flipped the same bits in both panels.}
\label{fig:faultmap}
\end{figure}

\begin{figure}[!t]
\centering
\includegraphics[width=0.72\textwidth]{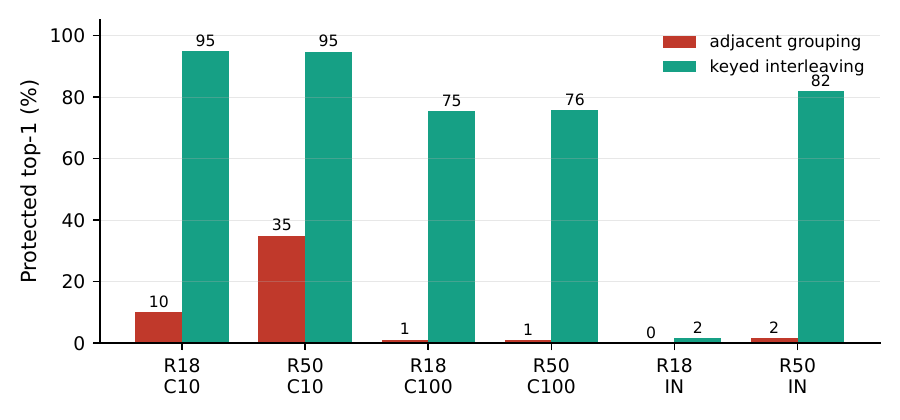}
\caption{Protected top-1 accuracy under the 500-flip PBS attack, adjacent grouping versus keyed interleaving, across six architecture and dataset settings (light code, $K=4$, 10\% coverage). Interleaving restores near-clean accuracy in every case except the coverage-limited ResNet-18 on ImageNet.}
\label{fig:intsummary}
\end{figure}

Keyed interleaving removes the concentration itself (Table~\ref{tab:attacks}, block 3b; Figures~\ref{fig:faultmap} and \ref{fig:intsummary}): with the identical flipped bits, uncorrectable codewords fall to zero on five of six settings and the maximum faults per codeword to 2 or 3, within $t=4$, as Theorem~\ref{thm:interleave} predicts, and accuracy returns to the clean range. The exception is ResNet-18 on ImageNet: every codeword is corrected but accuracy stays at 1.5\% because at 10\% coverage the unprotected 90\% of the model is itself corrupted by 500 flips, a coverage limit rather than a code failure (10 to 500 flip sweep in Table~\ref{tab:sweep_r18}, Appendix~\ref{app:attacks}). T-BFA (Table~\ref{tab:attacks}, block 3c) must edit the classifier weights, where 97\% of its flips land and which \sanai protects: ASR collapses from 93 to 98\% to 0 to 8\% (Figure~\ref{fig:tbfa}). The bit-level fault maps and the independent fault-map pass are in Appendix~\ref{app:attacks}.

\begin{figure}[!t]
\centering
\includegraphics[width=0.9\textwidth]{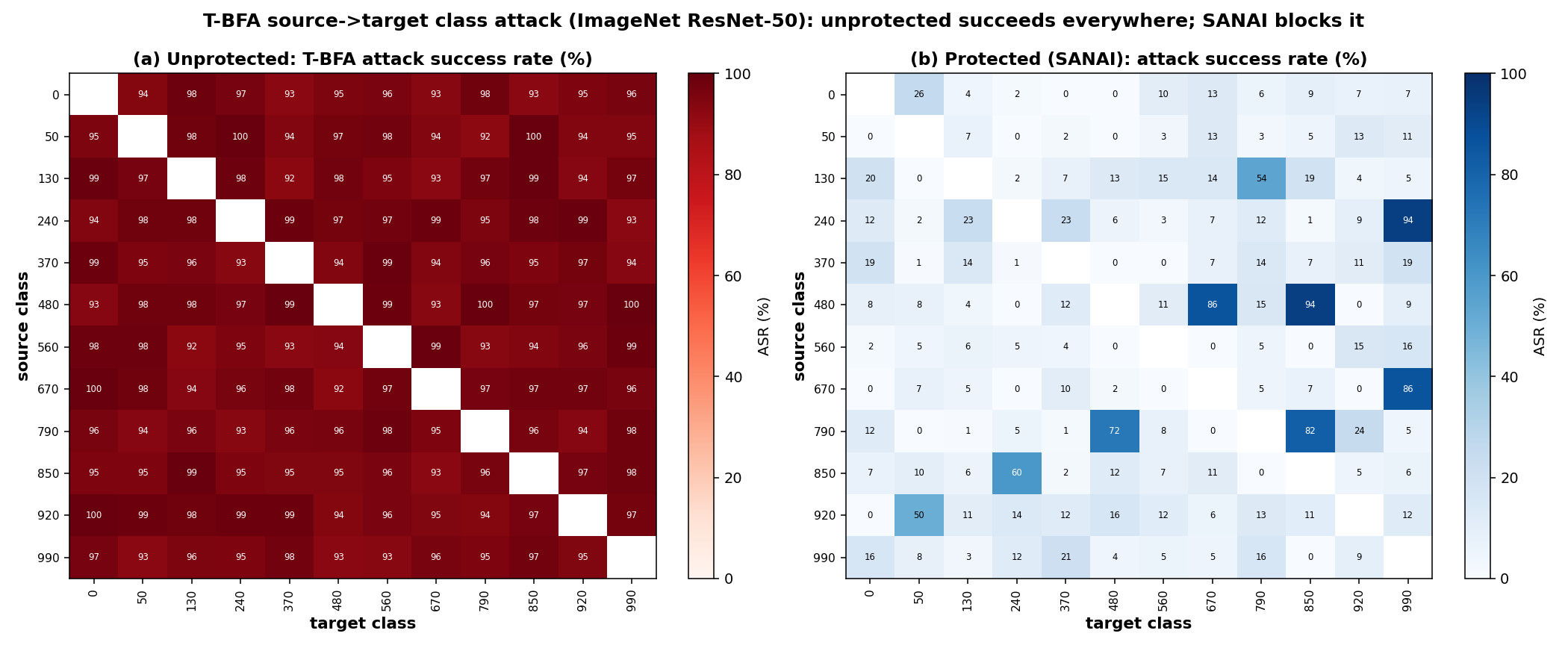}
\caption{T-BFA source-to-target attack matrix (ImageNet ResNet-50, 132 source-target pairs). Each cell is the attack success rate for forcing the row (source) class to the column (target) class. (a) Unprotected: the attack succeeds on essentially every pair (90 to 100\%). (b) Protected (\sanai): the success rate collapses to near zero on almost all pairs, with a few residual leaks, because the flips land in the protected classifier weights and are corrected before inference.}
\label{fig:tbfa}
\end{figure}

\subsection{Large Language Models}
\label{sec:llm}

\begin{figure}[!t]
\centering
\includegraphics[width=0.98\textwidth]{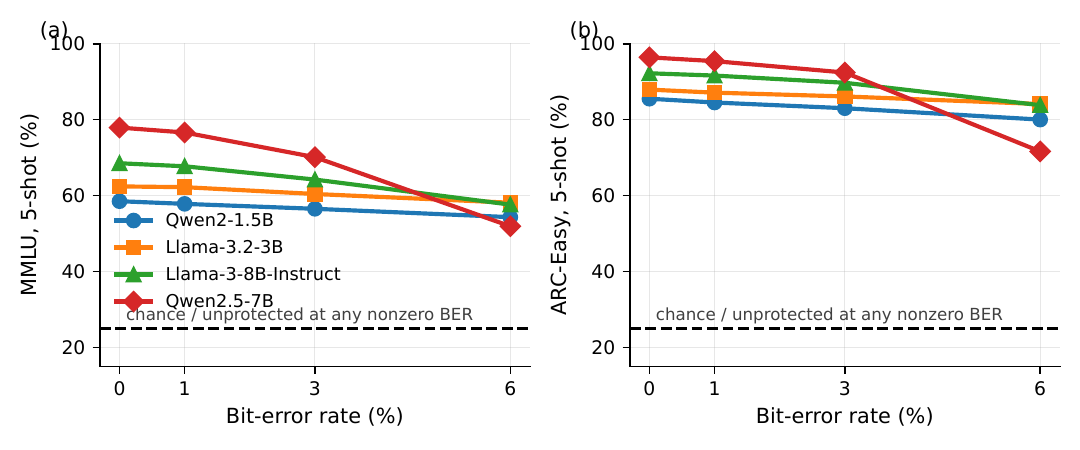}
\caption{Protected accuracy versus BER on (a) MMLU and (b) ARC-Easy, 5-shot, 2,000 questions, three seeds; every protected model stays well above chance ($\approx25\%$), whereas unprotected models fall onto the chance line at any nonzero BER.}
\label{fig:llmfig}
\end{figure}

\paragraph{Random bit errors.}
Table~\ref{tab:llm} (blocks 1a and 1b) applies the LLM operating point at 40\% coverage. Unprotected models collapse at any nonzero BER (perplexity at the $10^{6}$ ceiling below 3B, $10^{2}$ to $10^{5}$ above, accuracy at chance); protected models track clean through six percent in perplexity (Figure~\ref{fig:crossover}, right) and stay well above chance on MMLU and ARC-Easy (Figure~\ref{fig:llmfig}). Larger models are more robust once protected: Qwen2.5-7B, Llama-3.1-8B, and Llama-2-13B are unchanged to two decimals even at 8\%, whereas the two smallest models climb steeply at 8\%. Accuracy falls faster than perplexity at 6\% on the 7B and 8B because those runs stored part of the parity inside protected layers (the overflow regime of Section~\ref{sec:capacity}); with clean carriers an 8\% injection on Qwen2.5-7B gives perplexity 8.32 against $10^{6}$ unprotected. Llama-3-8B-Instruct holds 83.8\% ARC-Easy and Llama-3.2-3B 84.1\% at 6\% BER.

\begin{table}[!t]
\centering
\caption{LLM results at INT8, strong code, $K=4$, $G=12\le t$, zero added memory. Block 1: random BER at 40\% coverage, mean of three seeds, protected model; unprotected models sit at the $10^{6}$ perplexity ceiling (below 3B) or $10^{2}$ to $10^{5}$ (above) and at chance accuracy at every nonzero BER. The accuracy runs stored part of the parity in protected layers (overflow regime, Section~\ref{sec:capacity}), so their 6\% values are conservative. Block 2: AttentionBreaker with gradient ranking and keyed interleaving, every attack corrected to the exact clean value. Block 3: SilentStriker, attacked / \sanai. Rows trace to Master Tables~\ref{tab:master_l1} to \ref{tab:master_l3}.}
\label{tab:llm}\label{tab:ab}\label{tab:ss}
\footnotesize
\setlength{\tabcolsep}{4pt}
\renewcommand{\arraystretch}{1.1}
\begin{adjustbox}{max width=\textwidth}
\begin{tabular}{llrrrrr}
\toprule
\addlinespace[2.5pt]
\blk \multicolumn{7}{l}{\textbf{Block 1a. Random BER: protected WikiText-2 perplexity}} \\[2pt]
\textbf{Model} & \textbf{Params} & \textbf{0\%} & \textbf{1\%} & \textbf{3\%} & \textbf{6\%} & \textbf{8\%} \\
\midrule
\addlinespace[2pt]
DistilGPT-2 & 82M & 67.9 & 69.6 & 73.9 & 84.0 & 1746 \\
GPT-2 & 124M & 42.9 & 44.0 & 46.8 & 61.1 & 1909 \\
Qwen2-0.5B & 0.5B & 17.34 & 17.41 & 17.41 & 17.44 & 18.26 \\
Llama-3.2-1B & 1B & 13.68 & 13.83 & 14.13 & 14.70 & 25.32 \\
Qwen2-1.5B & 1.5B & 11.81 & 11.83 & 11.88 & 11.89 & 12.35 \\
Llama-3.2-3B & 3B & 10.36 & 10.44 & 10.54 & 10.74 & 11.73 \\
Qwen2.5-7B & 7B & 7.72 & 7.72 & 7.72 & 7.72 & 7.74 \\
Llama-3.1-8B & 8B & 7.95 & 7.95 & 7.95 & 7.95 & 8.04 \\
Llama-2-13B & 13B & 6.19 & 6.20 & 6.19 & 6.20 & 6.20 \\
\addlinespace[2pt]
\midrule
\addlinespace[3.5pt]
\blk \multicolumn{7}{l}{\textbf{Block 1b. Random BER: protected 5-shot MMLU / ARC-Easy (\%)}} \\[2pt]
\textbf{Model} & \textbf{Params} & \textbf{0\%} & \textbf{1\%} & \textbf{3\%} & \multicolumn{2}{c}{\textbf{6\%}} \\
\midrule
\addlinespace[2pt]
Qwen2-1.5B & 1.5B & 58.5 / 85.5 & 57.8 / 84.5 & 56.5 / 83.0 & \multicolumn{2}{c}{54.3 / 80.0} \\
Llama-3.2-3B & 3B & 62.4 / 87.9 & 62.2 / 87.1 & 60.4 / 86.1 & \multicolumn{2}{c}{58.1 / 84.1} \\
Llama-3-8B-Instruct & 8B & 68.5 / 92.2 & 67.7 / 91.6 & 64.2 / 89.7 & \multicolumn{2}{c}{57.6 / 83.8} \\
Qwen2.5-7B & 7B & 77.9 / 96.4 & 76.6 / 95.4 & 70.1 / 92.4 & \multicolumn{2}{c}{51.9 / 71.6} \\
\addlinespace[2pt]
\midrule
\addlinespace[3.5pt]
\blk \multicolumn{7}{l}{\textbf{Block 2. AttentionBreaker: MMLU (\%) and WikiText-2 perplexity}} \\[2pt]
\textbf{Model} & \textbf{Attack} & \textbf{Flips} & \textbf{Undef.\ MMLU} & \textbf{Undef.\ PPL} & \textbf{\sanai MMLU} & \textbf{\sanai PPL} \\
\midrule
\addlinespace[2pt]
Qwen2-0.5B & Top-10 & 10 & 25.8 & 1,294 & \textbf{45.6} & \textbf{17.65} \\
 & GenBFA & 26 & 23.8 & 11,935 & \textbf{45.6} & \textbf{17.65} \\
 & Adaptive & 16,056 & 27.0 & $>$$10^{6}$ & \textbf{45.6} & \textbf{17.65} \\
Llama-3.2-1B & GenBFA & 1,308 & 28.4 & 542 & \textbf{31.6} & \textbf{14.20} \\
 & Adaptive & 20,971 & 28.4 & 18,835 & \textbf{31.6} & \textbf{14.20} \\
Qwen2.5-7B & Top-500 & 500 & 25.0 & 31,047 & \textbf{79.2} & \textbf{7.74} \\
Llama-3-8B-Instruct & Top-10 & 10 & 27.0 & 64,978 & \textbf{71.0} & \textbf{10.89} \\
 & GenBFA & 1,018,082 & 24.4 & 504 & \textbf{71.0} & \textbf{10.89} \\
 & Adaptive & 1,174,405 & 23.6 & 916 & \textbf{71.0} & \textbf{10.89} \\
\addlinespace[2pt]
\midrule
\addlinespace[3.5pt]
\blk \multicolumn{7}{l}{\textbf{Block 3. SilentStriker (stealthy): attacked / \sanai}} \\[2pt]
\textbf{Model} & \textbf{Params} & \textbf{MMLU} & \textbf{ARC} & \textbf{HellaSwag} & \textbf{TruthfulQA} & \textbf{PPL} \\
\midrule
\addlinespace[2pt]
Qwen2-0.5B & 0.5B & 43.0 / \textbf{45.0} & 27.0 / \textbf{28.5} & 38.5 / \textbf{45.0} & 23.0 / \textbf{22.0} & 20.98 / \textbf{17.65} \\
Llama-3.2-1B & 1B & 22.0 / \textbf{31.0} & 30.5 / \textbf{31.5} & 49.0 / \textbf{52.5} & 25.0 / \textbf{24.0} & 21.02 / \textbf{14.20} \\
Qwen2.5-7B & 7B & 76.0 / \textbf{78.5} & 40.5 / \textbf{42.0} & 63.0 / \textbf{63.0} & 32.5 / \textbf{29.5} & 8.05 / \textbf{7.74} \\
Llama-3-8B-Instruct & 8B & 37.0 / \textbf{70.0} & 32.5 / \textbf{42.5} & 48.0 / \textbf{59.0} & 27.5 / \textbf{30.5} & 12.3 / \textbf{10.89} \\
\addlinespace[3pt]
\bottomrule
\end{tabular}
\end{adjustbox}
\end{table}

\paragraph{Optimized sign-bit attacks.}
\label{sec:llm-attacks}
AttentionBreaker \citep{das2024attentionbreaker} uses a gradient-and-magnitude score and an evolutionary search (GenBFA) to find a few weights whose sign bits collapse the model into gibberish; SilentStriker \citep{xu2025silentstriker} keeps text fluent while flipping the answer. We evaluate the published attacks, an \emph{adaptive} variant scored on the \sanai-corrected model, and a \emph{code-aware} attacker that knows $G$, $t$, and the layout (Table~\ref{tab:ab}, block 2). In every case \sanai restores the exact clean MMLU and perplexity with bit-identical weights. The adaptive attacker fails by construction: no set of sign flips raises the corrected loss above clean, so flipping all 16,056 population weights on Qwen2-0.5B, or $1.17\times10^{6}$ bits on the 8B, leaves the corrected model unchanged; with $G=12\le t$ no sign-bit attack places more than 12 errors in a codeword.

Defeating the code requires $t+1$ errors in one codeword, which with $G\le t$ means a second flipped plane on some weight. Master Table~\ref{tab:master_l3} prices it on Qwen2-0.5B: under the public adjacent layout, matching the damage of the 26-flip undefended attack needs roughly 600 layout-aware two-plane errors, a $23\times$ increase; with keyed interleaving and no key the errors scatter to at most three per codeword and \sanai restores the clean perplexity at every budget up to 1,300. A key-holding attacker can break codewords again, but each buys less damage (1,170 versus 37,635 at 650 errors).

SilentStriker is stealthy, so accuracy is the damage metric (outputs in Table~\ref{tab:outputs}, Appendix~\ref{app:llm}): on Llama-3-8B-Instruct, 160 flips drop MMLU from 70.0 to 37.0\%; \sanai flags the tamper and returns every metric to clean on all four models (Table~\ref{tab:ss}, block 3). Over 1 to 100 flips on Qwen2-0.5B, every attack is detected and corrected bit-exactly in 20 of 20 trials, with no false alarm. Two ablations matter (Master Table~\ref{tab:master_l2}): a weight-magnitude ranking never selects the small attention projections the attacks target, so the model collapses as if undefended, whereas the gradient-informed probe selects them from 10\% coverage; and $K=3$ ($G=16>t$) reopens sign-only overloads.

\subsection{Fine-Tuning Robustness and Ablations}
\label{sec:ablation}

\begin{figure}[!t]
\centering
\includegraphics[width=0.9\textwidth]{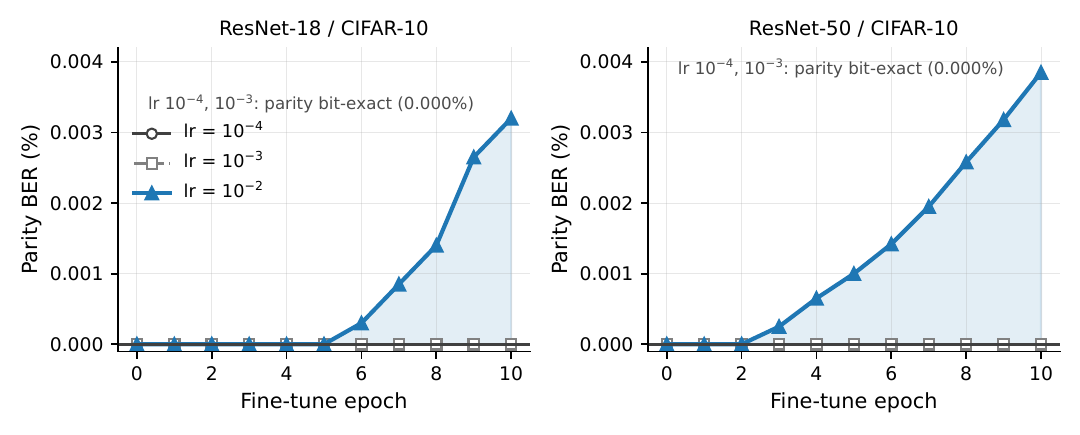}
\caption{Parity survival under fine-tuning: fraction of parity bits that no longer decode after each epoch on ResNet-18 (left) and ResNet-50 (right) on CIFAR-10. At learning rates $10^{-4}$ and $10^{-3}$ the parity stays bit-exact for all ten epochs; at $10^{-2}$ it drifts late and slightly (0.003 to 0.004\% at epoch ten).}
\label{fig:ft}
\end{figure}

\paragraph{Fine-tuning.}
Fine-tuning ResNet-18 and ResNet-50 on CIFAR-10 for ten epochs leaves the parity bit-exact at $10^{-4}$ and $10^{-3}$; only at $10^{-2}$ does the earliest convolution drift past $\Delta/2$ (0.003 to 0.004\% of parity bits; Figure~\ref{fig:ftdrift}, Appendix~\ref{app:finetune}), and the surviving parity still corrects a 12\% BER to 93.8\% and 94.7\% (Figure~\ref{fig:ft}; per-rate values in Table~\ref{tab:ft}, Appendix~\ref{app:finetune}).

\paragraph{Ablations.}
Appendix~\ref{app:ablation} (Table~\ref{tab:abl}) isolates each choice. \emph{Coverage:} 19.3\% to 52.8\% buys 0.5 points at 6\% and 2 at 12\% for $2.7\times$ the parity (rows V1.1, V1.24). \emph{Code:} on ResNet-50/ImageNet, $t=4$ ($f=3.1\%$) collapses by 6\% BER, $t=8$ ($6.3\%$) falls to 9.2\% at 12\%, $t=12$ ($9.4\%$) holds 77.3\%, and $m=8$, $t=20$ ($7.8\%$) trails $t=12$ despite the larger $t$ (Remark~\ref{rem:t}; rows V1.26 to V1.29). \emph{Planes:} $K=4$ lifts ResNet-50 on ImageNet from 76.5 to 80.9\% at 12\% BER while leaving ResNet-18 unchanged (rows V1.16 to V1.19). \emph{Replication:} $R\in\{1,2,3\}$ at capacity-limited coverage leaves protected accuracy unchanged (94.84, 94.80, 94.78 at 12\%), so $R=1$ is the default (rows V1.30 to V1.32). \emph{Precision:} INT6 ResNet-18/CIFAR-10 holds 93.6\% at 12\% (row V1.33); INT4 sits past the crossing of Eq.~\ref{eq:ceiling}.

\section{Hardware Realization}
\label{sec:hardware}

\begin{figure}[!t]
\centering
\includegraphics[width=0.95\textwidth]{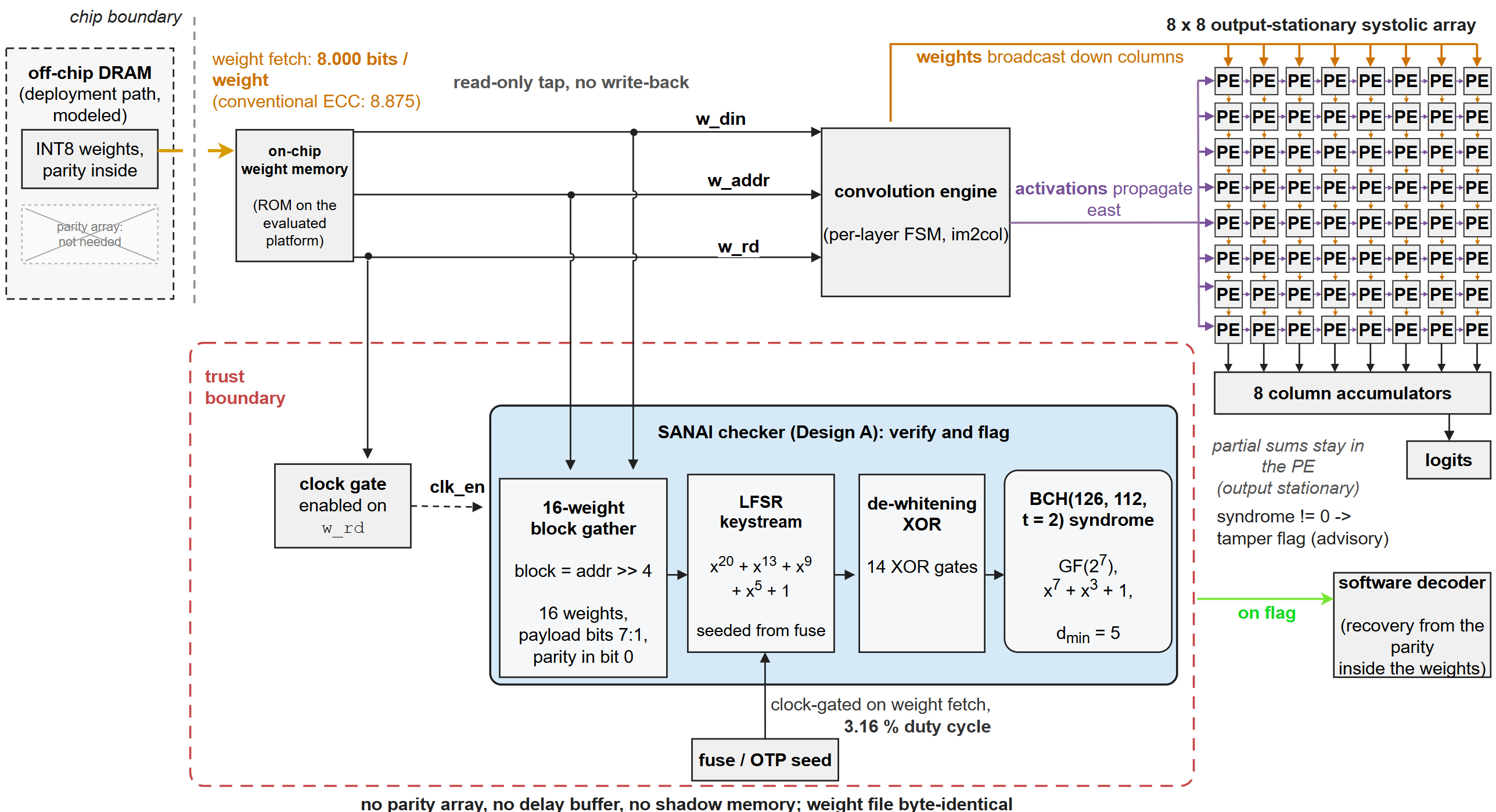}
\caption{\sanai Design A checker in the $8\times8$ output-stationary accelerator: a read-only tap on \texttt{w\_rd}, \texttt{w\_addr}, and \texttt{w\_din} between weight memory and the convolution engine, with \texttt{w\_rd} gating the clock and the other two feeding the 16-weight block gather, LFSR keystream, de-whitening XOR, and BCH$(126,112,t{=}2)$ syndrome. A nonzero syndrome raises an advisory tamper flag to the software decoder; there is no write-back, buffer, or parity array, and the weight file is byte-identical. The trust boundary encloses checker, clock gate, and seed.}
\label{fig:hwsys}
\end{figure}

We implement two streaming checkers on an $8\times8$ output-stationary systolic array, synthesized with Synopsys Design Compiler on gscl45nm (Figure~\ref{fig:hwsys}); both work on 16-weight blocks in fetch order, whiten the parity with a fuse-seeded keystream, and tap the weight path read-only, flagging a nonzero syndrome. The checker is model-agnostic: its area, timing, and energy per block do not depend on the network, and per-inference cost scales with the weight count; the evaluated workload, an INT8 CIFAR-10 CNN, fits the on-chip ROM and permits a bit-exact full-inference check.

\begin{proposition}[Hardware reduction of STDM]
\label{prop:hwreduce}
Let $L_s=1$, one bit per symbol, lattice step $\Delta$ equal to one quantization step, and dither $d=\tfrac{\Delta}{2}z$ with $z\in\{0,1\}$ a keystream bit. Then the embedding of Eq.~\ref{eq:stdm} writes $\beta\oplus z$ into the least significant bit of the carrier weight, the reader returns that bit XORed with $z$, and the perturbation is $|\delta w|\le\Delta/2$ exactly as in the general case.
\end{proposition}

\emph{Design A} is this proposition in silicon: each block is one BCH$(126,112,t{=}2)$ codeword over $\mathrm{GF}(2^{7})$ whose payload is the top seven bits of each weight and whose 14 parity bits occupy bit 0 of 14 weights, so any single flip is a weight-one codeword error and $d_{\min}=5$ detects it at every bit position, including the sign bit. \emph{Design B} computes the STDM projection of the block on an integer lattice; its detection is magnitude dependent (100\% at bit 6, 50\% at bit 5, 0\% at the sign bit, which moves the projection by exactly two lattice periods), which is why Design A is primary. Both coexist with the software code in the same weights and both detect and flag rather than correct: an in-line correcting variant needs a 140 KB shadow buffer, the whole weight file, and silently miscorrects 44 to 47\% of blocks above capacity, so the flag hands recovery to the software decoder. Design A costs 0.41\% area, 0.63\% dynamic power, zero stall cycles, and zero memory, and closes at 1.27 ns (Table~\ref{tab:hw}; detection rates, storage traffic, and sensitivity in Appendix~\ref{app:hardware}). Against a conventional per-block ECC checker synthesized under the same constraints (same code, parity from a separate SRAM port), Design A costs 13\% more logic than the detect-only variant but no parity array (1.22 MB for ResNet-18), closes at 1.27 ns against 2.51 ns, and is 40\% smaller and $2.6\times$ faster than the correcting variant (Table~\ref{tab:convecc}); the protected model is byte-identical to the original instead of 10.94\% larger and fetches 9.86\% fewer bits per weight, a modeled 195 $\mu$J per ResNet-18 inference at DRAM energies \citep{han2016eie}.

\begin{table}[!t]
\centering
\caption{Complete measured overhead (Synopsys DC V-2023.12-SP5, gscl45nm, 100 MHz constraint). Block A compares the full accelerator with and without a checker; block B characterizes the checkers themselves.}
\label{tab:hw}
\small
\setlength{\tabcolsep}{6pt}
\begin{tabular}{lrrr}
\toprule
\addlinespace[3pt]
\blk \multicolumn{4}{l}{\textbf{Block A. Full accelerator}} \\[3pt]
& \textbf{Unprotected} & \textbf{With Design A} & \textbf{With Design B} \\
\midrule
\addlinespace[2pt]
Full-top critical path (ns) & 30.38 & 31.14 & 30.99 \\
Dynamic power (mW) & 113.11 & 113.81 & 113.75 \\
Leakage (mW) & 7.581 & 7.586 & 7.599 \\
Weight memory added (bytes) & 0 & 0 & 0 \\
Stall cycles added & 0 & 0 & 0 \\
\addlinespace[3pt]
\midrule
\addlinespace[5pt]
\blk \multicolumn{4}{l}{\textbf{Block B. Checker alone}} \\[3pt]
& & \textbf{Design A} & \textbf{Design B} \\
\midrule
\addlinespace[2pt]
Area ($\mu$m$^2$) & & 5,763.94 & 7,084.55 \\
Area, clock-gated ($\mu$m$^2$) & & 5,824.95 & 7,173.72 \\
Area overhead over the accelerator (\%) & & 0.4118 & 0.5071 \\
Critical path (ns) & & 1.27 & 1.31 \\
Power (mW) & & 0.797 & 0.619 \\
Single-flip detection, bits 7 / 6 / 5 / 0 (\%) & & 100 / 100 / 100 / 100 & 0 / 100 / 50.4 / 1.4 \\
\addlinespace[3pt]
\bottomrule
\end{tabular}
\end{table}

\begin{table}[!t]
\centering
\caption{Design A against a measured conventional per-block ECC checker (same code and payload, same tool, library, and clock; Synopsys DC V-2023.12-SP5, gscl45nm, 100 MHz constraint).}
\label{tab:convecc}
\small
\begin{tabular}{lrrr}
\toprule
 & \textbf{Design A (ours)} & \textbf{Conv.\ detect-only} & \textbf{Conv.\ correcting} \\
\midrule
Parity location & bit 0 of the weights & separate SRAM & separate SRAM \\
Checker area ($\mu$m$^2$) & 5,763.94 & 5,092.84 & 9,646.46 \\
Critical path (ns) & \textbf{1.27} & 2.51 & 3.28 \\
Parity SRAM, CIFAR-10 CNN & \textbf{0 B} & 15,355 B & 15,355 B \\
Parity SRAM, ResNet-18 & \textbf{0 B} & 1.22 MB & 1.22 MB \\
Correction buffer & 0 & 0 & 16 B \\
Miscorrection at $t{+}1$ & flags & flags & 47.21\% \\
\bottomrule
\end{tabular}
\end{table}

\section{Discussion and Limitations}
\label{sec:discussion}

The curves of Section~\ref{sec:vision} use a localized model in which carriers stay clean, as for voltage scaling and RowHammer; under a uniform model that corrupts the parity-carrying layers too, keyed parity placement keeps ResNet-18/CIFAR-10 at 89.3\% at 12\% BER, against 78.8\% with sequential parity and 10.8\% unprotected (Table~\ref{tab:parity-int}, Appendix~\ref{app:ablation}). Proposition~\ref{prop:access} guards against an adversary who holds the weights but not the key; the remaining limit of any ECC is the code-aware multi-plane attacker (Section~\ref{sec:llm-attacks}), against which keyed interleaving raises the cost of each unit of damage. The scheme does not depend on the training framework: applied directly to exported ONNX graphs \citep{onnx2017}, the protected model runs unchanged in ONNX Runtime without custom operators (Appendix~\ref{app:onnx}). The protectable fraction is bounded by carrier capacity, so very low precisions such as INT4 sit past the crossing of Eq.~\ref{eq:ceiling} (Appendix~\ref{app:storage}), and the hardware checker detects and flags rather than corrects, handing recovery to the software decoder (Appendix~\ref{app:hardware}).

\section{Conclusion}
\label{sec:conclusion}

In summary, \sanai provides robust protection for sensitive model-weight bits with zero additional memory overhead. It achieves this by embedding parity bits directly into model weights and distributing them across a two-dimensional space using a scalable STDM scheme. \sanai is particularly effective in high-BER scenarios, where alternative embedding schemes either fail or incur substantial accuracy degradation. It maintains near-clean model accuracy at BERs of up to 12\%, corrects progressive, targeted, and sign-bit attacks, and reduces the hardware overhead of the checker to only 0.41\% in area. To the best of our knowledge, \sanai is the first \textbf{secure ECC} approach designed to defend against bit-flip attacks targeting critical model-weight bits with the goal of bypassing ECC protection. Furthermore, \sanai is scalable across diverse model architectures and sizes and incurs minimal hardware footprint.

\subsection*{AI Use Statement}
In this work, we used generative AI tools for editing and condensing the manuscript text and for checking LaTeX formatting. We have not used generative AI tools for generating research ideas, designing experiments, producing results, or writing experimental code; the remaining required disclosure tasks are not applicable to this work. All AI-assisted text was reviewed and verified by the authors, who take responsibility for the final content of this work, including text, claims, and artifacts.

\subsection*{Ethics Statement}
This work studies defenses against memory faults and bit-flip attacks on deployed neural networks. The attack implementations we use are re-implementations of published, code-available attacks and are evaluated only on public models and benchmarks; we release no new attack capability. The key-gated recovery mechanism is presented as fault tolerance with lightweight access control and is not claimed to be a cryptographic guarantee (Section~\ref{sec:discussion}).

\bibliography{arxiv}
\bibliographystyle{arxiv_style}

\appendix
\clearpage
\section*{Appendix}
\addcontentsline{toc}{section}{Appendix}

\section{Master Tables}
\label{app:master}

Every number in the main text is a row, or part of a row, of one of the master tables below (V1, V2, A1, L1, L2, L3). Each row is one run with its full configuration, so numbers for the same model that differ across main-text tables can be traced to the knob that differs. Cfg: S = strong code shortened from BCH$(127,50,t{=}12)$ with 88 parity bits; L = light code shortened from BCH$(127,99,t{=}4)$ with 32 parity bits. Blank cells were not run. Coverage is the protected weight fraction; $R$ is parity replication; Intl.\ marks keyed interleaving.

\begin{table}[h]
\centering
\caption{Master Table V1: vision runs under random BER with the strong code (INT8 unless noted), top-1 accuracy (\%). Rows V1.24 to V1.33 are the ablations of Section~\ref{sec:ablation}. ImageNet rows use layer-mode selection at a 40\% budget; unprotected ImageNet accuracy is below 1\% at 12\% BER for every model. $^{\dagger}$five-plane budget.}
\label{tab:master_v1}
\scriptsize
\setlength{\tabcolsep}{3pt}
\renewcommand{\arraystretch}{1.05}
\begin{adjustbox}{max width=\textwidth}
\begin{tabular}{l ll l c c c c rrrrrrr}
\toprule
\textbf{ID} & \textbf{Model} & \textbf{Data} & \textbf{Quant} & \textbf{Code} & $K$ & \textbf{Cov.} & $R$ & \textbf{0\%} & \textbf{1\%} & \textbf{3\%} & \textbf{6\%} & \textbf{8\%} & \textbf{10\%} & \textbf{12\%} \\
\midrule
V1.1 & ResNet-18 & C10 & PTQ & S & 3 & 19.3\% & 1 & 95.09 & 94.95 & 94.66 & 94.54 & 94.46 & 94.36 & 93.06 \\
V1.2 & ResNet-18 & C10 & Jacob & S & 3 & 19.5\% & 1 & 95.05 & 94.99 & 94.58 & 94.53 & 94.41 & 94.41 & 93.53 \\
V1.3 & ResNet-50 & C10 & PTQ & S & 3 & 19.2\% & 1 & 94.93 & 94.87 & 94.79 & 94.51 & 94.54 & 94.10 & 93.70 \\
V1.4 & VGG-16 & C10 & PTQ & S & 3 & 20.0\% & 1 & 90.20 & 90.12 & 89.98 & 89.87 & 89.72 & 89.76 & 89.19 \\
V1.5 & ResNet-18 & C100 & PTQ & S & 3 & 14.0\% & 1 & 77.05 & 77.30 & 77.10 & 76.82 & 76.69 & 76.11 & 75.86 \\
V1.6 & ResNet-18 & C10 & LSQ & S & 3 & 18.6\% & 1 & 94.98 & 93.26 & 90.78 & 78.60 & 76.25 & 63.41 & 27.38 \\
V1.7 & ResNet-18 & C10 & LSQ & S & 3 & 36.6\% & 1 & 94.98 & 94.86 & 94.52 & 93.26 & 92.73 & 91.26 & 85.84 \\
V1.8 & ResNet-50 & C100 & PTQ & S & 3 & 36.1\% & 1 & 75.77 & 75.64 & 75.36 & 75.32 & 75.41 & 75.36 & 74.71 \\
V1.9 & VGG-16 & C100 & PTQ & S & 3 & 23.7\% & 1 & 65.92 & 65.91 & 65.74 & 65.72 & 65.74 & 65.56 & 64.69 \\
V1.10 & EfficientNet-B0 & C100 & PTQ & S & 4 & 30.0\% & 1 & 62.86 & 63.29 & 62.49 & 61.96 & 61.46 & 59.97 & 53.22 \\
V1.11 & MaxViT-T & IN & PTQ & S & 4 & 40.9\% & 1 & 88.14 & 88.14 & 88.06 & 88.05 & 87.97 & 87.69 & 87.55 \\
V1.12 & RegNet-Y-16GF & IN & PTQ & S & 4 & $\approx$40\% & 1 & 85.78 & 85.66 & 85.50 & 85.05 & 84.64 & 84.62 & 84.48 \\
V1.13 & Wide ResNet-50-2 & IN & PTQ & S & 4 & $\approx$40\% & 1 & 84.00 & 83.91 & 83.97 & 84.06 & 83.91 & 83.58 & 83.58 \\
V1.14 & Swin-T & IN & PTQ & S & 4 & $\approx$40\% & 1 & 86.61 & 86.50 & 86.28 & 85.42 & 85.34 & 84.92 & 83.44 \\
V1.15 & ViT-B/16 & IN & PTQ & S & 5 & 43.5\% & 1 & 85.59 & 85.50 & 85.42 & 85.28 & 84.44 & 83.25 & 77.89 \\
V1.16 & ResNet-18 & IN & PTQ & S & 3 & 40.4\% & 1 & 76.14 & 76.05 & 76.14 & 75.95 & 75.70 & 75.78 & 75.70 \\
V1.17 & ResNet-18 & IN & PTQ & S & 4 & 40.4\% & 1 & 76.14 & 76.20 & 76.12 & 76.16 & 75.95 & 75.98 & 75.91 \\
V1.18 & ResNet-50 & IN & PTQ & S & 3 & 40.9\% & 1 & 82.83 & 81.83 & 81.03 & 79.58 & 78.58 & 79.05 & 76.45 \\
V1.19 & ResNet-50 & IN & PTQ & S & 4 & 40.9\% & 1 & 82.83 & 82.72 & 82.52 & 82.41 & 82.06 & 81.31 & 80.88 \\
\midrule
V1.24 & ResNet-18 & C10 & PTQ & S & 3 & 52.8\% & 1 & 95.06 & 95.00 & 94.91 & 95.02 & 94.84 & 94.95 & 95.02 \\
V1.25 & ResNet-18 & C10 & PTQ & S & 4 & 52.8\% & 1 & 95.06 & 95.02 & 95.00 & 95.05 & 94.94 & 94.95 & 94.89 \\
V1.26 & ResNet-50 & IN & PTQ & $t{=}4$ & 3 & 23.4\% & 1 & 82.83 & 81.53 & 64.30 & 2.61 & 0.34 & 0.00 & 0.05 \\
V1.27 & ResNet-50 & IN & PTQ & $t{=}8$ & 3 & 23.4\% & 1 & 82.83 & 82.31 & 80.95 & 78.41 & 65.55 & 33.28 & 9.17 \\
V1.28 & ResNet-50 & IN & PTQ & $t{=}12$ & 3 & 23.4\% & 1 & 82.83 & 82.09 & 81.03 & 80.73 & 80.09 & 78.86 & 77.30 \\
V1.29 & ResNet-50 & IN & PTQ & $t{=}20$, $m{=}8$ & 3 & 23.4\% & 1 & 82.83 & 82.34 & 81.33 & 80.88 & 79.84 & 78.11 & 75.67 \\
V1.30 & ResNet-50 & C10 & PTQ & S & 3 & 39.8\% & 1 & 94.86 & 94.88 & 94.69 & 94.97 & 94.67 & 94.75 & 94.84 \\
V1.31 & ResNet-50 & C10 & PTQ & S & 3 & 20.4\% & 2 & 94.86 & 94.91 & 94.86 & 94.86 & 94.89 & 94.83 & 94.80 \\
V1.32 & ResNet-50 & C10 & PTQ & S & 3 & 15.1\% & 3 & 94.86 & 94.92 & 94.88 & 94.95 & 94.84 & 94.97 & 94.78 \\
V1.33 & ResNet-18 & C10 (INT6) & PTQ & S & 4 & 52.8\% & 1 & 95.08 & 95.00 & 94.86 & 94.73 & 94.56 & 94.20 & 93.64 \\
\bottomrule
\end{tabular}
\end{adjustbox}
\end{table}

\begin{table}[h]
\centering
\caption{Master Table V1b: ResNet-18 / CIFAR-10 under weight-level sensitivity bins (block A; high = top 1\% of weights by attack-damage proxy, mid = next 8\%; V1.35 had host bits exposed and 11 overflow carriers) and under the uniform fault model that corrupts the carrier layers at the same BER (block B; strong code, $K=3$, 19\% coverage). Protected (unprotected) top-1 (\%).}
\label{tab:master_v1b}
\scriptsize
\setlength{\tabcolsep}{5pt}
\renewcommand{\arraystretch}{1.05}
\begin{tabular}{l l c c c c c c}
\toprule
\addlinespace[2pt]
\blk \multicolumn{8}{l}{\textbf{Block A. Weight-level sensitivity bins}} \\[2pt]
\textbf{ID} & \textbf{Bins protected} & \textbf{Layers} & \textbf{$K$} & \textbf{Cov.} & \textbf{3\%} & \textbf{6\%} & \textbf{T-BFA ASR unprot.\ / prot.} \\
\midrule
V1.34 & high only & 14 & 4 & 24.8\% & 94.95 (69.16) & 95.01 (15.03) & 95.3 / 4.5 \\
V1.35 & high + half mid & 16 & 4 & 56.5\% & 93.32 (66.92) & 79.11 (17.06) & 95.3 / 8.4 \\
\addlinespace[2pt]
\midrule
\addlinespace[4pt]
\blk \multicolumn{8}{l}{\textbf{Block B. Uniform fault model: carrier layers corrupted at the same BER}} \\[2pt]
\textbf{ID} & \textbf{Parity placement} & & \textbf{$K$} & \textbf{Cov.} & \textbf{0\%} & \textbf{6\%} & \textbf{12\%} \\
\midrule
V1.36 & interleaved & & 3 & 19\% & 95.06 & 93.92 (10.25) & \textbf{89.34} (10.78) \\
V1.37 & sequential & & 3 & 19\% & 95.06 & 93.83 (10.25) & 78.83 (10.78) \\
\addlinespace[2pt]
\bottomrule
\end{tabular}
\end{table}

\begin{table}[h]
\centering
\caption{Master Table V2: vision runs under the burst protocol of \citet{kamineni2026e3} with the strong code (block A), under random BER with the light code at the attack operating point (block B), and under key-gated recovery with $L_s=2$ (block C). Top-1 (\%); unprotected in parentheses where measured.}
\label{tab:master_v2}
\scriptsize
\setlength{\tabcolsep}{3.5pt}
\renewcommand{\arraystretch}{1.05}
\begin{adjustbox}{max width=\textwidth}
\begin{tabular}{l ll c c rrrrr}
\toprule
\blk \multicolumn{10}{l}{\textbf{Block A. Burst protocol, strong code}} \\[1pt]
\textbf{ID} & \textbf{Model} & \textbf{Data} & \textbf{$K$} & \textbf{Cov.} & \multicolumn{2}{c}{\textbf{0\%}} & \textbf{1.5\%} & \textbf{3\%} & \textbf{6\%} \\
\midrule
V2.1 & ResNet-18 & C10 & 3 & 20\% & \multicolumn{2}{c}{95.09} & 94.9 & 94.7 & 94.5 \\
V2.2 & ResNet-50 & C10 & 3 & 20\% & \multicolumn{2}{c}{94.93} & 94.9 & 94.8 & 94.5 \\
V2.3 & MobileNet-V2 & C10 & 4 & 20\% & \multicolumn{2}{c}{79.27} & 78.2 & 76.6 & 75.3 \\
V2.4 & EfficientNet-B0 & C10 & 4 & 20\% & \multicolumn{2}{c}{84.23} & 83.8 & 83.0 & 81.4 \\
V2.5 & ResNet-18 & IN & 3 & 20\% & \multicolumn{2}{c}{76.14} & 75.5 & 75.1 & 73.3 \\
V2.6 & ResNet-50 & IN & 3 & 20\% & \multicolumn{2}{c}{82.83} & 82.2 & 81.6 & 80.5 \\
V2.7 & MobileNet-V2 & IN & 4 & 20\% & \multicolumn{2}{c}{78.39} & 78.4 & 78.1 & 77.8 \\
V2.8 & EfficientNet-B0 & IN & 4 & 20\% & \multicolumn{2}{c}{83.42} & 82.5 & 80.6 & 76.2 \\
\multicolumn{10}{l}{Extended to 12\% burst BER on five settings: V2.1 93.1, V2.2 93.7, V2.3 67.1, V2.4 75.0, V2.6 62.6.} \\
\midrule
\addlinespace[5pt]
\blk \multicolumn{10}{l}{\textbf{Block B. Random BER, light code, keyed interleaving; protected (unprotected)}} \\[1pt]
\textbf{ID} & \textbf{Model} & \textbf{Data} & \textbf{$K$} & \textbf{Cov.} & \textbf{0\%} & \textbf{1\%} & \textbf{3\%} & \textbf{6\%} & \textbf{12\%} \\
\midrule
V2.9 & ResNet-18 & C10 & 4 & 10\% & 95.1 & 95.1 (87.5) & 94.9 (52.5) & 94.8 (20.4) & 94.7 (10.0) \\
V2.10 & ResNet-18 & IN & 4 & 10\% & 76.1 & 75.5 (0.2) & 73.4 (0.0) & 72.4 (0.0) & 46.2 (0.0) \\
V2.11 & ResNet-50 & IN & 4 & 10\% & 82.8 & 82.6 (0.5) & 81.8 (0.8) & 79.9 (0.0) & 78.1 (1.2) \\
V2.12 & MobileNet-V2 & IN & 4 & 10\% & 78.4 & 77.2 (0.0) & 72.6 (0.3) & 64.1 (0.0) & 6.3 (0.1) \\
V2.13 & EfficientNet-B0 & IN & 4 & 10\% & 83.4 & 82.1 (0.0) & 79.7 (0.0) & 73.3 (0.1) & 37.1 (0.3) \\
\midrule
\addlinespace[5pt]
\blk \multicolumn{10}{l}{\textbf{Block C. Key-gated recovery, strong code, $K=3$}} \\[1pt]
\textbf{ID} & \textbf{Model} & \textbf{Data} & \textbf{Key} & \textbf{Cov.} & \textbf{0\%} & \textbf{1.5\%} & \textbf{3\%} & \textbf{6\%} & \textbf{12\% / 20\%} \\
\midrule
V2.14 & ResNet-18 & C10 & correct & 19.3\% & 95.09 & 94.89 & 94.66 & 94.54 & 93.06 / 44.10 \\
V2.15 & ResNet-18 & C10 & wrong & 19.3\% & 94.88 & 77.43 & 44.58 & 22.91 & 10.90 / 9.98 \\
V2.16 & ResNet-18 & IN & correct & 20\% & 76.14 & 75.50 & 75.08 & 73.28 & 59.42 / 0.02 \\
V2.17 & ResNet-18 & IN & wrong & 20\% & 67.30 & 0.00 & 0.02 & 0.00 & 0.62 / 0.02 \\
\bottomrule
\end{tabular}
\end{adjustbox}
\end{table}

\begin{table}[h]
\centering
\caption{Master Table A1: vision runs under adversarial bit-flip attacks (INT8). Block A: PBS with keyed interleaving (ResNets: light code, layer-mode selection; compact networks: strong code, weight-mode selection). Block B: the same 500 PBS flips under adjacent and keyed grouping. Block C: T-BFA with the strong code. Block D: fault-map pass on the compact networks. cw$>t$ counts uncorrectable codewords and max/cw is the largest fault count in one codeword; $^{\dagger}$weight-mode selection. Row A1.15 (100 flips) is from an earlier run.}
\label{tab:master_a1}
\scriptsize
\setlength{\tabcolsep}{5pt}
\renewcommand{\arraystretch}{1.08}
\begin{tabular}{l l l c c c c c c}
\toprule
\blk \multicolumn{9}{l}{\textbf{Block A. PBS, 500 flips, keyed interleaving}} \\
\textbf{ID} & \textbf{Model} & \textbf{Data} & \textbf{Code} & \textbf{Cov.} & \textbf{Unprot.\ (\%)} & \textbf{Prot.\ (\%)} & \textbf{cw$>t$} & \textbf{max/cw} \\
\midrule
A1.1 & ResNet-18 & C10 & light, $t{=}4$ & 10\% & 10.06 & 94.9 & 0 & 3 \\
A1.2 & ResNet-50 & C10 & light, $t{=}4$ & 10\% & 35.00 & 94.8 & 0 & 3 \\
A1.3 & MobileNet-V2 & C10 & strong, $t{=}12$ & 10\% & 9.94 & 64.1 & 1$^{\dagger}$ & 5$^{\dagger}$ \\
A1.4 & EfficientNet-B0 & C10 & strong, $t{=}12$ & 10\% & 10.00 & 10.3 & 0$^{\dagger}$ & 3$^{\dagger}$ \\
A1.5 & ResNet-18 & C100 & light, $t{=}4$ & 10\% & 1.16 & 75.5 & 0 & 2 \\
A1.6 & ResNet-50 & C100 & light, $t{=}4$ & 10\% & 0.94 & 75.8 & 0 & 2 \\
A1.7 & MobileNet-V2 & C100 & strong, $t{=}12$ & 10\% & 1.02 & 35.0 & 0$^{\dagger}$ & 2$^{\dagger}$ \\
A1.8 & EfficientNet-B0 & C100 & strong, $t{=}12$ & 10\% & 0.98 & 60.9 & 3$^{\dagger}$ & 8$^{\dagger}$ \\
\midrule
\blk \multicolumn{9}{l}{\textbf{Block B. Adjacent / keyed grouping, same flipped bits, light code $t{=}4$}} \\
\textbf{ID} & \textbf{Model} & \textbf{Data} & \textbf{Flips} & \textbf{Cov.} & \multicolumn{2}{c}{\textbf{Prot.\ (\%) adj / keyed}} & \textbf{cw$>t$} & \textbf{max/cw} \\
\midrule
A1.9 & ResNet-18 & C10 & 500 & 10\% & \multicolumn{2}{c}{10.06 / 94.86} & 22 / 0 & 14 / 3 \\
A1.10 & ResNet-50 & C10 & 500 & 10\% & \multicolumn{2}{c}{35.00 / 94.81} & 14 / 0 & 15 / 3 \\
A1.11 & ResNet-18 & C100 & 500 & 10\% & \multicolumn{2}{c}{1.16 / 75.48} & 26 / 0 & 13 / 2 \\
A1.12 & ResNet-50 & C100 & 500 & 10\% & \multicolumn{2}{c}{0.94 / 75.77} & 49 / 0 & 10 / 2 \\
A1.13 & ResNet-18 & IN & 500 & 10\% & \multicolumn{2}{c}{0.00 / 1.53} & 14 / 0 & 12 / 2 \\
A1.14 & ResNet-50 & IN & 500 & 10\% & \multicolumn{2}{c}{1.70 / 81.88} & 9 / 5 & 12 / 2 \\
A1.15 & ResNet-18 & C10 & 100 & 10\% & \multicolumn{2}{c}{66.0 / 95.1} & 3 / 0 & 12 / 2 \\
\midrule
\blk \multicolumn{9}{l}{\textbf{Block C. Targeted attack (T-BFA), strong code, high-bin selection}} \\
\textbf{ID} & \textbf{Model} & \textbf{Data} & \textbf{Pair} & \textbf{Cov.} & \multicolumn{2}{c}{\textbf{Flips}} & \textbf{ASR unprot.} & \textbf{ASR prot.} \\
\midrule
A1.17 & ResNet-18 & C10 & $3\to5$ & 24.8\% & \multicolumn{2}{c}{15} & 95.3 / 95.7 & 4.5 / 4.9 \\
A1.18 & ResNet-18 & C100 & $90\to70$ & 10\% & \multicolumn{2}{c}{10 to 20} & 93.0 & 0.0 \\
A1.19 & ResNet-18 & C100 & $90\to80$ & 10\% & \multicolumn{2}{c}{10 to 20} & 94.0 & 0.0 \\
A1.20 & ResNet-50 & IN & 132-pair mean & 10\% & \multicolumn{2}{c}{10 to 20} & $\sim$98 & $\sim$8 \\
\midrule
\blk \multicolumn{9}{l}{\textbf{Block D. Fault-map pass, compact networks, 500 flips, light code $t{=}4$}} \\
\textbf{ID} & \textbf{Model} & \textbf{Data} & \textbf{Flips} & \multicolumn{3}{c}{\textbf{Layout}} & \textbf{cw$>t$} & \textbf{max/cw} \\
\midrule
A1.21 & MobileNet-V2 & C10 & 500 & \multicolumn{3}{c}{adjacent / keyed} & 42 / 1 & 10 / 5 \\
A1.22 & MobileNet-V2 & C100 & 500 & \multicolumn{3}{c}{adjacent / keyed} & 35 / 0 & 10 / 2 \\
A1.23 & EfficientNet-B0 & C10 & 500 & \multicolumn{3}{c}{adjacent / keyed} & 17 / 0 & 14 / 3 \\
A1.24 & EfficientNet-B0 & C100 & 500 & \multicolumn{3}{c}{adjacent / keyed} & 23 / 3 & 24 / 8 \\
\bottomrule
\end{tabular}
\end{table}

\begin{table}[h]
\centering
\caption{Master Table L1: LLM runs under random BER (INT8, strong code, $K=4$, 40\% coverage, magnitude ranking, adjacent grouping, three seeds). P = protected, U = unprotected (perplexity capped at $10^{6}$); perplexity rows give mean $\pm$ std, accuracy rows are 5-shot on 2,000 questions to 6\% BER. Perplexity runs keep carriers clean; accuracy runs let parity overflow into protected layers (Section~\ref{sec:capacity}), so the 7B and 8B accuracy at 6\% is conservative.}
\label{tab:master_l1}
\scriptsize
\setlength{\tabcolsep}{3pt}
\renewcommand{\arraystretch}{1.05}
\begin{adjustbox}{max width=\textwidth}
\begin{tabular}{l l l c ccccc}
\toprule
\textbf{ID} & \textbf{Model} & \textbf{Metric} & \textbf{P/U} & \textbf{0\%} & \textbf{1\%} & \textbf{3\%} & \textbf{6\%} & \textbf{8\%} \\
\midrule
L1.1 & DistilGPT-2 & PPL W2 & P & 67.86 & 69.59 $\pm$ 0.85 & 73.93 $\pm$ 1.22 & 84.03 $\pm$ 2.27 & 1746 $\pm$ 1346 \\
L1.2 &  & PPL W103 & P & 61.64 & 63.16 $\pm$ 0.68 & 67.39 $\pm$ 1.23 & 76.77 $\pm$ 2.96 & 1721 $\pm$ 1358 \\
L1.3 &  & PPL W2 & U & 67.86 & $10^{6}$ & $10^{6}$ & $10^{6}$ & $10^{6}$ \\
L1.4 & GPT-2 & PPL W2 & P & 42.91 & 44.00 $\pm$ 0.32 & 46.77 $\pm$ 0.37 & 61.07 $\pm$ 4.87 & 1909 $\pm$ 667 \\
L1.5 &  & PPL W103 & P & 38.43 & 39.34 $\pm$ 0.04 & 42.26 $\pm$ 0.30 & 52.99 $\pm$ 2.91 & 1695 $\pm$ 407 \\
L1.6 &  & PPL W2 & U & 42.91 & $10^{6}$ & $10^{6}$ & $10^{6}$ & $10^{6}$ \\
L1.7 & Qwen2-0.5B & PPL W2 & P & 17.34 & 17.41 $\pm$ 0.03 & 17.41 $\pm$ 0.04 & 17.44 $\pm$ 0.03 & 18.26 $\pm$ 0.25 \\
L1.8 &  & PPL W103 & P & 15.22 & 15.27 $\pm$ 0.02 & 15.29 $\pm$ 0.03 & 15.33 $\pm$ 0.03 & 15.94 $\pm$ 0.15 \\
L1.9 &  & PPL W2 & U & 17.34 & $10^{6}$ & $10^{6}$ & $10^{6}$ & $10^{6}$ \\
L1.10 & Qwen2-1.5B & PPL W2 & P & 11.81 & 11.83 $\pm$ 0.00 & 11.88 $\pm$ 0.01 & 11.89 $\pm$ 0.02 & 12.35 $\pm$ 0.18 \\
L1.11 &  & PPL W103 & P & 10.92 & 10.94 $\pm$ 0.00 & 10.97 $\pm$ 0.00 & 10.97 $\pm$ 0.02 & 11.35 $\pm$ 0.15 \\
L1.12 &  & PPL W2 & U & 11.81 & $10^{6}$ & $10^{6}$ & $10^{6}$ & $10^{6}$ \\
L1.13 &  & MMLU / ARC-E & P & 58.5 / 85.5 & 57.8 $\pm$ 0.3 / 84.5 $\pm$ 0.19 & 56.5 $\pm$ 0.4 / 83.0 $\pm$ 0.37 & 54.3 $\pm$ 0.5 / 80.0 $\pm$ 0.41 & (to 6\%) \\
L1.14 & Llama-3.2-1B & PPL W2 & P & 13.68 & 13.83 $\pm$ 0.04 & 14.13 $\pm$ 0.08 & 14.70 $\pm$ 0.30 & 25.32 $\pm$ 9.44 \\
L1.15 &  & PPL W103 & P & 12.66 & 12.77 $\pm$ 0.04 & 13.06 $\pm$ 0.05 & 13.54 $\pm$ 0.21 & 23.98 $\pm$ 9.58 \\
L1.16 &  & PPL W2 & U & 13.68 & $9.5{\times}10^{5}$ & $10^{6}$ & $10^{6}$ & $10^{6}$ \\
L1.17 & Llama-3.2-3B & PPL W2 & P & 10.36 & 10.44 $\pm$ 0.01 & 10.54 $\pm$ 0.02 & 10.74 $\pm$ 0.03 & 11.73 $\pm$ 0.14 \\
L1.18 &  & PPL W103 & P & 9.73 & 9.78 $\pm$ 0.01 & 9.88 $\pm$ 0.02 & 10.03 $\pm$ 0.05 & 10.94 $\pm$ 0.17 \\
L1.19 &  & PPL W2 & U & 10.36 & $2.2{\times}10^{5}$ & $5.4{\times}10^{5}$ & $5.8{\times}10^{5}$ & $5.8{\times}10^{5}$ \\
L1.20 &  & MMLU / ARC-E & P & 62.4 / 87.9 & 62.2 $\pm$ 0.3 / 87.1 $\pm$ 0.29 & 60.4 $\pm$ 0.1 / 86.1 $\pm$ 0.54 & 58.1 $\pm$ 0.7 / 84.1 $\pm$ 0.16 & (to 6\%) \\
L1.21 & Qwen2.5-7B & PPL W2 & P & 7.72 & 7.72 $\pm$ 0.00 & 7.72 $\pm$ 0.00 & 7.72 $\pm$ 0.00 & 7.74 $\pm$ 0.02 \\
L1.22 &  & PPL W2 & U & 7.72 & $3.9{\times}10^{3}$ & $2.7{\times}10^{4}$ & $6.6{\times}10^{4}$ & $4.1{\times}10^{5}$ \\
L1.23 &  & MMLU / ARC-E & P & 77.9 / 96.4 & 76.6 $\pm$ 0.4 / 95.4 $\pm$ 0.14 & 70.1 $\pm$ 0.5 / 92.4 $\pm$ 0.60 & 51.9 $\pm$ 0.9 / 71.6 $\pm$ 1.51 & (to 6\%) \\
L1.24 & Llama-3-8B-Instruct & MMLU / ARC-E & P & 68.5 / 92.2 & 67.7 $\pm$ 0.4 / 91.6 $\pm$ 0.04 & 64.2 $\pm$ 0.5 / 89.7 $\pm$ 0.29 & 57.6 $\pm$ 0.8 / 83.8 $\pm$ 0.40 & (to 6\%) \\
L1.25 & Llama-3.1-8B & PPL W2 & P & 7.95 & 7.95 $\pm$ 0.00 & 7.95 $\pm$ 0.00 & 7.95 $\pm$ 0.00 & 8.04 $\pm$ 0.11 \\
L1.26 &  & PPL W2 & U & 7.95 & $3.9{\times}10^{2}$ & $8.2{\times}10^{4}$ & $3.1{\times}10^{4}$ & $6.7{\times}10^{4}$ \\
L1.27 & Llama-2-13B & PPL W2 & P & 6.19 & 6.20 $\pm$ 0.00 & 6.19 $\pm$ 0.00 & 6.20 $\pm$ 0.00 & 6.20 $\pm$ 0.01 \\
L1.28 &  & PPL W2 & U & 6.19 & $3.8{\times}10^{2}$ & $3.0{\times}10^{3}$ & $6.3{\times}10^{3}$ & $7.8{\times}10^{3}$ \\
\midrule
\multicolumn{9}{l}{Unprotected MMLU / ARC-E is 24 to 26\% (four-way chance) on every model at every nonzero BER.} \\
\bottomrule
\end{tabular}
\end{adjustbox}
\end{table}

\begin{table}[h]
\centering
\caption{Master Table L2: LLM runs under AttentionBreaker (INT8, strong code, $K=4$, $G=12$). Rank: grad = gradient-weighted magnitude, mag = weight magnitude; Intl.: adj = adjacent codewords, key = keyed interleaving; ``In'' = fraction of flipped bits inside the protected set. The largest error count in one codeword, where logged, never exceeded $t=12$ (Qwen2-0.5B rows 1 to 12; Llama-3.2-1B adaptive 5; Llama-3-8B GenBFA and adaptive $\le12$). Row L2.8 is the ranking ablation. In the coverage ablation (L2.23 to L2.25) the attacked layer enters the protected set at 30\% coverage, where recovery is exact. GPT-2 and DistilGPT-2 report perplexity only.}
\label{tab:master_l2}
\scriptsize
\setlength{\tabcolsep}{3.2pt}
\renewcommand{\arraystretch}{1.05}

\begin{tabular}{l l l r c c c rr rr c}
\toprule
\textbf{ID} & \textbf{Model} & \textbf{Attack} & \textbf{Flips} & \textbf{Rank} & \textbf{Cov.} & \textbf{Intl.} & \textbf{U MMLU} & \textbf{U PPL} & \textbf{\sanai MMLU} & \textbf{\sanai PPL} & \textbf{In} \\
\midrule
L2.1 & Qwen2-0.5B & Top-3 & 3 & grad & 40\% & key & 45.2 & 20.9 & 45.6 & 17.65 & 3/3 \\
L2.2 &  & Top-10 & 10 & grad & 40\% & key & 25.8 & 1,294 & 45.6 & 17.65 & 10/10 \\
L2.3 &  & Top-28 & 28 & grad & 40\% & key & 26.2 & 4,429 & 45.6 & 17.65 & 28/28 \\
L2.4 &  & Top-100 & 100 & grad & 40\% & key & 26.2 & 964,460 & 45.6 & 17.65 & 100/100 \\
L2.5 &  & Top-500 & 500 & grad & 40\% & key & 27.0 & $>10^{6}$ & 45.6 & 17.65 & 500/500 \\
L2.6 &  & GenBFA & 26 & grad & 40\% & key & 23.8 & 11,935 & 45.6 & 17.65 & 26/26 \\
L2.7 &  & Adaptive & 16,056 & grad & 40\% & key & 27.0 & $>10^{6}$ & 45.6 & 17.65 & all \\
L2.8 &  & Top-10 & 10 & mag & 40\% & key & 25.8 & 1,294 & 25.8 & 1,294 & 0/10 \\
L2.9 &  & Top-10 & 10 & grad & 10\% & adj & 25.8 & 1,294 & 45.6 & 17.65 & 10/10 \\
L2.10 & Llama-3.2-1B & Top-10 & 10 & grad & 40\% & key & 28.6 & 16.2 & 31.6 & 14.20 & all \\
L2.11 &  & Top-100 & 100 & grad & 40\% & key & 25.0 & 143 & 31.6 & 14.20 & all \\
L2.12 &  & GenBFA & 1,308 & grad & 40\% & key & 28.4 & 542 & 31.6 & 14.20 & all \\
L2.13 &  & Adaptive & 20,971 & grad & 40\% & key & 28.4 & 18,835 & 31.6 & 14.20 & all \\
L2.14 & Qwen2.5-7B & Top-3 & 3 & grad & 40\% & key & 23.5 & 7.84 & 79.2 & 7.74 & all \\
L2.15 &  & Top-28 & 28 & grad & 40\% & key & 23.5 & 12.97 & 79.2 & 7.74 & all \\
L2.16 &  & Top-100 & 100 & grad & 40\% & key & 25.5 & 1,497 & 79.2 & 7.74 & all \\
L2.17 &  & Top-500 & 500 & grad & 40\% & key & 25.0 & 31,047 & 79.2 & 7.74 & all \\
L2.18 & Llama-3-8B-Inst. & Top-3 & 3 & grad & 40\% & key & 63.8 & 17.0 & 71.0 & 10.89 & all \\
L2.19 &  & Top-10 & 10 & grad & 40\% & key & 27.0 & 64,978 & 71.0 & 10.89 & all \\
L2.20 &  & Top-100 & 100 & grad & 40\% & key & 26.0 & 3,474 & 71.0 & 10.89 & all \\
L2.21 &  & GenBFA & 1,018,082 & grad & 40\% & key & 24.4 & 504 & 71.0 & 10.89 & all \\
L2.22 &  & Adaptive & 1,174,405 & grad & 40\% & key & 23.6 & 916 & 71.0 & 10.89 & all \\
L2.26 & GPT-2 & Top-3 & 3 & grad & 40\% & key & n/a & 2,064 & n/a & 43.10 & all \\
 &  & Top-10 & 10 & grad & 40\% & key & n/a & 2,926 & n/a & 43.10 & all \\
 &  & Top-28 & 28 & grad & 40\% & key & n/a & 3,147 & n/a & 43.10 & all \\
 &  & Top-100 & 100 & grad & 40\% & key & n/a & 3,798 & n/a & 43.10 & all \\
L2.27 & DistilGPT-2 & Top-3 & 3 & grad & 40\% & key & n/a & 616 & n/a & 67.87 & all \\
 &  & Top-10 & 10 & grad & 40\% & key & n/a & 2,661 & n/a & 67.87 & all \\
 &  & Top-28 & 28 & grad & 40\% & key & n/a & 8,219 & n/a & 67.87 & all \\
 &  & Top-100 & 100 & grad & 40\% & key & n/a & 5,343 & n/a & 67.87 & all \\
\midrule
\blk \multicolumn{12}{l}{\textbf{Coverage ablation, Llama-3-8B-Instruct; the attack targets \texttt{layers.1.mlp.down\_proj}}} \\
\midrule
L2.23 & Llama-3-8B-Inst. & Top-10 & 10 & grad & 10\% & key & 27.0 & 64,978 & 27.0 & 64,978 & 0/10 \\
L2.24 &  & Top-10 & 10 & grad & 20\% & key & 27.0 & 64,978 & 27.0 & 64,978 & 0/10 \\
L2.25 &  & Top-10 & 10 & grad & 30\% & key & 27.0 & 64,978 & \textbf{71.0} & \textbf{10.89} & 10/10 \\
\bottomrule
\end{tabular}

\end{table}

\begin{table}[h]
\centering
\caption{Master Table L3: LLM runs under the code-aware attacker, SilentStriker, and the detection sweep (INT8, strong code, $K=4$). Code-aware rows are two-plane attacks that place 13 errors per targeted codeword and list WikiText-2 perplexity after \sanai at each error budget; blind = interleaved layout unknown to the attacker. SilentStriker rows give attacked / \sanai MMLU and perplexity and the number of the 160 attacked codewords whose syndrome flagged (other metrics in Table~\ref{tab:ss}); every model recovers with zero weights wrong. Detection rows are 20 trials of random sign flips inside the protected set on Qwen2-0.5B at 10\% coverage; the clean model raised zero false alarms over $1.19\times10^{7}$ codewords.}
\label{tab:master_l3}
\scriptsize
\setlength{\tabcolsep}{2.6pt}
\renewcommand{\arraystretch}{1.05}

\begin{tabular}{l l >{\raggedright\arraybackslash}p{0.10\textwidth} >{\raggedright\arraybackslash}p{0.18\textwidth} c c >{\raggedright\arraybackslash}p{0.255\textwidth} c}
\toprule
\textbf{ID} & \textbf{Model} & \textbf{Attack} & \textbf{Budget} & \textbf{Cov.} & \textbf{Intl.} & \textbf{Result} & \textbf{Detected / exact} \\
\midrule
L3.1 & Qwen2-0.5B & Code-aware & 13 / 65 / 130 / 260 / 650 / 1,300 & 40\% & adj & PPL 18.1 / 28.0 / 48.1 / 769 / 37,635 / 93,819 & yes / no \\
L3.2 & & Code-aware & same budgets & 40\% & blind & PPL 17.65 at every budget (max/cw $\le3$) & yes / yes \\
L3.3 & & Code-aware & same budgets & 40\% & key & PPL 18.4 / 34.9 / 50.4 / 137 / 1,170 / 312,778 & yes / no \\
L3.4 & Llama-3.2-1B & Code-aware & 130 / 260 / 650 / 1,300 / 2,600 & 40\% & adj & PPL 14.9 / 15.8 / 19.1 / 23.0 / n.a. & yes / no \\
L3.5 & & Code-aware & same budgets & 40\% & key & PPL 15.0 / 16.1 / 23.9 / 48.7 / 129.5 & yes / no \\
\midrule
L3.6 & Qwen2-0.5B & SilentStriker & 160 flips & 60\% & key & MMLU 43.0 / 45.0, PPL 20.98 / 17.65, 146 of 160 codewords flagged & yes / yes \\
L3.7 & Llama-3.2-1B & SilentStriker & 160 flips & 60\% & key & MMLU 22.0 / 31.0, PPL 21.02 / 14.20, 157 of 160 codewords flagged & yes / yes \\
L3.8 & Qwen2.5-7B & SilentStriker & 160 flips & 60\% & key & MMLU 76.0 / 78.5, PPL 8.05 / 7.74, 160 of 160 codewords flagged & yes / yes \\
L3.9 & Llama-3-8B-Inst. & SilentStriker & 160 flips & 60\% & key & MMLU 37.0 / 70.0, PPL 12.3 / 10.89 & yes / yes \\
\midrule
L3.10 & Qwen2-0.5B & Sign flips & 1 & 10\% & key & corrected bit-exactly (max/cw 1) & 20/20 / 20/20 \\
L3.11 & & Sign flips & 2 & 10\% & key & corrected bit-exactly & 20/20 / 20/20 \\
L3.12 & & Sign flips & 5 & 10\% & key & corrected bit-exactly & 20/20 / 20/20 \\
L3.13 & & Sign flips & 10 & 10\% & key & corrected bit-exactly & 20/20 / 20/20 \\
L3.14 & & Sign flips & 20 & 10\% & key & corrected bit-exactly & 20/20 / 20/20 \\
L3.15 & & Sign flips & 50 & 10\% & key & corrected bit-exactly & 20/20 / 20/20 \\
L3.16 & & Sign flips & 100 & 10\% & key & corrected bit-exactly & 20/20 / 20/20 \\
\bottomrule
\end{tabular}

\end{table}

\section{Algorithm and Embedding Details}
\label{app:algo}

Algorithm~\ref{alg:main} states \sanai end to end. Figure~\ref{fig:embedding} illustrates the two-scale lattice used when $B$ parity bits are embedded per symbol: a group of $B$ bits is read as a level in $[0,2^{B}-1]$, the coarse step $\Delta2^{B}$ selects the cell, the fine step $\Delta$ places one sub-level per value inside each cell, and decoding rounds to the nearest fine sub-level, $\hat\beta=\mathrm{round}((p-d)/\Delta)\bmod2^{B}$. Setting $L_s=1$ recovers plain QIM capacity of one bit per carrier weight while keeping key dependence; raising $L_s$ averages $L_s$ independent low-bit perturbations, trading capacity for robustness of the hidden parity.

\begin{algorithm}[h]
\SetAlgoLined
\DontPrintSemicolon
\small
\KwIn{Quantized weights $\{\theta_\ell\}_{\ell=1}^{L}$ at $b$ bits; coverage $\rho$; planes $K$; parent code BCH$(n,k,t)$ over $\mathrm{GF}(2^m)$ with $n=2^m-1$, storing $p$ parity bits per $k_s$-bit message; group size $G$; lattice step $\Delta$; spread $L_s$; key $\kappa$; probe rate $p_0$}
\KwOut{Protected weights $\theta'$ (offline); restored weights $\hat\theta$ (read time)}
\tcp{Phase 1: protect, offline, run once}
\ForEach{layer $\ell$}{$S_\ell \leftarrow A_{\mathrm{clean}} - \mathbb{E}_{\epsilon\sim\mathrm{Bern}(p_0)}[A(\theta_\ell\oplus\epsilon)]$ \tcp*{Eq.~\ref{eq:sens}}}
sort layers so that $S_{\ell_{(1)}}\ge\cdots\ge S_{\ell_{(L)}}$; $\mathcal{P}\leftarrow$ largest prefix with $\sum_{\ell\in\mathcal{P}}|\theta_\ell|\le\rho W$ \tcp*{Eq.~\ref{eq:select}}
$\mathcal{C}\leftarrow$ remaining layers, least sensitive first; \quad $\pi\leftarrow\mathrm{KeyedPerm}(\mathcal{P},\kappa)$ \tcp*{Eq.~\ref{eq:interleave}}
\ForEach{codeword $c$ of $G$ weights from $\pi(\mathcal{P})$}{$M_c\leftarrow\lfloor\theta_c/2^{\,b-K}\rfloor$; \; $\pi_c\leftarrow\mathrm{BCH}_{(n,k,t)}(M_c)$}
$\Pi\leftarrow\pi_{c_1}\Vert\pi_{c_2}\Vert\cdots$; \textbf{assert} $|\Pi|\le W_cB/L_s$ \tcp*{Eq.~\ref{eq:ceiling}}
$(v,d)\leftarrow\mathrm{PRNG}(\kappa)$ per carrier group \;
\ForEach{parity symbol $\beta\in\Pi$ with carrier group $w\subset\mathcal{C}$}{$p\leftarrow w\cdot v+d$; \; snap to $p^{\star}$; \; $w\leftarrow w+\frac{p^{\star}-p}{L_s}v$ \tcp*{Eq.~\ref{eq:stdm}, $|\delta w|\le\Delta/2$}}
$\theta'\leftarrow\{\theta_\ell\}$ \tcp*{$|s(\theta')|=|s(\theta)|$: zero added memory}
\tcp{Phase 2: restore, read time}
$\tilde\theta\leftarrow$ read from faulty memory; \; $(v,d)\leftarrow\mathrm{PRNG}(\kappa)$ \;
\ForEach{carrier group $w'\subset\mathcal{C}$}{$\hat\beta\leftarrow\lfloor(w'\cdot v+d)/(\Delta/2)\rfloor\bmod2$}
unpack $\{\hat\pi_c\}$ from $\hat\Pi$ \;
\ForEach{codeword $c$ of $\pi(\mathcal{P})$}{$\hat M_c\leftarrow\mathrm{BCH\_Decode}(\tilde M_c,\hat\pi_c)$; \; $\hat\theta_c\leftarrow\hat M_c\cdot2^{\,b-K}+(\tilde\theta_c\bmod2^{\,b-K})$; \; flag if syndrome nonzero}
\Return $\hat\theta\leftarrow\pi^{-1}(\{\hat\theta_c\})\cup\{\tilde\theta_\ell\}_{\ell\notin\mathcal{P}}$ \;
\caption{\sanai: sensitivity-guided bit-plane ECC with keyed interleaving and cross-layer STDM parity hiding}
\label{alg:main}
\end{algorithm}

\paragraph{Complexity.}
Let $C_{\mathrm{eval}}$ be the cost of one validation pass. The protect phase costs $O(L\,C_{\mathrm{eval}})$ for probing, $O(L\log L)$ for selection, $O(N_{\mathrm{cw}}nt)$ for BCH encoding, and $O(W_c)$ for STDM embedding. The restore phase costs $O(W_c)$ for de-projection, $O(N_{\mathrm{cw}}(nt+t^{2}))$ for Berlekamp-Massey decoding \citep{massey1969shift}, and $O(W_pK)$ for splicing, linear in model size and independent of the fault rate. The mixed-integer programs of prior embedded ECC are NP-hard in general and need commercial solvers whose runtime grows sharply with layer count.

\section{Proofs}
\label{app:proofs}

\paragraph{Proposition~\ref{prop:survival}.}
The decoder corrects any pattern of at most $t$ errors and fails on any pattern of more, so the failure event is $\{X>t\}$ with $X\sim\mathrm{Bin}(n,p)$, which gives the binomial tail. Writing $X/n$ as the mean of $n$ independent Bernoulli$(p)$ variables and applying Hoeffding's inequality \citep{hoeffding1963probability} to the deviation $f-p>0$ gives $\Pr[X/n>f]\le\exp(-2n(f-p)^{2})$. \qed

\paragraph{Lemma (random-fault correctability).}
Let each codeword hold $GK$ protected bits flipped independently with probability $p$. Then $f_c\sim\mathrm{Bin}(GK,p)$ and $\Pr[f_c>t]=\sum_{i>t}\binom{GK}{i}p^{i}(1-p)^{GK-i}$, which for $GKp\ll t$ is $O((GKp)^{t+1})$. With $GK=96$, $t=4$, $p=0.06$ the per-codeword failure probability is small and the model-level accuracy is preserved, matching the flat protected curves in Master Table~\ref{tab:master_v1}.

\paragraph{Lemma (adversarial concentration).}
A targeted attacker with a flip budget $F$ maximizes damage by maximizing $\max_c f_c$. Under adjacent grouping all $F$ flips of a sensitive region can be placed in $\lceil F/(GK)\rceil$ codewords, so one codeword may reach $f_c=\min(F,GK)\gg t$. This is the measured behaviour: 500 PBS flips drive 29 codewords above $t=4$ with a peak of 27 (Table~\ref{tab:attacks}).

\paragraph{Proposition~\ref{prop:access}.}
(a) With the correct key the reader forms $w'\cdot v+d=p^{\star}+\delta$ where $\delta$ is the read-time perturbation of the projection; $p^{\star}$ lies on the lattice selected by $\beta$ and the two lattices are separated by $\Delta/2$, so rounding returns $\beta$ whenever $|\delta|<\Delta/2$. (b) For an independent pseudo-random $v'\in\{-1,+1\}^{L_s}$, $\mathbb{E}[v\cdot v']=0$ and $w'\cdot v'$ is a sum of terms whose signs are independent of $\beta$; the parity of its rounded level is a Bernoulli$(\tfrac12)$ variable independent across groups, so all $P_{\mathrm{bits}}$ bits are recovered with probability $2^{-P_{\mathrm{bits}}}$. \qed

\paragraph{Theorem~\ref{thm:interleave}.}
Since $\pi$ is uniform and unknown to the attacker, the codeword of each chosen weight is a uniformly random bin among $M$, independent across flips up to the without-replacement correction, which only tightens the bound. The maximum load of $F$ balls in $M$ bins is $O(F/M+\log M/\log\log M)$ in expectation \citep{raab1998balls}. For $F\ll M\log M$ the first term vanishes and the second is $O(1)$ for any practical $M$, so $\max_cf_c\le t$ with high probability. \qed

\paragraph{Worked example of the interleaving bound.}
Take the light code, $t=4$ and $G=24$ weights per codeword, and a model of $M=1000$ codewords. An attacker who knows the code parameters flips $F=10$ bits in one physical region, intending to overload one codeword. Under adjacent grouping all ten land in that codeword, $f_c=10>t$, and it fails; the attacker wins with $t+1=5$ flips. Under a keyed permutation the ten flips fall into the $M$ codewords like balls into bins: $\mathbb{E}[f_c]=F/M=0.01$ and $\max_c f_c\approx F/M+\ln M/\ln\ln M\approx3.6\le t$, so every codeword is corrected although ten bits were flipped. Setting the bound above $t$ gives $F/M\gtrsim0.4$, so the blind attacker needs on the order of 400 flips before any codeword fails, against 5 under adjacent grouping. Without the key the attacker cannot direct a flip into a chosen codeword: the probability of landing $t+1$ blind flips in one specific codeword is about $M^{-5}=10^{-15}$. Master Table~\ref{tab:master_a1}, block B, is the measured counterpart: 500 real PBS flips, 22 uncorrectable codewords under adjacent grouping and none under interleaving.

\paragraph{Sign-bit bound.}
A sign-bit attack alters one bit plane per weight; each weight contributes its sign bit to exactly one codeword, so the attack injects at most $G$ errors per codeword. If $G\le t$ every sign-bit configuration lies within the correction radius and is repaired exactly. Defeating a codeword therefore requires a second error on some weight, that is a second flipped plane. \qed

\paragraph{Proposition~\ref{prop:hwreduce}.}
At $L_s=1$ the direction $v\in\{-1,+1\}$ is a scalar and $p=vw+d$ is an affine bijection of the integer weight. With $\Delta$ equal to one quantization step the two interleaved lattices are the even and odd integer levels, so snapping $p$ to the lattice selected by $\beta$ sets the parity of the integer level, that is the LSB of the stored weight. The dither $d=\tfrac{\Delta}{2}z$ shifts the lattice by half a step, which toggles which lattice is labelled $\beta$ and complements the written bit when $z=1$; the composition writes $\beta\oplus z$. Reading inverts the same affine map and takes the result modulo 2, recovering $\beta\oplus z$, and XOR with $z$ returns $\beta$. Rounding to the nearest level tolerates any displacement below $\Delta/2$. \qed

\section{Qualitative Comparison of Weight-Protection Schemes}
\label{app:related}

Table~\ref{tab:qual} places \sanai beside the schemes discussed in Section~\ref{sec:related} along the axes that decide whether a protection is deployable on an edge accelerator: added memory, where the parity lives, the granularity at which protection is applied, the fault tolerance actually demonstrated, the accuracy cost of the protection itself, whether retraining is needed, whether the parity is bound to a key, whether it survives fine-tuning, and which architectures it was shown on. Three patterns stand out. First, every scheme that adds no memory stores its parity next to the data it protects, so a burst that hits a block takes weight and check bits together; \sanai is the only one whose parity is moved cross-layer into carriers chosen for their tolerance. Second, no prior scheme binds parity to a key, so a copied model can always be repaired by whoever holds it; the key-seeded STDM embedding is what turns fault tolerance into an access-control primitive. Third, the demonstrated fault tolerance of the embedded schemes stops at a six percent BER, and none report behaviour under fine-tuning or on transformers or language models. The last row summarizes what the rest of the paper measures.

\begin{table}[h]
\centering
\caption{Qualitative comparison of weight-protection schemes. $^{*}$ marks the strongest prior embedded scheme. \sanai is the only scheme that keeps parity in-weight and moves it cross-layer, decoupling the parity from what it protects.}
\label{tab:qual}
\scriptsize
\setlength{\tabcolsep}{3.5pt}
\renewcommand{\arraystretch}{1.12}
\begin{adjustbox}{max width=\textwidth}
\begin{tabular}{lccccccccc}
\toprule
\textbf{Scheme} & \textbf{Mem.} & \textbf{Parity} & \textbf{Granul.} & \textbf{Fault tol.} & \textbf{Acc.\ drop} & \textbf{Retrain} & \textbf{Key} & \textbf{Fine-tune} & \textbf{Arch.} \\
\midrule
Conventional ECC \citep{reagen2018ares} & $\sim$12.5\% & External & Word & High, 1 to 2 bit & Negligible & No & None & Unaffected & Agnostic \\
Weight scaling \citep{hong2019terminal} & Low & None & Tensor & Low at high BER & $>3\%$ & No & None & Weakens & CNN \\
NN-ECC \citep{koppula2019eden} & Low & In-weight & Layer & High, learned & Low & Full & None & Re-embed & CNN \\
Weight throttling \citep{he2020defending} & Low & None & Clipped & Saturates & Moderate & Yes & None & Drift-sensitive & CNN \\
Selective prot.\ \citep{rakin2019bit} & Low & External & Weight & Uneven & Mod.\ high BER & No & None & Not rep. & CNN \\
Weight nulling \citep{qin2017robustness} & Low & None & Zeroing & Fails at 6\% & High & No & None & Not rep. & CNN \\
Opportunistic \citep{burel2021zero} & Low & In-weight & Block & Gaps & High at 6\% & No & None & Not rep. & CNN \\
Value-aware \citep{lee2022value} & Low & In-weight & Magnitude & Medium & Moderate & No & None & Not rep. & CNN \\
E3-CODE$^{*}$ \citep{kamineni2026e3} & Low & In-weight & Layer-uniform & High to 6\% & Low to 6\% & No & None & Not rep. & CNN \\
\midrule
\hl
\textbf{\sanai (ours)} & \textbf{Zero} & \textbf{Cross-layer} & \textbf{Adaptive BP} & \textbf{High, 10 to 12\%} & \textbf{0.97 to 0.99} & \textbf{No} & \textbf{Key-gated} & \textbf{Parity BER 0} & \textbf{CNN, ViT, LLM} \\
\bottomrule
\end{tabular}
\end{adjustbox}
\end{table}

\section{Experimental Details}
\label{app:setup}

\begin{table}[h]
\centering
\caption{Evaluated vision architectures. Params is the ImageNet reference count; the INT8 footprint in MB equals the parameter count in millions. CIFAR-adapted variants of the first three models ($3\times3$ stride-1 stem, no early max-pooling) hold 11.16M, 23.52M, and 14.73M weights.}
\label{tab:arch}
\small
\begin{adjustbox}{max width=\linewidth}
\begin{tabular}{llcc}
\toprule
\textbf{Model} & \textbf{Family} & \textbf{Params (M)} & \textbf{Datasets} \\
\midrule
ResNet-18 / ResNet-50 \citep{he2016deep} & Residual CNN & 11.7 / 25.6 & C10, C100, IN \\
VGG-16 \citep{simonyan2014very} & Plain CNN & 138.4 & C10, C100 \\
MobileNet-V2 \citep{sandler2018mobilenetv2} & Depthwise CNN & 3.5 & C10, C100, IN \\
EfficientNet-B0 \citep{tan2019efficientnet} & Depthwise CNN & 5.3 & C10, C100, IN \\
Wide ResNet-50-2 \citep{zagoruyko2016wide} & Residual CNN & 68.8 & IN \\
RegNet-Y-16GF \citep{radosavovic2020designing} & Design-space CNN & 83.4 & IN \\
Swin-T \citep{liu2021swin} & Hierarchical ViT & 28.2 & IN \\
MaxViT-T \citep{tu2022maxvit} & Hybrid ViT & 30.4 & IN \\
ViT-B/16 \citep{dosovitskiy2020image} & Plain ViT & 85.8 & IN \\
\bottomrule
\end{tabular}
\end{adjustbox}
\end{table}

\begin{table}[h]
\centering
\caption{Operating points of the one mechanism. Both codes are shortened from parents over $\mathrm{GF}(2^7)$ ($x^7+x+1$, $n=127$); ``48+88'' means 48 message plus 88 stored parity bits per codeword. $G$ is weights per codeword, $f=t/n$ the correctable fraction of the parent, $\rho$ the coverage. The sign-bit guarantee $G\le t$ holds only for the LLM point. All points add zero memory.}
\label{tab:ops}
\small
\setlength{\tabcolsep}{5pt}
\begin{adjustbox}{max width=\linewidth}
\begin{tabular}{llcccccl}
\toprule
\textbf{Regime} & \textbf{Code} & $K$ & $G$ & $f$ & $\rho$ & \textbf{Interleave} & \textbf{Used in} \\
\midrule
Random BER, vision & strong: parent BCH$(127,50,12)$, 48+88 bits & 3 (4 depthwise, ImageNet) & 16 ($>t$) & 9.4\% & 20\% (40\% ImageNet) & optional & Sec.~\ref{sec:vision}, \ref{sec:sota}, \ref{sec:security} \\
Attacks, vision & light: parent BCH$(127,99,4)$, 96+32 bits & 4 & 24 ($>t$) & 3.1\% & 10\% & keyed & Sec.~\ref{sec:attacks} \\
LLMs & strong: parent BCH$(127,50,12)$, 48+88 bits & 4 & 12 ($\le t$) & 9.4\% & 10 to 60\% & keyed (attacks) & Sec.~\ref{sec:llm} \\
\bottomrule
\end{tabular}
\end{adjustbox}
\end{table}

\paragraph{Quantization and injection.}
PTQ uses per-output-channel scales for convolutions and per-tensor scales for linear layers. QAT checkpoints (Jacob, LSQ) store their exact training-time scales, so integer tensors are recovered losslessly before protection. Faults are injected into the two's-complement byte representation of the integer weights. Each BER point uses a fixed random seed for vision; LLM random-BER results average three seeds (0, 1, 2). CIFAR models are trained to convergence; ImageNet models use torchvision pretrained weights, with clean INT8 top-1 of 76.1 to 83.4\%.

\paragraph{Sensitivity proxies.}
For vision the probe of Eq.~\ref{eq:sens} at $p_0=2\%$ ranks layers. For LLMs three proxies were compared: weight-$\ell_1$ magnitude, gradient-weighted magnitude $\sum|g\odot w|$ from one backward pass over four 5-shot MMLU prompts, and the attack's own per-layer loss probe. The two gradient-informed proxies choose 118 small layers (39.1\% of weights) on Qwen2-0.5B where magnitude chooses 32 large ones (39.0\%); Appendix~\ref{app:llm} shows only the gradient-informed proxies protect the attacked layers. Token-embedding and output-projection matrices are excluded as carriers and from injection.

\paragraph{Attack re-implementations.}
PBS and T-BFA follow the public implementations \citep{rakin2019bit,rakin2021tbfa} on the INT8 tensors; T-BFA uses 90 source-target pairs on CIFAR-10 and 132 on ImageNet. AttentionBreaker follows the public code: the attack loss is the causal-LM cross-entropy over a batch of four 5-shot MMLU prompts; the per-weight score is $\alpha\widehat{|g|}+(1-\alpha)\widehat{|w|}$ with $\alpha=0.5$ min-max normalized per layer; every attention and MLP projection is ranked by the loss reached after flipping the sign bit of its top $p\%$ weights, $p\in\{10^{-3},10^{-2},0.1,1,5\}$; the population space is the top 2\% of the most fragile layer; GenBFA minimizes the flip set subject to $\mathcal{L}\ge7$ with tournament selection, uniform crossover, 1 to 9\% drop-mutation, and 100 generations of 40 candidates. All flips are on bit 7. We re-implement on \sanai's own INT8 tensors rather than on \texttt{bitsandbytes} LLM.int8() \citep{dettmers2022llmint8} so that every flip lands on exactly the integer \sanai encodes. MMLU for the attack experiments uses 500 questions from 20 subjects and WikiText-2 perplexity 20k test tokens.

\section{Portability to a Deployment Format}
\label{app:onnx}

To confirm that \sanai does not depend on the training framework, we apply it directly to ONNX graphs \citep{onnx2017}: the INT8 weight initializers of the exported model are the only inputs to the scheme, and the protected file runs unchanged in ONNX Runtime \citep{onnxruntime2018} with no custom operator or runtime hook. On the CIFAR-10 SimpleCNN (static QDQ INT8, 88.15\% clean), a BCH$(63,39,t{=}4)$ code over $\mathrm{GF}(2^{6})$, shorter than the paper's family to match the small classifier, protects the top four planes of the classifier, and its parity is hidden in the convolution layers by integer-lattice STDM ($\Delta=8$, one bit per 16-weight block) at zero file growth (154,589 bytes before and after) and zero accuracy cost; protecting all eight planes costs 0.18 points. Twenty sign flips and a 1\% random BER are fully detected (18 of 18 and 40 of 40 affected codewords flagged, none falsely) and corrected bit-exactly. On DistilGPT-2 (weight-only per-channel INT8, 76.4 WikiText-2 perplexity under this quantizer and token set, hence not directly comparable with Table~\ref{tab:llm}), we protect the sign and top magnitude planes of the most gradient-sensitive projection (\texttt{h.0.attn.c\_proj}) and, relaxing the carrier exclusion of Section~\ref{sec:setup}, place the 884,736 parity bits in the token-embedding table, the largest and least sensitive tensor in the graph, again at zero file growth and a 0.04 perplexity cost. Fifty gradient-targeted sign flips that raise perplexity to 3,080 are flagged (38 of 38 affected codewords) and reversed exactly, and a 0.1\% BER over all planes is brought from 258 to 77.7 perplexity, near clean, by restoring the two protected planes alone. In both cases every corrupted codeword was flagged and every clean codeword passed, so the sensitivity-aware, zero-overhead protection and the detect-and-flag behaviour carry over unchanged to a standard deployment format.

\section{Storage Accounting and the Precision Squeeze}
\label{app:storage}

\begin{table}[h]
\centering
\caption{Left: storage cost at the efficient operating point (INT8). Middle: absolute counts in millions; $^{a}$CIFAR-100. Right: fit for the ImageNet backbones, protected fraction, hidden parity as a fraction of model bits, parity bits per model weight, and carrier layers (insensitive / protected-layer overflow); the Swin-T, RegNet, and Wide ResNet rows are from the earlier logs pending re-extraction. Added memory is zero in every row.}
\label{tab:storage}
\scriptsize
\setlength{\tabcolsep}{3pt}
\begin{minipage}[t]{0.24\linewidth}
\begin{adjustbox}{max width=\linewidth}
\begin{tabular}{llcc}
\toprule
\textbf{Arch} & \textbf{Data} & \textbf{Cov.} & \textbf{b/wt} \\
\midrule
ResNet-18 & C10 & 19.3\% & 1.06 \\
ResNet-18 & C100 & 14.0\% & 0.77 \\
ResNet-50 & C10 & 19.2\% & 1.05 \\
VGG-16 & C10 & 20.0\% & 1.10 \\
\bottomrule
\end{tabular}
\end{adjustbox}
\end{minipage}\hfill
\begin{minipage}[t]{0.42\linewidth}
\begin{adjustbox}{max width=\linewidth}
\begin{tabular}{lccccc}
\toprule
\textbf{Arch} & \textbf{Lay.} & \textbf{Wts} & \textbf{Bits} & \textbf{Prot.\ wts} & \textbf{Parity bits} \\
\midrule
ResNet-18 & 13/20 & 11.16 & 89.3 & 2.12 (19.0\%) & 11.64 (13.0\%) \\
ResNet-18$^{a}$ & 11/20 & 11.17 & 89.4 & 1.56 (14.0\%) & 8.62 (9.6\%) \\
ResNet-50 & 28/53 & 23.52 & 188.2 & 4.52 (19.2\%) & 24.78 (13.2\%) \\
VGG-16 & 10/16 & 14.73 & 117.8 & 2.95 (20.0\%) & 16.18 (13.7\%) \\
\bottomrule
\end{tabular}
\end{adjustbox}
\end{minipage}\hfill
\begin{minipage}[t]{0.31\linewidth}
\begin{adjustbox}{max width=\linewidth}
\begin{tabular}{lcccc}
\toprule
\textbf{Model} & \textbf{Prot.} & \textbf{Parity} & \textbf{b/wt} & \textbf{Carr.} \\
\midrule
MaxViT-T & 40.9\% & 37.5\% & 3.00 & 132 \\
Swin-T & 39.5\% & 36.2\% & 2.90 & 35 / 0 \\
RegNet-Y-16GF & 40.0\% & 36.6\% & 2.93 & 70 / 1 \\
Wide RN-50-2 & 37.3\% & 34.2\% & 2.73 & 21 / 0 \\
ViT-B/16 ($K{=}5$) & 43.5\% & 53.2\% & 4.26 & 26 / 8 \\
\bottomrule
\end{tabular}
\end{adjustbox}
\end{minipage}
\end{table}

Counting the full code expansion $1+p/k_s=1+88/48$ instead of the systematic parity fraction $p/k_s$ would predict about 1.7 bits per weight at the default operating point and contradict the measured 1.05 to 1.10; the systematic property is what keeps the cost low. For the LLM random-BER experiments, all three small models hide the full payload at 3.5 to 4.0 bits per weight from 17 to 32 insensitive carrier layers (Qwen2-0.5B: 24 insensitive plus 8 overflow; GPT-2: 32 plus 8; DistilGPT-2: 17 plus 1), and the attack-regime configuration at 10\% coverage uses about 9.2\% of model bits with no overflow.

\begin{figure}[h]
\centering
\begin{minipage}[t]{0.47\textwidth}\centering
\includegraphics[width=\linewidth]{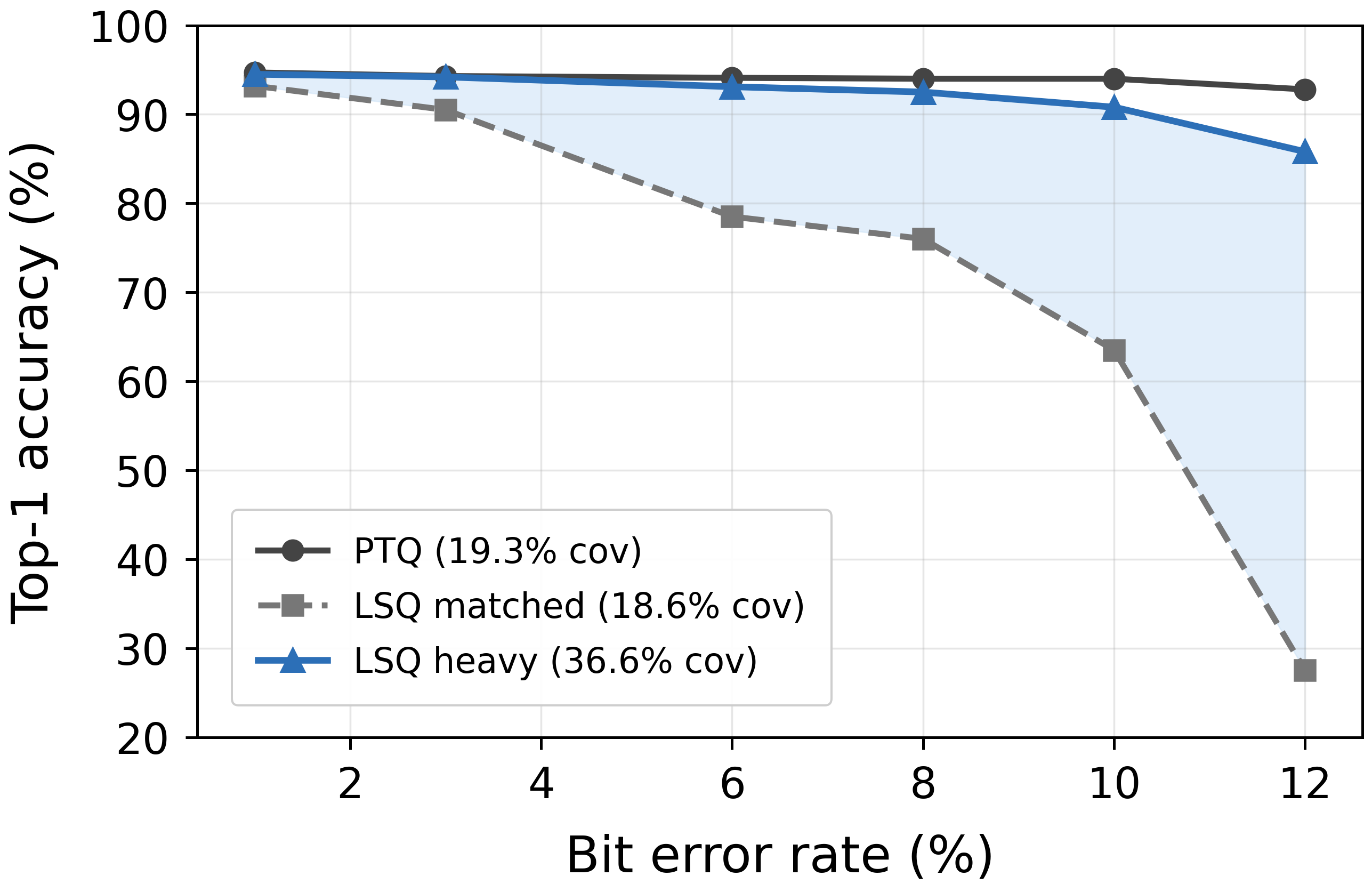}
\end{minipage}\hfill
\begin{minipage}[t]{0.47\textwidth}\centering
\includegraphics[width=\linewidth]{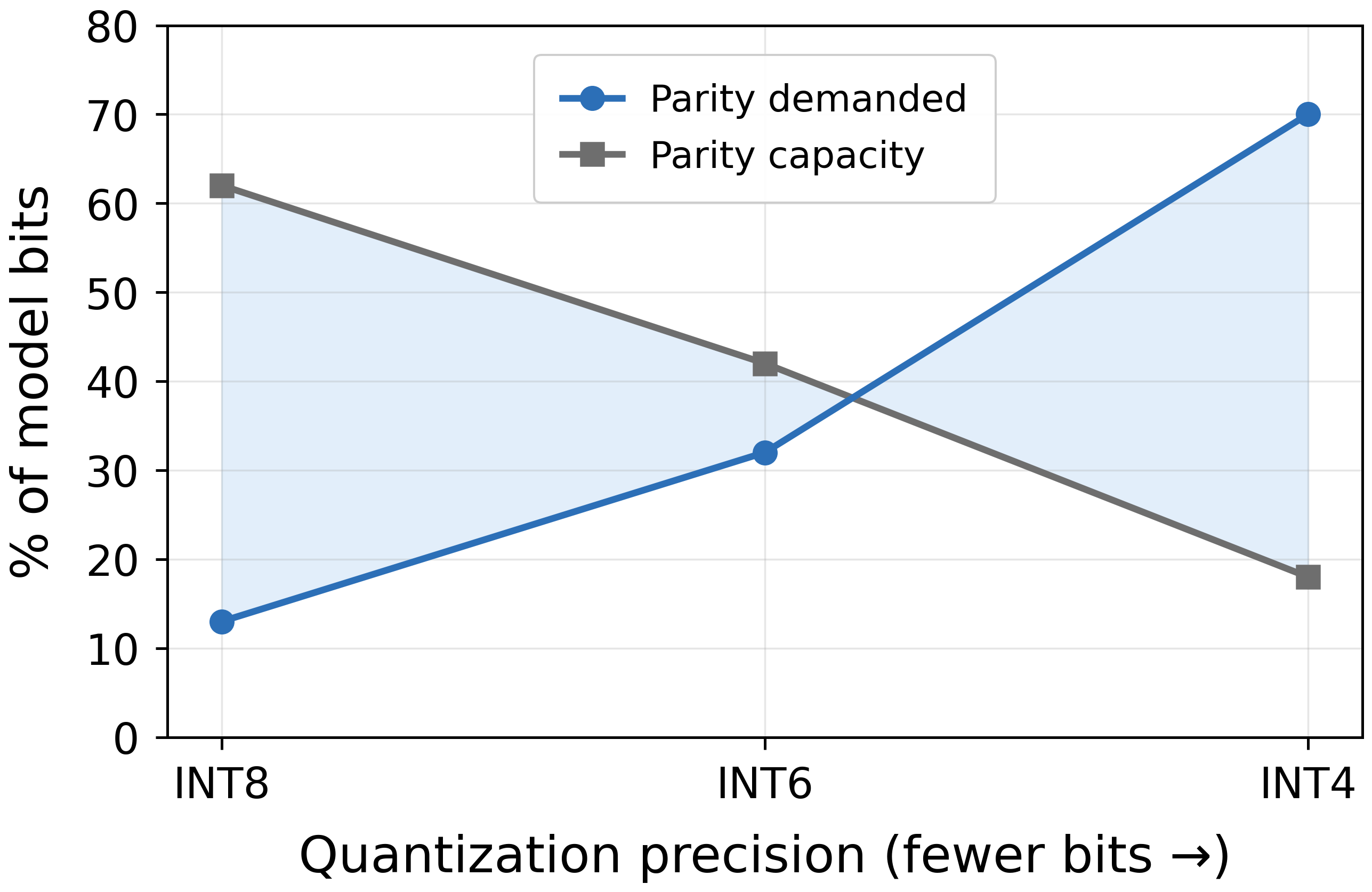}
\end{minipage}
\caption{Left: quantizer ablation; LSQ needs roughly $2.5\times$ the protection budget of PTQ. Right: the precision squeeze; parity demand rises while carrier capacity falls, and configurations past the crossing cannot be protected at zero added memory.}
\label{fig:squeeze}
\end{figure}

\paragraph{Precision squeeze.}
Two effects act together as precision falls. Lower precision makes more layers fragile, so the sensitive set grows and demand rises; the same shrinking bit budget leaves fewer robust carriers with spare low-order bits, so the capacity $W_ch$ falls. Figure~\ref{fig:squeeze} (right) plots both: demand rises from about 13\% of model bits at INT8 to about 70\% at INT4 while capacity falls from about 62\% to 18\%, and past the crossing no assignment of parity to carriers satisfies Eq.~\ref{eq:ceiling}. INT6 on CIFAR-10 remains feasible (a small budget degrades to about 15\% at 12\% BER, 35\% coverage with $K=4$ recovers to about 86\%); INT6 on ImageNet and INT4 sit past the crossing.

\section{Extended Vision Results}
\label{app:vision}

\begin{table}[h]
\centering
\caption{Full comparison with embedded ECC under burst errors: top-1 accuracy at 1.5, 3, and 6\% BER for all six methods and eight settings. Norm.\ is accuracy normalized to each method's own software baseline; baseline values as reported by \citet{kamineni2026e3}.}
\label{tab:baseline_full}
\scriptsize
\setlength{\tabcolsep}{2.4pt}
\renewcommand{\arraystretch}{1.05}
\begin{adjustbox}{max width=\textwidth}
\begin{tabular}{llc rrr rrr rrr rrr rrr rrr}
\toprule
& & \textbf{SW} & \multicolumn{3}{c}{\textbf{W.\ Nulling}} & \multicolumn{3}{c}{\textbf{Opp.\ Parity}} & \multicolumn{3}{c}{\textbf{Value-Aware}} & \multicolumn{3}{c}{\textbf{E3-CODE}} & \multicolumn{3}{c}{\textbf{E3-CODE-H}} & \multicolumn{3}{c}{\textbf{\sanai (ours)}} \\
\cmidrule(lr){4-6}\cmidrule(lr){7-9}\cmidrule(lr){10-12}\cmidrule(lr){13-15}\cmidrule(lr){16-18}\cmidrule(lr){19-21}
\textbf{Data} & \textbf{Model} & \textbf{Acc.} & 1.5 & 3 & 6 & 1.5 & 3 & 6 & 1.5 & 3 & 6 & 1.5 & 3 & 6 & 1.5 & 3 & 6 & 1.5 & 3 & 6 \\
\midrule
\multirow{4}{*}{C10}
& ResNet-18 & 82.0 & 82.0 & 19.0 & 0.05 & 81.7 & 81.4 & 10.9 & 82.0 & 81.2 & 77.5 & 81.8 & 81.8 & 78.8 & 82.0 & 81.9 & 81.3 & 94.9 & 94.7 & \textbf{94.5} \\
& ResNet-50 & 85.2 & 85.0 & 54.1 & 5.5 & 84.8 & 84.0 & 9.1 & 85.0 & 84.7 & 79.5 & 85.1 & 84.6 & 83.4 & 85.1 & 84.8 & 84.0 & 94.9 & 94.8 & \textbf{94.5} \\
& MobileNet-V2 & 83.6 & 83.0 & 8.8 & 0.0 & 82.9 & 79.5 & 9.1 & 83.0 & 81.3 & 41.2 & 83.0 & 82.4 & 59.4 & 83.6 & 83.0 & 76.8 & 78.2 & 76.6 & 75.3 \\
& EfficientNet-B0 & 82.8 & 82.0 & 2.9 & 0.0 & 80.3 & 64.8 & 9.8 & 82.0 & 77.9 & 22.6 & 82.3 & 78.6 & 55.6 & 82.7 & 82.3 & 80.1 & 83.8 & 83.0 & \textbf{81.4} \\
\midrule
\multirow{4}{*}{IN}
& ResNet-18 & 69.5 & 69.3 & 10.0 & 0.4 & 69.2 & 67.3 & 0.1 & 69.3 & 68.2 & 59.0 & 69.3 & 68.9 & 65.3 & 69.4 & 69.3 & 67.7 & 75.5 & 75.1 & \textbf{73.3} \\
& ResNet-50 & 76.1 & 75.9 & 18.8 & 0.4 & 75.8 & 73.8 & 0.0 & 75.9 & 75.1 & 65.1 & 75.9 & 75.8 & 70.0 & 76.1 & 75.9 & 74.7 & 82.2 & 81.6 & \textbf{80.5} \\
& MobileNet-V2 & 71.5 & 69.9 & 5.5 & 0.0 & 65.8 & 58.1 & 0.1 & 69.9 & 66.0 & 20.9 & 69.9 & 66.7 & 36.9 & 71.2 & 70.5 & 65.5 & 78.4 & 78.1 & \textbf{77.8} \\
& EfficientNet-B0 & 77.0 & 73.7 & 1.1 & 0.0 & 74.1 & 48.8 & 0.1 & 73.7 & 56.8 & 0.5 & 75.7 & 70.9 & 1.8 & 76.6 & 75.5 & 64.0 & 82.5 & 80.6 & \textbf{76.2} \\
\midrule
\hl
\multicolumn{2}{l}{\textbf{Norm.}} & 1.00 & 0.98 & 0.19 & 0.01 & 0.97 & 0.88 & 0.06 & 0.98 & 0.94 & 0.58 & 0.99 & 0.97 & 0.72 & 0.99 & 0.99 & 0.94 & \textbf{0.99} & \textbf{0.99} & \textbf{0.97} \\
\bottomrule
\end{tabular}
\end{adjustbox}
\end{table}

\paragraph{Critical BER.}
Figure~\ref{fig:critical} reports the improvement in critical BER relative to the unprotected model on ResNet-50/ImageNet. \sanai reaches $6\times10^{-2}$, $300{,}000\times$ the unprotected $2\times10^{-7}$ and $4\times$ beyond E3-CODE-H at $1.5\times10^{-2}$. The same ordering holds on EfficientNet-B0, where our $3\times10^{-2}$ exceeds E3-CODE-H at $7\times10^{-3}$ and value-aware parity at $4\times10^{-4}$. Under random BER at the attack operating point (rows V2.9 to V2.13 of Master Table~\ref{tab:master_v2}) the compact architectures hold through 9\% but drop at 12\%, consistent with the lower $f=3.1\%$ of the short $t=4$ code; the long $t=12$ code of the random-BER operating point is the right choice for that regime, which is why Table~\ref{tab:ops} keeps both.

\section{Extended Attack Results on Vision Models}
\label{app:attacks}

\begin{figure}[h]
\centering
\includegraphics[width=0.8\textwidth]{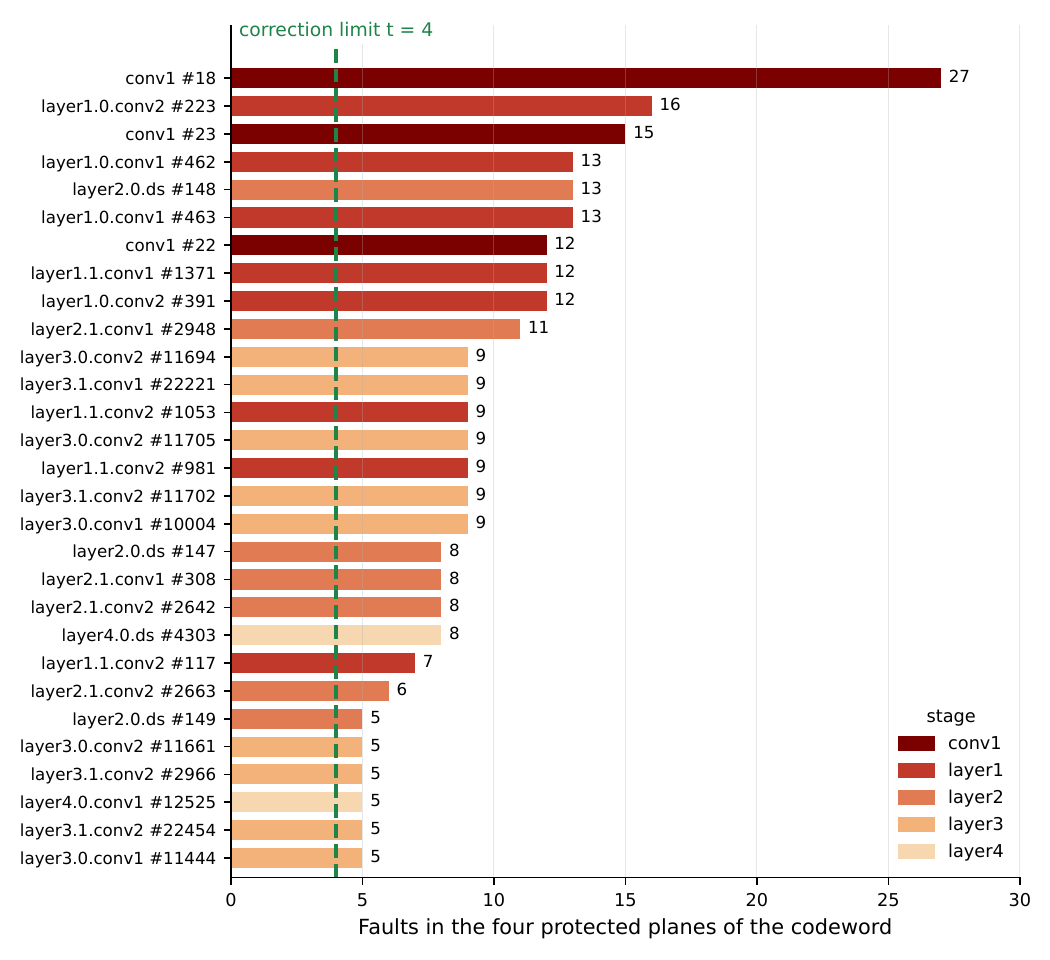}\\[6pt]
\includegraphics[width=0.98\textwidth]{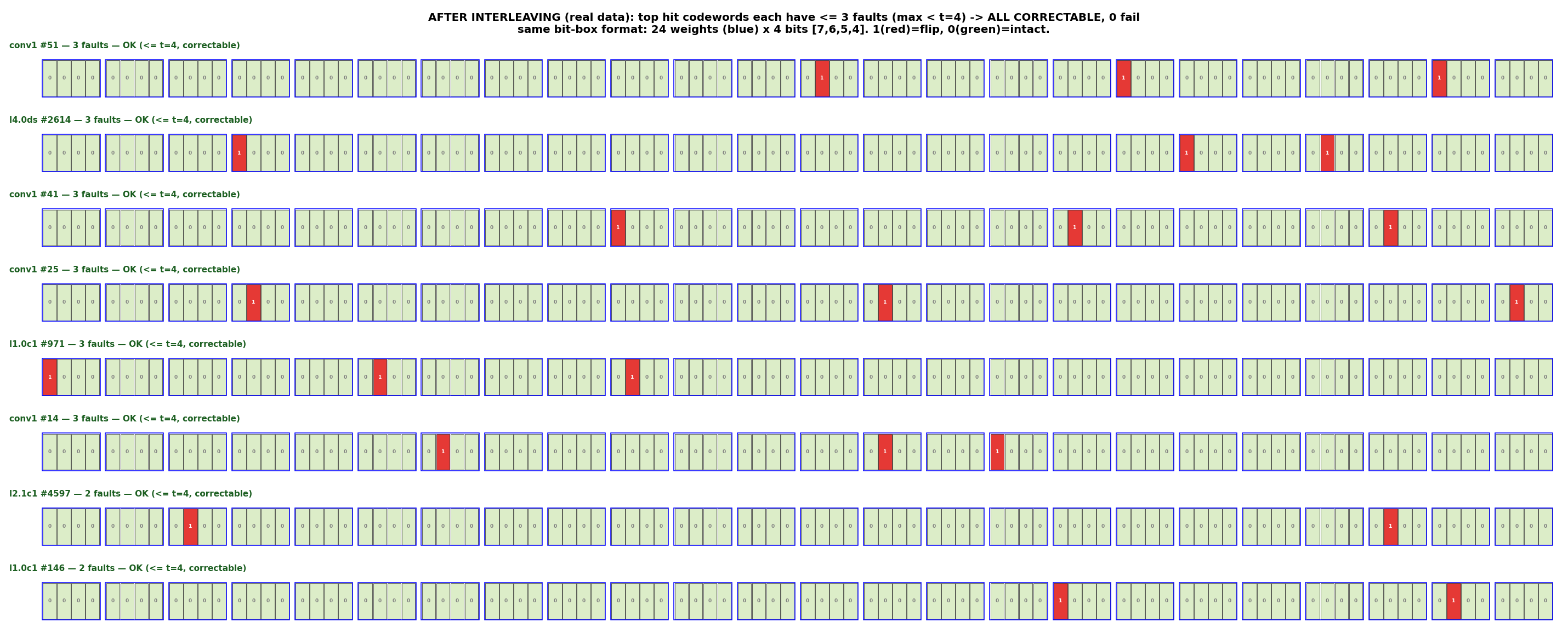}
\caption{Fault-map pass over the 500-flip PBS attack on ResNet-18/CIFAR-10 (light code, $t=4$, 24 weights per codeword, faults counted in the four protected planes). Top: all 29 codewords that exceed $t$ under adjacent grouping, ranked by fault count and labelled by layer and codeword index; the worst, \texttt{conv1} \#18, takes 27 faults, and the failures sit in the early convolutions where PBS concentrates. Bottom: bit-level map of the most heavily hit codewords after keyed interleaving; each row is a codeword of 24 weights with its four protected bits (7, 6, 5, 4), a highlighted cell is a flipped bit. Faults are scattered so that no weight carries more than one flip and no codeword holds more than 3 faults, all correctable.}
\label{fig:bitbox}
\end{figure}

Figure~\ref{fig:interleave} illustrates the mechanism; Figure~\ref{fig:faultmap} shows the measured per-codeword fault map behind Table~\ref{tab:attacks} and Figure~\ref{fig:intsummary} the accuracy across all six settings. An independent fault-map pass over the same ResNet-18/CIFAR-10 attack, counting only faults in the four protected planes with its own weight selection, gives the same picture with larger counts (29 uncorrectable codewords, the worst with 27 faults; 0 and a maximum of 3 after interleaving) and adds three observations: the flips concentrate on bits 7 and 6 and almost never touch bit 4, so the attack targets exactly the planes the code protects; the failing codewords sit in \texttt{conv1}, \texttt{layer1}, and \texttt{layer2}, where PBS focuses; and the worst codeword fails because nearly every one of its 24 weights is hit on bit 6 or 7, the concentration that the keyed permutation breaks (Figure~\ref{fig:bitbox}). Under adjacent grouping at 500 PBS flips on ResNet-18/CIFAR-10, 22 codewords exceed $t=4$ with a peak of 14; under interleaving the identical faults spread across many codewords, none exceeding 3. Across the six settings (Master Table~\ref{tab:master_a1}, block B) adjacent grouping leaves 9 to 49 codewords uncorrectable and accuracy at 0 to 35\%, while keyed interleaving leaves none (five on ResNet-50 ImageNet) and restores 94.9, 94.8, 75.5, 75.8, and 81.9\%. ResNet-18 on ImageNet is the one setting where interleaving corrects every codeword yet accuracy does not return (1.5\%): at 10\% coverage and 500 flips the unprotected 90\% of the model is corrupted, so correcting the protected codewords is insufficient; higher coverage or a lower flip budget resolves it. On the compact networks (Master Table~\ref{tab:master_a1}, block D) interleaving removes 90 to 100\% of the uncorrectable codewords (42 to 1, 35 to 0, 17 to 0, 23 to 3) and cuts the peak fault count from 10 to 24 down to 2 to 8; the residual one to three codewords on MobileNet-V2 and EfficientNet-B0 are all in the classifier layer, on which the attack concentrates entirely. The bit-level map of the most heavily hit interleaved codewords (Figure~\ref{fig:bitbox}) shows no weight carrying more than one flip. T-BFA succeeds almost always unprotected (90 to 100\% ASR with 10 to 20 flips), 97\% of which land on \texttt{fc.weight}; under \sanai the ASR falls to 0\% on most of the 132 ImageNet pairs, with a few residual leaks (Figure~\ref{fig:tbfa}).

\begin{figure}[h]
\centering
\includegraphics[width=0.95\textwidth]{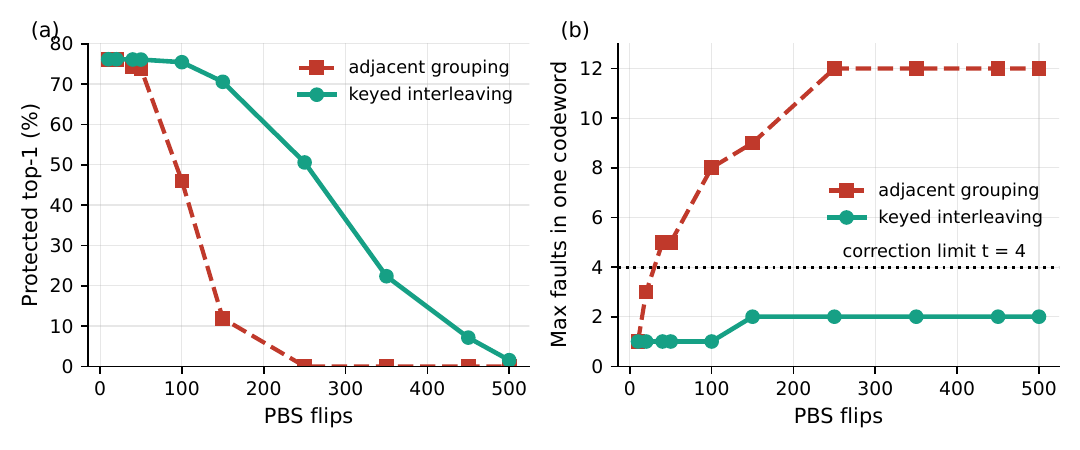}
\caption{ResNet-18/ImageNet flip sweep. Left: protected accuracy versus PBS flips; keyed interleaving holds far longer than adjacent grouping. Right: largest fault count in one codeword; interleaving stays at 1 to 2, below $t=4$, at every attack strength while adjacent grouping rises to 12. Where the interleaved accuracy still drops at high flip counts the per-codeword count is still below $t$, so the loss comes from the unprotected fraction (a coverage limit), not from interleaving.}
\label{fig:sweep_r18}
\end{figure}

\paragraph{Attack-strength sweep on ImageNet: interleaving versus coverage.}
To separate the effect of keyed interleaving from the effect of coverage, we sweep the PBS attack on ResNet-18/ImageNet from 10 to 500 flips (light code, $K=4$, 10\% coverage) and record, for adjacent and keyed grouping, the protected accuracy and the largest fault count in any codeword (Table~\ref{tab:sweep_r18}, Figure~\ref{fig:sweep_r18}). Two things follow. First, interleaving never fails at the codeword level: the interleaved maximum stays at 1 or 2 faults, below $t=4$, and no codeword becomes uncorrectable at any strength, whereas under adjacent grouping the maximum passes $t$ by 40 flips, reaches 12 by 250, and up to 14 codewords are uncorrectable. Second, interleaving extends the usable range substantially: at 100 flips adjacent grouping has fallen to 46.0\% while interleaved is at 75.4\%, and at 250 flips adjacent is at 0\% while interleaved retains 50.6\%. The interleaved accuracy does decline at very high budgets (1.5\% at 500 flips), but at those points every protected codeword is still corrected; the loss comes from the 90\% of the model left unprotected at 10\% coverage, which the attack corrupts directly once the budget is large. On this model the bottleneck at high attack strength is therefore coverage, not interleaving, which is why the ImageNet operating point of Section~\ref{sec:vision} uses 40\%.

\begin{table}[h]
\centering
\caption{ResNet-18/ImageNet PBS flip sweep (INT8, light code $t=4$, $K=4$, 10\% coverage): protected top-1 (\%), largest fault count in one codeword, and number of uncorrectable codewords, adjacent versus keyed grouping, same flipped bits.}
\label{tab:sweep_r18}
\small
\setlength{\tabcolsep}{6pt}
\begin{tabular}{rcccccc}
\toprule
 & \multicolumn{2}{c}{\textbf{Top-1 (\%)}} & \multicolumn{2}{c}{\textbf{Max faults / cw}} & \multicolumn{2}{c}{\textbf{Codewords $>t$}} \\
\textbf{Flips} & Adj. & Keyed & Adj. & Keyed & Adj. & Keyed \\
\midrule
10 & 76.14 & 76.14 & 1 & 1 & 0 & 0 \\
20 & 76.14 & 76.14 & 3 & 1 & 0 & 0 \\
40 & 74.19 & 76.08 & 5 & 1 & 2 & 0 \\
50 & 73.81 & 76.06 & 5 & 1 & 2 & 0 \\
100 & 45.98 & \textbf{75.41} & 8 & 1 & 3 & 0 \\
150 & 11.88 & \textbf{70.55} & 9 & 2 & 3 & 0 \\
250 & 0.00 & \textbf{50.55} & 12 & 2 & 10 & 0 \\
350 & 0.00 & 22.34 & 12 & 2 & 11 & 0 \\
450 & 0.00 & 7.12 & 12 & 2 & 14 & 0 \\
500 & 0.00 & 1.53 & 12 & 2 & 14 & 0 \\
\bottomrule
\end{tabular}
\end{table}

\section{Extended LLM Results}
\label{app:llm}

Figure~\ref{fig:llmfig} plots the protected MMLU and ARC-Easy accuracy of Table~\ref{tab:master_l1} against BER: every protected model stays well above the 25\% chance line through 6\%, while the unprotected models fall onto it at the first nonzero BER; the 6\% points of the 7B and 8B are conservative for the overflow reason given in the table caption.

\paragraph{Layer ranking and coverage.}
On Qwen2-0.5B under 10 sign flips (undefended 25.8\% / 1,294), a weight-$\ell_1$ ranking leaves the attacked \texttt{layers.1.self\_attn.o\_proj} unprotected and the model collapses as if undefended (row L2.8 of Master Table~\ref{tab:master_l2}), whereas gradient-weighted magnitude and the attack-probe loss both select it and recover 45.6\% / 17.65 at every coverage from 10\% to 40\% and under both layouts (rows L2.2, L2.9). On Llama-3-8B-Instruct the attack concentrates on \texttt{layers.1.mlp.down\_proj}; at 10\% and 20\% coverage that layer sits just outside the selected set and the recovered MMLU stays at the attacked 27.0, at 30\% it enters and recovery is exact (71.0; rows L2.23 to L2.25). This is a coverage property, not a code failure: whenever the attacked layer is protected, recovery is bit-exact, and the minimum coverage that guarantees this grows with model size.

\paragraph{Code-aware attacker on Llama-3.2-1B.}
The 1B model yields little to the code-aware attacker (rows L3.4 and L3.5 of Master Table~\ref{tab:master_l3}): 1,300 two-plane errors on the adjacent layout raise the corrected perplexity only from 14.20 to 23.0, the blind attacker on the interleaved layout achieves nothing, and the keyed attacker needs 2,600 errors to reach 129.5.

\paragraph{Qualitative outputs.}
Table~\ref{tab:outputs} shows generated answers from Llama-3-8B-Instruct in four states. AttentionBreaker collapses generation into repeated-token garbage; SilentStriker keeps the answer fluent but changes it, dropping the correct option or drifting off the question; after \sanai the output is bit-identical to the clean model in both cases, because the restored weights are bit-identical to clean. Qwen2-0.5B behaves the same way under the 100-flip attack: the attacked model emits code fragments, digit runs, or characters from another script, and after \sanai the generation is the clean model's generation again.

\begin{table}[h]
\centering
\caption{Representative Llama-3-8B-Instruct INT8 outputs before attack, under AttentionBreaker and SilentStriker, and after \sanai recovery. AttentionBreaker causes repeated-token corruption, while SilentStriker preserves fluency but alters the response. \sanai restores the clean output exactly.}
\label{tab:outputs}
\scriptsize
\renewcommand{\arraystretch}{1.05}
\setlength{\tabcolsep}{4pt}
\begin{tabular}{p{0.15\linewidth}p{0.19\linewidth}p{0.17\linewidth}p{0.20\linewidth}p{0.19\linewidth}}
\toprule
\textbf{Prompt} & \textbf{Before attack} & \textbf{AttentionBreaker} & \textbf{SilentStriker} & \textbf{After \sanai} \\
\midrule
What is true for a type-Ia supernova? &
A type-Ia supernova is a specific astronomical event. Here are the options: A) explosion of a massive star; B) merger of two white dwarfs; C) collapse of a neutron star; \ldots &
\texttt{<repeated-token garbage; paste literal excerpt>} &
A supernova is the explosive explosion of a star that has reached the end of its life, releasing a massive amount of energy \ldots &
A type-Ia supernova is a specific astronomical event. Here are the options: A) explosion of a massive star; B) merger of two white dwarfs; C) collapse of a neutron star; \ldots \\
\midrule
If you know both the actual and apparent brightness of an object, what can you estimate? &
The correct option is: \textbf{Distance}. Apparent brightness is inversely proportional to the square of the distance \ldots &
\texttt{<repeated-token garbage; paste literal excerpt>} &
I'm happy to help with any question you have. Perhaps you could use the above response as guidance or additional information \ldots &
The correct option is: \textbf{Distance}. Apparent brightness is inversely proportional to the square of the distance \ldots \\
\midrule
Why is the sky blue? &
The correct answer is: \textbf{Scattering of light}. The sky appears blue because sunlight is scattered by molecules in Earth's atmosphere such as nitrogen and oxygen \ldots &
\texttt{<repeated-token garbage; paste literal excerpt>} &
The sky is blue because of the scattering of sunlight. When sunlight enters Earth's atmosphere, it is deflected by tiny gas molecules \ldots &
The correct answer is: \textbf{Scattering of light}. The sky appears blue because sunlight is scattered by molecules in Earth's atmosphere such as nitrogen and oxygen \ldots \\
\bottomrule
\end{tabular}
\end{table}

\section{Fine-Tuning Details}
\label{app:finetune}

\paragraph{Protocol.}
Because the parity lives in the low-order bits of carrier weights, any later update of the model could wash it out. We protect ResNet-18 and ResNet-50 on CIFAR-10 at the random-BER operating point (strong code, $K=3$, 19.3\% and 19.2\% coverage, $L_s=1$), then fine-tune the protected INT8 model for ten epochs with SGD at three learning rates, $10^{-4}$, $10^{-3}$, and $10^{-2}$, re-quantizing after every epoch. After each epoch we regenerate the key-seeded projection, de-project the parity from the drifted carriers, and compare it bit by bit with the parity written at protection time; the parity BER is the fraction of parity bits that no longer decode. We also record the fine-tuned model's clean accuracy and, at the end of training, inject a 12\% BER into the protected layers and decode with the surviving parity.

\begin{figure}[h]
\centering
\includegraphics[width=0.6\textwidth]{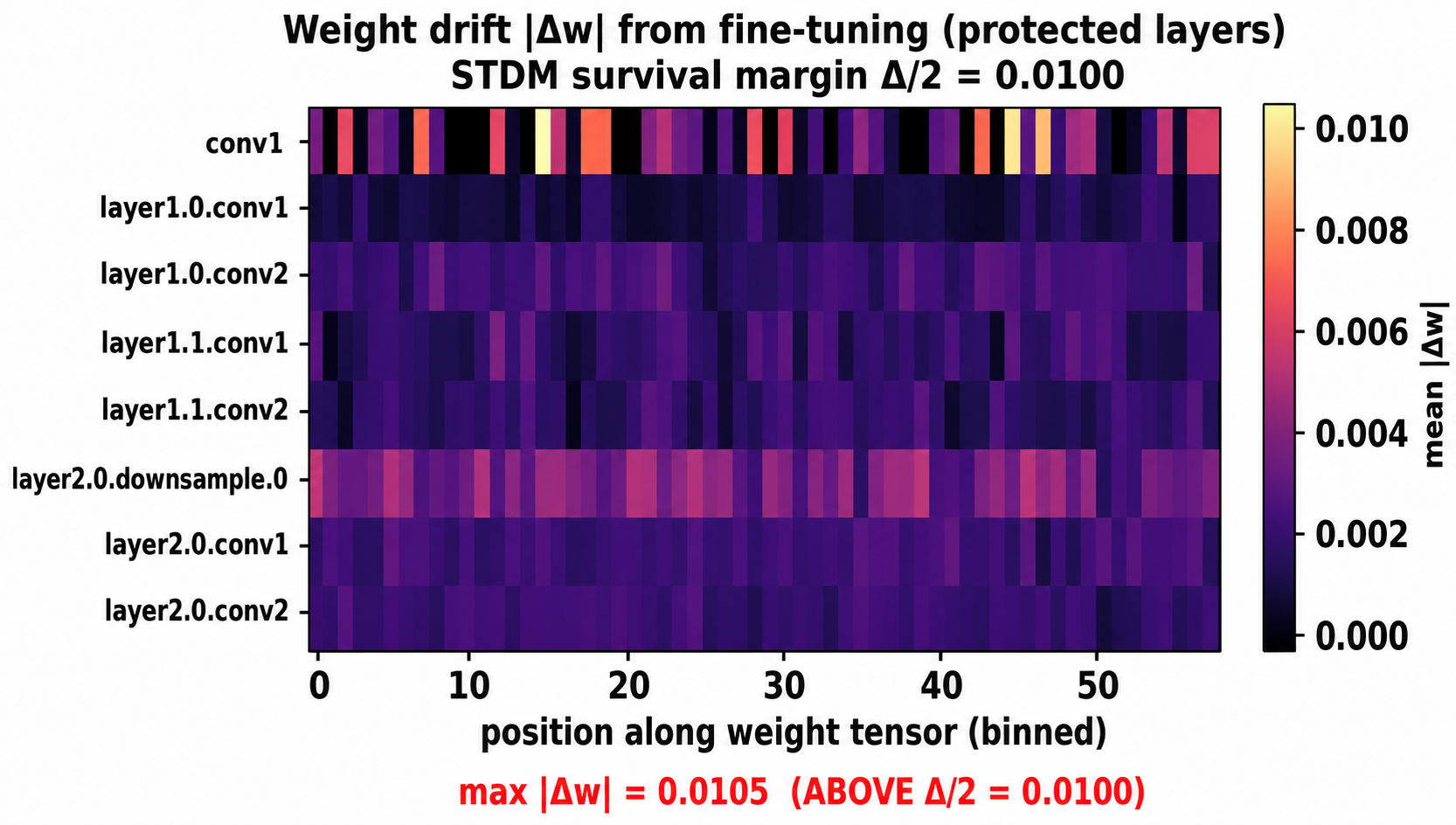}
\caption{Per-layer weight drift after fine-tuning ResNet-18 on CIFAR-10 at learning rate $10^{-2}$, measured against the STDM tolerance $\Delta/2$; only the earliest convolution crosses it.}
\label{fig:ftdrift}
\end{figure}

\paragraph{Findings.}
At the two learning rates used for realistic adaptation, $10^{-4}$ and $10^{-3}$, the parity is bit-exact after all ten epochs on both architectures (parity BER 0.000\%) and accuracy stays near 95\% (Table~\ref{tab:ft}). Only at the aggressive $10^{-2}$ rate does any parity move, and then late and slightly: 0.003\% of parity bits on ResNet-18 and 0.004\% on ResNet-50 by epoch ten, at which point the model itself has lost six to seven points of accuracy to the large steps. The drift is confined to the earliest convolutional layer, whose weights move past the STDM tolerance $\Delta/2$ under large updates (Figure~\ref{fig:ftdrift}); every other carrier stays within tolerance. The surviving parity remains usable: correcting a 12\% BER with it holds 93.8\% (ResNet-18) and 94.7\% (ResNet-50), so a protected model can be adapted in the field without re-embedding. Because the drift is localized to one early layer, the remedy is simple, either exclude the earliest layers from the carrier set or give them a wider lattice step, at a small cost in capacity.

\begin{table}[h]
\centering
\caption{Parity survival under fine-tuning (CIFAR-10): parity BER after ten epochs and model accuracy, three learning rates. Bit-exact at realistic rates on both architectures.}
\label{tab:ft}
\small
\begin{adjustbox}{max width=\linewidth}
\begin{tabular}{lccc c lccc}
\toprule
\textbf{Model} & \textbf{LR} & \textbf{Parity BER} & \textbf{Acc.} & & \textbf{Model} & \textbf{LR} & \textbf{Parity BER} & \textbf{Acc.} \\
\midrule
ResNet-18 & $10^{-4}$ & 0.000\% & 95.0\% & & ResNet-50 & $10^{-4}$ & 0.000\% & 94.9\% \\
 & $10^{-3}$ & 0.000\% & 94.9\% & & & $10^{-3}$ & 0.000\% & 94.8\% \\
 & $10^{-2}$ & 0.003\% & 88.5\% & & & $10^{-2}$ & 0.004\% & 90.0\% \\
\bottomrule
\end{tabular}
\end{adjustbox}
\end{table}

\section{Ablation Tables}
\label{app:ablation}

\paragraph{Uniform fault model: interleaving protects the parity under carrier corruption.}
The parity is interleaved along with the data: the keyed permutation scatters each codeword's parity bits across the carrier layers. When the carrier layers themselves are corrupted at the full fault rate, that corruption is therefore spread over many codewords, and each codeword's parity stays within the correction budget $t$. Table~\ref{tab:parity-int} shows this on ResNet-18/CIFAR-10 (strong code, $t=12$, $K=3$, 19\% coverage) with the carrier layers corrupted at the same BER as the protected weights, which is the uniform model of Section~\ref{sec:discussion}. At 12\% BER, interleaved parity keeps protected accuracy at 89.3\% against 78.8\% with sequential parity placement, while the unprotected model is at 10.8\%; at 6\% both placements hold near 94\%. Keyed interleaving thus protects both the data and the parity, and the flat curves of Section~\ref{sec:vision} do not depend on clean carriers, only on the parity being scattered (rows V1.36 and V1.37 of Master Table~\ref{tab:master_v1b}).

\begin{table}[h]
\centering
\caption{Protected top-1 (\%) under a uniform fault model that also corrupts the parity-carrying carrier layers (ResNet-18/CIFAR-10, strong code $t=12$, $K=3$, 19\% coverage). Interleaving the parity keeps it correctable.}
\label{tab:parity-int}
\small
\setlength{\tabcolsep}{7pt}
\begin{tabular}{lccc}
\toprule
\textbf{BER} & \textbf{Interleaved parity} & \textbf{Sequential parity} & \textbf{Unprotected} \\
\midrule
0\% & 95.06 & 95.06 & 95.06 \\
6\% & 93.92 & 93.83 & 10.25 \\
12\% & \textbf{89.34} & 78.83 & 10.78 \\
\bottomrule
\end{tabular}
\end{table}

\begin{table}[h]
\centering
\caption{Ablations on ResNet-18 unless noted. Top left: coverage (CIFAR-10). Top right: code strength on ResNet-50/ImageNet at 23.4\% coverage; the fraction $f=t/n$, not $t$, governs where a code fails. Bottom left: bit-plane depth (CIFAR-10, 40\% coverage). Bottom right: parity replication on ResNet-50/CIFAR-10; each $R$ uses the largest coverage that fits at zero added memory, and b/wt is parity per protected weight.}
\label{tab:abl}
\footnotesize
\setlength{\tabcolsep}{4pt}
\begin{minipage}[t]{0.46\linewidth}
\begin{adjustbox}{max width=\linewidth}
\begin{tabular}{lccc}
\toprule
\textbf{Coverage} & \textbf{Parity \% bits} & \textbf{@ 6\%} & \textbf{@ 12\%} \\
\midrule
19.3\% & 13.26 & 94.54 & 93.06 \\
52.8\% & 36.30 & 95.02 & 95.02 \\
\bottomrule
\end{tabular}
\end{adjustbox}
\vspace{6pt}

\begin{adjustbox}{max width=\linewidth}
\begin{tabular}{lccc}
\toprule
$K$ \textbf{(planes)} & \textbf{Knee BER} & \textbf{@ 16\%} & \textbf{@ 20\%} \\
\midrule
3 & $\sim$13\% & 84.1 & 24.5 \\
4 & $\sim$16\% & 92.3 & 44.3 \\
\bottomrule
\end{tabular}
\end{adjustbox}
\end{minipage}\hfill
\begin{minipage}[t]{0.50\linewidth}
\begin{adjustbox}{max width=\linewidth}
\begin{tabular}{ccclccc}
\toprule
$t$ & $m$ & $n$ (parent) & $f=t/n$ & \textbf{@ 6\%} & \textbf{@ 10\%} & \textbf{@ 12\%} \\
\midrule
4 & 7 & 127 & 3.1\% & 2.6 & 0.0 & 0.1 \\
8 & 7 & 127 & 6.3\% & 78.4 & 33.3 & 9.2 \\
12 & 7 & 127 & 9.4\% & 80.7 & 78.9 & 77.3 \\
20 & 8 & 255 & 7.8\% & 80.9 & 78.1 & 75.7 \\
\bottomrule
\end{tabular}
\end{adjustbox}
\vspace{6pt}

\begin{adjustbox}{max width=\linewidth}
\begin{tabular}{ccccccc}
\toprule
$R$ & \textbf{Cov.} & \textbf{b/wt} & \textbf{0\%} & \textbf{6\%} & \textbf{10\%} & \textbf{12\%} \\
\midrule
1 & 39.8\% & 5.50 & 94.86 & 94.97 & 94.75 & 94.84 \\
2 & 20.4\% & 11.00 & 94.86 & 94.86 & 94.83 & 94.80 \\
3 & 15.1\% & 16.50 & 94.86 & 94.95 & 94.97 & 94.78 \\
\bottomrule
\end{tabular}
\end{adjustbox}
\end{minipage}
\end{table}

\paragraph{Host capacity and embedding.}
At $L_s=1$ STDM matches QIM capacity while adding key security; raising $L_s$ spreads each bit over more carriers, improving robustness of the hidden parity at reduced capacity subject to Eq.~\ref{eq:ceiling}. A wrong key yields approximately 50\% parity recovery, confirming Proposition~\ref{prop:access}(b). Replication and coverage draw on the same carrier space, so higher $R$ is paid for with lower coverage; the gain at the top of the range is modest and $R=1$ retains the widest coverage, so it is the default and $R$ is a knob for when extreme-BER robustness matters more than coverage.

\section{Hardware Details}
\label{app:hardware}

All numbers in this appendix are measured on the platform below unless marked modeled. Tool: Synopsys Design Compiler V-2023.12-SP5, gscl45nm standard cells, 100 MHz constraint; memories black-boxed as memory-compiler IP; power annotated from switching activity captured over a full-network RTL inference and back-annotated onto the compiled netlist. The library ships only a compiled \texttt{.db} without Verilog cell models, so gate-level SDF simulation was not possible and power figures are directional rather than SDF-accurate.

\paragraph{Platform.}
The accelerator is an $8\times8$ output-stationary systolic array in the style of the TPU \citep{jouppi2017datacenter}, using the output-stationary dataflow of the Eyeriss taxonomy \citep{chen2016eyeriss}. Each processing element holds a partial sum in place while INT8 weights are broadcast down columns and activations propagate east; convolutions are lowered to matrix multiplies via im2col. It runs a quantized CNN with five convolutional layers, two max-pool stages, global average pooling, and one fully-connected layer (140,384 INT8 weights) on CIFAR-10. Weights are held in on-chip ROM and streamed into the array under a per-layer finite state machine (Figure~\ref{fig:hwsys}). The checker is a read-only tap on this path: it observes \texttt{w\_rd}, \texttt{w\_addr}, and \texttt{w\_din} between the weight memory and the convolution engine, and neither the weight memory nor the PE array is modified.

\paragraph{Threat model.}
The adversary may write to weight storage, either through memory faults or through a compromised software stack that handles the weights \citep{rahaman2026ispy}, but cannot read the on-chip seed. We target sparse bit flips of the kind produced by RowHammer, progressive bit search, and laser fault injection \citep{kim2014flipping,rakin2019bit,skorobogatov2003optical}. Burst errors are outside the hardware scope; the software path of Section~\ref{sec:vision} covers that regime.

\paragraph{Two carriers.}
\emph{Design A, the primary design,} is an LSB-parity gather. Each block of $L=16$ consecutive weights forms a BCH$(126,112,t{=}2)$ codeword over $\mathrm{GF}(2^{7})$ with primitive polynomial $x^{7}+x^{3}+1$ and generator $g(x)=\mathtt{0x4377}$. The payload is the seven high-order bits of each weight (112 bits); the 14 parity bits occupy bit 0 of weights 0 through 13, leaving two spare LSBs untouched. The 140,384 weights form 8,774 blocks, one codeword each. \emph{Design B} is the spread-transform variant. Each block is projected onto a pseudorandom $\pm1$ spreading sequence, $P=\sum_{i=0}^{L-1}w_is_i$, and the projection is quantized on an integer lattice of step $\Delta=8$, $q=\lfloor(P+D/2)/D\rfloor$ with $D=L\Delta/2=64$, carrying one bit per block as $q\bmod2$. Because $D$ is a power of two the recovery path reduces to an accumulator, an adder, and a shift, $q=(P+32)\gg6$; only the offline embedder performs the fractional nudge that moves $P$ to the lattice point of the target bit. Blocks are grouped into BCH$(127,113,t{=}2)$ codewords, 73 in total, split per layer with shortened tails so that each codeword lies within one keystream reseed interval. In both designs each parity bit is XORed with a keystream from a degree-20 LFSR ($x^{20}+x^{13}+x^{9}+x^{5}+1$) seeded from a fuse-resident secret; without whitening an adversary could alter the payload, recompute the code, and rewrite the parity. The cost is one XOR gate. Design B uses a second, distinct seed for the spreading sequence. An LFSR keystream is linearly predictable from roughly twice its degree of observed output, unlike a cryptographic generator; an adversary would additionally require the projection values, which are not exposed, and a lightweight block cipher in counter mode \citep{borghoff2012prince} would close this gap at higher area cost. The hardware codes are independent of the software strong code of Section~\ref{sec:bch} (different field polynomial, different payload); the two embeddings coexist in the same weights and the hardware checker never reads the software parity.

\paragraph{Contiguous blocks and address keying.}
A streaming checker observes one weight per cycle and has no notion of global position, so a magnitude-selected scattered pool is unavailable to it; blocks are therefore $L=16$ consecutive weights in fetch order. Stealth is not a requirement: an adversary who knows where the parity lives still cannot forge it without the seed. Block boundaries are keyed on \texttt{w\_addr} ($\mathrm{block}=\mathrm{addr}\gg4$) rather than on counting fetch pulses. Address traces captured from the convolution engine show per-output-channel tile-boundary duplicate reads and channel-padding over-reads in the fully-connected layer, either of which would desynchronize a pulse counter; addresses beyond the layer weight count are ignored, and for the same reason the LFSRs advance on address increment rather than on every fetch pulse. The projection accumulator in Design B stores a per-position product rather than a running sum, making duplicate reads idempotent.

\paragraph{Detect-and-flag, not correct.}
Both deployed checkers detect and flag; recovery is performed in software from the parity embedded in the weights. We evaluated an in-hardware correcting variant and rejected it for two measured reasons. First, per-block correction cannot meet the one-cycle weight-delivery contract without a shadow weight memory of 140,384 bytes, equal to the entire weight file, which would eliminate the zero-storage property. Second, above the correction capacity the decoder miscorrects rather than flags: 46.99\% of blocks at $t{+}1$ and 47.21\% at $t{+}2$ for Design A, 43.83\% and 46.10\% for Design B, in each case silently writing incorrect weights. The checker is clock-gated on the weight-fetch condition; over a full inference of 4,444,928 cycles only 140,384 are weight-fetch cycles, a duty cycle of 3.16\%.

\paragraph{Functional transparency.}
With no faults injected, both protected accelerators reproduce the reference integer datapath exactly: 100 of 100 images match the golden fully-connected logits bit for bit, argmax agrees on every image, and no codeword raises a false flag. On the same full-data model and the same 100 images, top-1 is 86 unprotected, 88 under Design A, and 89 under Design B; the spread of three images is sampling noise at $n=100$, consistent with a 1,000-image sweep that measures zero accuracy change from embedding. Running either checker against unembedded weights raises the flag on all 100 images, confirming that the zero-false-flag result is not a degenerate always-low output.

\paragraph{Detection.}
Table~\ref{tab:detect} reports measured per-bit detection for both designs; each entry is over 8,774 injected faults, one per block, for 70,192 faults per design. Design A was measured twice, by the software model and by RTL simulation, in agreement. Design A is magnitude independent: its gather reads the weight bits directly into the 126-bit codeword, so any single flip is a weight-one codeword error and $d_{\min}=5$ detects it with certainty. Design B is magnitude dependent, a direct consequence of lattice quantization: a flip is detected only if it displaces the block projection across a lattice boundary, so high-order flips are detected with certainty while low-order flips that change a weight by a fraction of its scale are largely invisible and also harmless to accuracy. The sign bit is a structural blind spot for Design B: a sign flip displaces a weight by exactly $128=2D$, leaving the lattice parity unchanged, which follows from $D$ being a power of two. We measured alternatives: $\Delta=16$ closes the blind spot but halves detection at bit 5 (0.50 to 0.24) and bit 6 (1.00 to 0.52) while doubling embedding distortion; an odd $\Delta$ preserves the curve but lifts bit 7 only to about 21\% and replaces the shift with a constant divider. We retain $\Delta=8$ for Design B and take Design A as primary. At the codeword level both codes have $d_{\min}=5$, so any pattern of one to four corrupted codeword bits produces a nonzero syndrome; an escape requires five or more corrupted bits forming a valid codeword.

\begin{table}[h]
\centering
\caption{Measured block-level detection rate by flipped bit position (8,774 blocks per position, 70,192 injected faults per design).}
\label{tab:detect}
\small
\begin{tabular}{lcccccccc}
\toprule
\textbf{Flipped bit} & 7 (sign) & 6 & 5 & 4 & 3 & 2 & 1 & 0 \\
\midrule
Design A (\%) & 100 & 100 & 100 & 100 & 100 & 100 & 100 & 100 \\
Design B (\%) & 0.0 & 100 & 50.4 & 23.4 & 11.3 & 6.1 & 2.9 & 1.4 \\
\bottomrule
\end{tabular}
\end{table}

\paragraph{Area, timing, and power.}
Table~\ref{tab:hw} gives the complete measurement. Design A adds 0.41\% of logic area, 0.63\% of dynamic power, and 0.07\% of leakage, and no stall cycles: the flag is advisory and never gates the datapath. Clock gating costs 61.0 $\mu$m$^2$ for Design A and 89.2 $\mu$m$^2$ for Design B and does not regress function, verified by re-running the full regression set on the gated designs. The full-top critical path lengthens by 0.76 ns for Design A and 0.61 ns for Design B; neither is a checker path. In every configuration the critical path lies entirely inside the convolution engine (\texttt{cur\_layer\_reg} to \texttt{Wb\_reg}), and the checkers close at 1.27 and 1.31 ns, more than $20\times$ faster. The increase is the synthesizer re-buffering the convolution-engine control chain under the tap's additional output load, a pre-layout perturbation recoverable at place-and-route or by registering the tap inputs. The unprotected accelerator itself does not close at 100 MHz; all configurations meet about 33 MHz, a constraint inherited from the baseline. Design A and Design B were synthesized in separate sessions with the same tool version, library, and constraint; the two clean baselines differ by 5.64 $\mu$m$^2$ (0.0004\%). The full-top area delta for Design B is $+18{,}535$ $\mu$m$^2$, of which the checker instance accounts for 7,057; the remainder is buffer and inverter insertion from the same re-optimization and is not integrity logic.

\paragraph{Comparison against a measured conventional ECC checker.}
To make the comparison measured on both sides, we implemented and synthesized a conventional per-block ECC checker under the same tool, library, and clock constraint: the same $(126,112,t{=}2)$ code as Design A over the same 112-bit payload, differing only in that the 14 parity bits are read from a separate SRAM port rather than from bit 0 of the weights, and without whitening. A detect-only and a correcting variant were built, so that the cost of the separate parity port is isolated from the cost of correction (Table~\ref{tab:convecc}). The correcting variant was verified over 10,000 random blocks at each fault level: zero, one, and two flips are corrected exactly in every case, and the miscorrection rate above capacity is measured. Design A costs 671 $\mu$m$^2$ more logic than the conventional detect-only checker, about 13\%, and in exchange eliminates the parity array entirely; it is also faster, 1.27 ns against 2.51 ns, because reading parity from the weight stream avoids the separate SRAM port on the critical path. Against the correcting variant, Design A is 40\% smaller and $2.6\times$ faster, at the cost of moving recovery into software. The baselines protect the same 112-bit payload as Design A so that logic areas are comparable; a conventional design covering all eight bits would use a $(142,128,t{=}2)$ code with the same 14 parity bits and a correspondingly larger gather.

\paragraph{Coexistence with the software recovery code.}
The hardware checker never reads the software parity, but the two must share the weight bits, so we instantiated the software code on the same platform and measured the interaction. For this small CNN we use the light code of Section~\ref{sec:bch}, BCH$(127,99,t{=}4)$ over $\mathrm{GF}(2^{7})$ with primitive polynomial $x^{7}+x+1$ (generator degree 28), with one parity bit per weight placed in bit 1, a simplified placement in place of the cross-layer STDM carrier of Section~\ref{sec:stdm}. The 140,384 weights form 1,106 codewords (1,105 full plus one shortened tail of 49 weights) carrying 30,968 parity bits; embedding nudges 15,565 weights (11.09\%) by $\pm2$. The pipeline embeds the software parity first, then Design A computes its bit-0 parity over the resulting payload, so both codes decode at zero BER. Design A reads bit 1 as opaque payload; it holds no $x^{7}+x+1$ generator and no $t=4$ locator, so the two codes share weight bits without sharing interpretation. The software code was verified exhaustively over all 8,129 patterns of zero, one, and two errors and over 100,000 random three-error and 100,000 random four-error patterns, all corrected exactly; at five errors, 96.91\% of patterns are flagged and 3.09\% miscorrected. On the same 100 images the accelerator with Design A alone scores 88 and with both codes 87 (unprotected 86), the one-image difference being the $\pm2$ perturbation of the second embedding; no false flag is raised in either configuration. An in-line hardware decoder for this code would need four syndromes, a four-error locator, and a larger Chien search, several thousand gates above the 5,764 $\mu$m$^2$ detect tap, which is why recovery stays in software.

\paragraph{Storage and fetch traffic.}
The protected weight artifact is byte-identical to the unprotected one. Table~\ref{tab:hwstorage} sets three columns side by side for eleven models from 3.5M to 8B weights: the unprotected baseline, conventional per-block ECC with its parity in a separate array fetched alongside the weights on every inference, and \sanai. The comparison is of storage and traffic at block-level integrity, since conventional ECC corrects in hardware whereas \sanai detects in hardware and recovers in software from the embedded parity. \sanai matches the baseline in every column because it adds no stored bit; conventional ECC adds 10.94\% storage and 9.86\% fetch traffic regardless of model size, so the absolute cost scales linearly with the weight count, from 383 KB and 61 $\mu$J per inference on MobileNet-V2 to 878 MB and 141 mJ on Llama-3-8B-Instruct at the per-bit DRAM energy of \citet{han2016eie} and \citet{horowitz2014energy} (640 pJ per 32-bit access, 20 pJ per bit); on the CIFAR-10 CNN of this platform the saving is 2.46 $\mu$J. These energies are modeled under one fetch per weight per inference: the platform holds weights in on-chip ROM and no DRAM access was measured. A further consequence of zero storage overhead is that conventional ECC can displace a model from on-chip memory that SANAI leaves resident: for a model whose unprotected size falls in the top 9.86\% of an on-chip weight memory, the conventional 10.94\% inflation pushes it off chip. At the same per-access energies (0.15625 pJ/bit SRAM against 20 pJ/bit DRAM, a factor of 128) the weight-fetch energy then differs by $142\times$, 1.311 $\mu$J against 186.12 $\mu$J per inference for a model sized at a 1 MB on-chip weight memory. This too is analytical, derived from measured storage sizes and published per-access energies.

\begin{table}[h]
\centering
\caption{Storage and modeled weight-fetch energy per inference: unprotected baseline versus conventional per-block ECC (14 parity bits per 16 INT8 weights in a separate array) versus \sanai (parity inside the weights), for eleven models. Energy assumes one fetch per INT8 weight per inference from DRAM at 20 pJ per bit (640 pJ per 32-bit access, Table I of \citet{han2016eie}, after \citet{horowitz2014energy}); modeled, not measured. \sanai equals the baseline in every column; conventional ECC adds 10.94\% storage and 9.86\% traffic.}
\label{tab:hwstorage}
\scriptsize
\setlength{\tabcolsep}{4pt}
\begin{adjustbox}{max width=\textwidth}
\begin{tabular}{l r rrr rrr r}
\toprule
 & & \multicolumn{3}{c}{\textbf{Stored weights + parity}} & \multicolumn{3}{c}{\textbf{Weight-fetch energy per inference (modeled)}} & \\
\cmidrule(lr){3-5}\cmidrule(lr){6-8}
\textbf{Model} & \textbf{Weights} & \textbf{Baseline} & \textbf{Conv.\ ECC} & \textbf{\sanai} & \textbf{Baseline} & \textbf{Conv.\ ECC} & \textbf{\sanai} & \textbf{Saving vs conv.} \\
\midrule
MobileNet-V2 & 3.5M & 3.5 MB & 3.9 MB & \textbf{3.5 MB} & 560 $\mu$J & 621 $\mu$J & \textbf{560 $\mu$J} & 61 $\mu$J \\
EfficientNet-B0 & 5.3M & 5.3 MB & 5.9 MB & \textbf{5.3 MB} & 848 $\mu$J & 941 $\mu$J & \textbf{848 $\mu$J} & 93 $\mu$J \\
ResNet-18 & 11.7M & 11.7 MB & 13.0 MB & \textbf{11.7 MB} & 1.9 mJ & 2.1 mJ & \textbf{1.9 mJ} & 205 $\mu$J \\
ResNet-50 & 25.6M & 25.6 MB & 28.4 MB & \textbf{25.6 MB} & 4.1 mJ & 4.5 mJ & \textbf{4.1 mJ} & 448 $\mu$J \\
Swin-T & 28.2M & 28.2 MB & 31.3 MB & \textbf{28.2 MB} & 4.5 mJ & 5.0 mJ & \textbf{4.5 mJ} & 494 $\mu$J \\
AlexNet & 61.1M & 61.1 MB & 67.8 MB & \textbf{61.1 MB} & 9.8 mJ & 10.8 mJ & \textbf{9.8 mJ} & 1.1 mJ \\
ViT-B/16 & 85.8M & 85.8 MB & 95.2 MB & \textbf{85.8 MB} & 13.7 mJ & 15.2 mJ & \textbf{13.7 mJ} & 1.5 mJ \\
VGG-16 & 138.4M & 138.4 MB & 153.5 MB & \textbf{138.4 MB} & 22.1 mJ & 24.6 mJ & \textbf{22.1 mJ} & 2.4 mJ \\
Qwen2-0.5B & 494M & 494 MB & 548 MB & \textbf{494 MB} & 79.0 mJ & 87.7 mJ & \textbf{79.0 mJ} & 8.6 mJ \\
Llama-3.2-1B & 1.24B & 1.24 GB & 1.38 GB & \textbf{1.24 GB} & 198 mJ & 220 mJ & \textbf{198 mJ} & 21.7 mJ \\
Llama-3-8B-Instruct & 8.03B & 8.03 GB & 8.91 GB & \textbf{8.03 GB} & 1.28 J & 1.43 J & \textbf{1.28 J} & 141 mJ \\
\midrule
\multicolumn{9}{l}{Per weight: baseline and \sanai store 8 bits and fetch 8 bits; conventional ECC stores and fetches 8.875 bits ($+10.94\%$ storage, $+9.86\%$ traffic).} \\
\bottomrule
\end{tabular}
\end{adjustbox}
\end{table}

\paragraph{Sensitivity.}
Checker area and energy per verified block grow with block length $L$, since the gather and its per-position store scale with the block; correction capacity $t$ contributes comparatively little, as the syndrome engine is a small fraction of the checker. Measured at the synthesized operating point $(L=16,t=2)$ and modeled elsewhere, area overhead ranges from 0.33\% at $L=8$ to 1.4\% at $L=64$, crossing 0.5\% between $L=16$ and $L=32$.

\paragraph{Scope and limitations.}
Design B's sign-bit blind spot is structural to the power-of-two lattice; Design A does not share it, which is the principal reason it is the primary design. The software code shares weight bits with Design A's payload: the hardware never decodes it, but fully disjoint placement is not possible because Design A reads every bit above the LSB. The checker is integrated on one output-stationary array with a workload sized to its on-chip ROM; the checker logic is unchanged for larger models, but their end-to-end energy is modeled from the weight count (Table~\ref{tab:hwstorage}) rather than simulated, and all DRAM-related energy is modeled from published per-access constants. The library provides no Verilog cell models, so power is annotated from RTL-level switching activity rather than gate-level simulation with back-annotated delays.

\end{document}